\documentclass[twoside,twocolumn,9pt]{article}
\usepackage{extsizes}
\usepackage[super,sort&compress,comma]{natbib} 
\usepackage[version=3]{mhchem}
\usepackage[left=1.5cm, right=1.5cm, top=1.785cm, bottom=2.0cm]{geometry}
\usepackage{balance}
\usepackage{times,mathptmx}
\usepackage{sectsty}
\usepackage{graphicx} 
\usepackage{lastpage}
\usepackage[format=plain,justification=justified,singlelinecheck=false,font={stretch=1.125,small,sf},labelfont=bf,labelsep=space]{caption}
\usepackage{float}
\usepackage{fancyhdr}
\usepackage{fnpos}
\usepackage[english]{babel}
\addto{\captionsenglish}{%
  
}
\usepackage{array}
\usepackage{droidsans}
\usepackage{charter}
\usepackage[T1]{fontenc}
\usepackage[usenames,dvipsnames]{xcolor}
\usepackage{setspace}
\usepackage[compact]{titlesec}

\usepackage{epstopdf}

\definecolor{cream}{RGB}{222,217,201}
\usepackage{multirow}
\usepackage{hyperref}
\hypersetup{
	colorlinks=true,
	linkcolor=blue,
	filecolor=magenta,      
	urlcolor=blue,
	citecolor=blue
}
\usepackage[normalem]{ulem}

\newcommand{\rate}[2]{$#1\times 10^{#2}$~s$^{-1}$}

\begin{document}

\pagestyle{fancy}
\thispagestyle{plain}
\fancypagestyle{plain}{

\fancyhead[C]{\includegraphics[width=18.5cm]{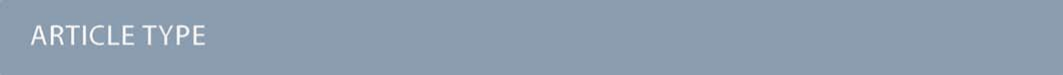}}
\fancyhead[L]{\hspace{0cm}\vspace{1.5cm}\includegraphics[height=30pt]{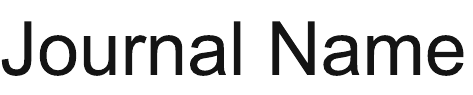}}
\fancyhead[R]{\hspace{0cm}\vspace{1.7cm}\includegraphics[height=55pt]{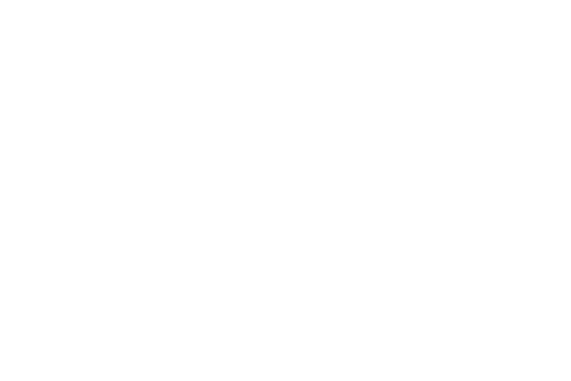}}
\renewcommand{\headrulewidth}{0pt}
}

\makeFNbottom
\makeatletter
\renewcommand\LARGE{\@setfontsize\LARGE{15pt}{17}}
\renewcommand\Large{\@setfontsize\Large{12pt}{14}}
\renewcommand\large{\@setfontsize\large{10pt}{12}}
\renewcommand\footnotesize{\@setfontsize\footnotesize{7pt}{10}}
\makeatother

\renewcommand{\thefootnote}{\fnsymbol{footnote}}
\renewcommand\footnoterule{\vspace*{1pt}%
\color{cream}\hrule width 3.5in height 0.4pt \color{black}\vspace*{5pt}} 
\setcounter{secnumdepth}{5}

\makeatletter 
\renewcommand\@biblabel[1]{#1}            
\renewcommand\@makefntext[1]%
{\noindent\makebox[0pt][r]{\@thefnmark\,}#1}
\makeatother 
\renewcommand{\figurename}{\small{Fig.}~}
\sectionfont{\sffamily\Large}
\subsectionfont{\normalsize}
\subsubsectionfont{\bf}
\setstretch{1.125} 
\setlength{\skip\footins}{0.8cm}
\setlength{\footnotesep}{0.25cm}
\setlength{\jot}{10pt}
\titlespacing*{\section}{0pt}{4pt}{4pt}
\titlespacing*{\subsection}{0pt}{15pt}{1pt}

\fancyfoot{}
\fancyfoot[LO,RE]{\vspace{-7.1pt}\includegraphics[height=9pt]{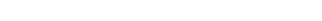}}
\fancyfoot[CO]{\vspace{-7.1pt}\hspace{13.2cm}\includegraphics{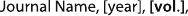}}
\fancyfoot[CE]{\vspace{-7.2pt}\hspace{-14.2cm}\includegraphics{RF}}
\fancyfoot[RO]{\footnotesize{\sffamily{1--\pageref{LastPage} ~\textbar  \hspace{2pt}\thepage}}}
\fancyfoot[LE]{\footnotesize{\sffamily{\thepage~\textbar\hspace{3.45cm} 1--\pageref{LastPage}}}}
\fancyhead{}
\renewcommand{\headrulewidth}{0pt} 
\renewcommand{\footrulewidth}{0pt}
\setlength{\arrayrulewidth}{1pt}
\setlength{\columnsep}{6.5mm}
\setlength\bibsep{1pt}

\makeatletter 
\newlength{\figrulesep} 
\setlength{\figrulesep}{0.5\textfloatsep} 

\newcommand{\topfigrule}{\vspace*{-1pt}%
\noindent{\color{cream}\rule[-\figrulesep]{\columnwidth}{1.5pt}} }

\newcommand{\botfigrule}{\vspace*{-2pt}%
\noindent{\color{cream}\rule[\figrulesep]{\columnwidth}{1.5pt}} }

\newcommand{\dblfigrule}{\vspace*{-1pt}%
\noindent{\color{cream}\rule[-\figrulesep]{\textwidth}{1.5pt}} }

\makeatother

\twocolumn[
  \begin{@twocolumnfalse}
\vspace{3cm}
\sffamily
\begin{tabular}{m{4.5cm} p{13.5cm} }

\includegraphics{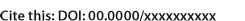} & \noindent\LARGE{\textbf{Nitrogen-vacancy centre in lonsdaleite: a novel nanoscale sensor?$^\dag$}} \\
\vspace{0.3cm} & \vspace{0.3cm} \\

 & \noindent\large{Anjay Manian,$^{\ast}$\textit{$^{a}$}\textit{$^{b}$}\textit{$^{c}$}\textit{$^{d}$} Mitchell O. de Vries,\textit{$^{e}$}\textit{$^{f}$} Daniel Stavrevski,\textit{$^{e}$} Qiang Sun,\textit{$^{e}$} Salvy P. Russo,\textit{$^{a}$} and Andrew D. Greentree $^{\ast}$\textit{$^{e}$}\textit{$^{g}$}} \\

\includegraphics{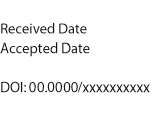} & \noindent\normalsize{Hexagonal diamond, often called lonsdaleite, is an exotic allotrope of carbon, predicted to be harder than cubic (conventional) diamond with a wider bandgap. Due to its pure sp$^3$ bonded lattice, it should be expected to host sub-bandgap defect centres (colour centres).  Here we perform \textit{ab initio} modeling of nitrogen-vacancy (NV) colour centres in hexagonal diamond nanocrystals; for both the neutral and negatively charged species (NV$^0$ and NV$^-$). We identify three distinct configurations for the NV center: two of which are analogous to NV in diamond, and one which is a configuration that can only exist in the hexagonal form. The diamond-like NV systems comprise three symmetry equivalent centers which reside on the same carbon plane, and one defect that is split across two planes and replaces a carbon-carbon bond. There is an additional NV centre where the N and V each have four nearest neighbour carbon atoms. The presence of this latter configuration would provide an unambiguous determination of the hexagonal nature of lonsdaleite. Quantum chemical analysis show all derivatives to be thermochemically stable, and each with their own unique photophysical properties, spectral profiles, and magneto-optical characteristics. By assuming that the ground state properties of the NV$^-$ in hexagonal diamond are comparable to those of NV$^-$ in cubic diamond, albeit with increased strain, we predict ground state fine structure splitting for two of the centres of 2.74~GHz and 4.56~MHz, compared with 2.87~GHz for cubic diamond. The possibility of optically detected magnetic resonance with NV$^-$ in lonsdaleite would provide a new carbon-based quantum sensing system, and an unambiguous method to resolve outstanding issues around the structure of lonsdaleite as hexagonal diamond.} 

\end{tabular}

 \end{@twocolumnfalse} \vspace{0.6cm}

  ]

\renewcommand*\rmdefault{bch}\normalfont\upshape
\rmfamily
\section*{}
\vspace{-1cm}


\footnotetext{\textit{$^{a}$~ARC Centre of Excellence in Exciton Science, School of Science, RMIT, Australia, 3001.}}
\footnotetext{\textit{$^{b}$~School of Science and Molecular Horizons, University of Wollongong, Australia, 2522.}}
\footnotetext{\textit{$^{c}$~ARC Centre of Excellence in Quantum Biotechnology, University of Wollongong, Australia, 2522.}}
\footnotetext{\textit{$^{d}$~amanian@uow.edu.au}}
\footnotetext{\textit{$^{e}$~ARC Centre of Nanoscale Biophotonics, School of Science, RMIT, Australia, 3001}}
\footnotetext{\textit{$^{f}$~Quantum Machines Unit, Okinawa Institute of Science and Technology Graduate University, Onna, Okinawa 904-0495, Japan.}}
\footnotetext{\textit{$^{g}$~andrew.greentree@rmit.edu.au}}

\footnotetext{\dag~Electronic Supplementary Information (ESI) available: [Electronic supplementary information contains further details as to how the nanocrystals are defined, a brief exploration of any possible basis set dependence, examination into the effects of hydrogen contamination, tabulated vertical excitation energies and their corresponding configuration state functions, visualised $\alpha$, $\beta$, and electron-paired natural transition orbitals, and the optimised coordinates for in-plane diamond, and each of the three lonsdaleite derivatives studied in this work, within the frozen hydrogen approximation.]. See DOI: 00.0000/00000000.}



\section{Introduction}
Optically active defects in wide bandgap semiconductors are emerging as one of the most exciting frontiers for quantum device physics\cite{uwbsc,ADSIWBGSC,qdbd}. The most commonly discussed solid-state host for these centres is diamond\cite{qcfnswdnvc,mcfqtboccid}, although there is also significant interest in other hosts including hexagonal boron nitride\cite{hroiaduqeihbn,qsaiwsdihbn}, silicon carbide \cite{deVries2021,Castelletto2020,Lohrmann2017}, alumina\cite{cdidciaaweosl}, zinc oxide \cite{MGK+2012} and titania\cite{sidcotfca}.

The most well studied solid-state defect centre for quantum applications, is the negatively-charged nitrogen-vacancy (NV$^-$) colour centre in diamond \cite{DMD+2013}.  This is due to long-lived spin lifetimes at room temperature, convenient optical initialisation and readout \cite{JGP+2004}, and sensitivity to magnetic \cite{TCC+2008}, electric \cite{DFD+2011}, thermal \cite{ABL+2010}, and strain fields \cite{OLM+2014}. See also \citenum{AGP2011} \& \citenum{BSB+2020} for reviews of NV diamond based quantum technology. 

\begin{figure*}[tb]
    \centering
    \includegraphics[width=\textwidth]{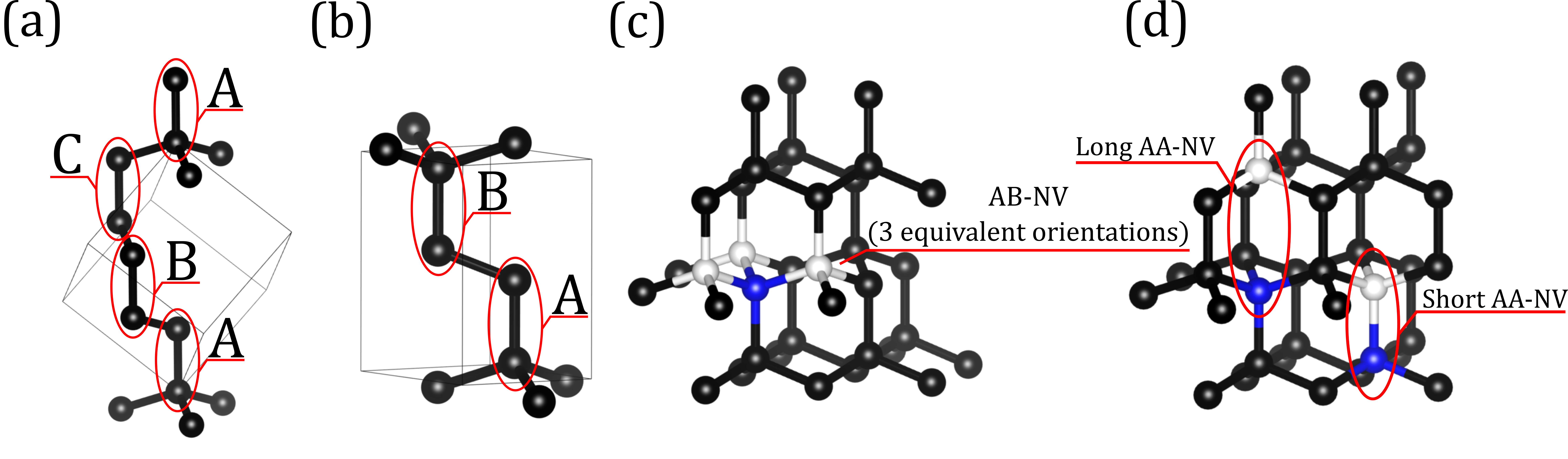}
    \caption{Schematic representations of the stacking configurations for (a) diamond and (b) lonsdaleite, with differing stacking sites labeled. The different NV species in lonsdaleite are also shown, with carbon, nitrogen, and vacancies shown in black, blue, and white respectively. Bonds between vacancies and atoms are preserved to permit easier visualisation. (c) Three symmetry-equivalent defects configurations, AB-NV, where the NV defect exists between A and B stacking sites. (d) Two unique AA-NV defect configurations, where both the N and V exist in the same stacking site. We refer to these as long (left) and short (right) AA-NV.}
    \label{imageDefectSchematic}
\end{figure*}

Lonsdaleite, or hexagonal diamond, is an exciting allotrope of carbon. Discovered in meteorites in 1967 \cite{FM67} often as nanocomposites with nanodiamond\cite{lifatcdadneaadm}, it is now beginning to be manufactured \cite{MWS+2020,HSK+2002,BBS+1995,BP67}, including methods for conversion from graphite to lonsdaleite \cite{gafshdbhpgp,nfodalbscog,ddspaa,lfiporptdg}. and there is hope that in the near future, large scale quantities of hexagonal diamond will be available for research and device applications\cite{lfwnt,soathdwc,nhdffgc}.

The properties of hexagonal diamond are extreme. Like cubic diamond it is an all sp$^3$ bonded material. It is predicted to be harder than cubic diamond\cite{hdanfoc,soathdwc} due to the shorter bond length, and to have a wider band gap\cite{coebohacd,lamsadtd}. So far it has proven difficult to experimentally confirm such predictions due to the lack of availability of large sizes and large quantities of lonsdaleite, and indeed this has led to controversy over the identification of lonsdaleite as hexagonal diamond\cite{lahpod,scondst60gpa,Nemeth2014,SMS2015}. However, it remains interesting to consider whether lonsdaleite may \textit{also} be a host for quantum defects, and in particular the NV centre, due to their superficial similarities between hexagonal and cubic diamond. If so, would such defects not be useful for quantum applications?

For the purposes of this study, we will assume that lonsdaleite is indeed hexagonal  diamond.  Recently, we have found evidence for a photo-stable colour centre in naturally formed meteoritic lonsdaleite\cite{Stavrevski2024}. Sun \textit{et al.}\cite{faothd110sfnvbqs} performed the first modeling of a nitrogen-vacancy (NV) centre in lonsdaleite, and significantly their model predicts stability of the negatively charged state as close as 12~\AA{}, indicating that it may well be superior to NV in diamond. More recently, Abdelghafar \textit{et al.}\cite{aDFTsonvcil,ACA2025} report on the difference between neutral and negatively charged NV centers (NV$^0$ and NV$^-$) and report on splitting in the zero-phonon line due to a shift in symmetry. 

Here we employ density functional theory to model the NV centre in hexagonal diamond. Employing the \textsc{Orca} software package\cite{ORCA,ORCA1}, our modelling demonstrates thermodynamic stability of three nonequivalent NV centre subspecies: three equivalent forms in the hexagonal plane, and two nonequivalent cross-plane centres. By comparing the bond length and derived properties of these centres with NV in cubic diamond (c-NV), we predict crystal field splittings of 407~THz, 290~THz and 261~THz for the three inequivalent NV$^-$ orientations, as well as spontaneous emission lifetimes of 7.74~$\mu$s, 64~$\mu$s, and 442~ns, and zero-phonon line wavelengths of 670~nm, 1078~nm, and 1016~nm. NV$^0$ was modelled alongside NV$^-$, with predicted crystal field splittings of 541~THz, 487~THz and 348~THz, alongside spontaneous emission lifetimes of 416~ns, 14~$\mu$s, and 907~ns, and zero-phonon line wavelengths of 500~nm, 656~nm, and 617~nm. These results should provide guidance for future experimental formation and investigations of NV in lonsdaleite.  The presence of three nonequivalent NV orientations is significant for vector magnetometry, where the degeneracy between the possible orientations is lifted even at zero magnetic field. This is different from c-NV where all orientations are equivalent at zero magnetic field.  Such results may be useful for more precise vector magnetic field determinations as well as ameliorating effects of confounding signatures, for example temperature effects.

This paper is organised as follows. We first introduce our notation for the NV colour centres. Following a presentation of the modeling techniques used throughout this work, we analyse the four cubic-diamond-like NV centres, as well as the emergent, additional NV centre. We study both the NV$^0$ and NV$^-$ charge states. Resolution of spectral properties and thermodynamic stability is included in this analysis. We then draw analogies between hexagonal NV and cubic NV centres to make preliminary predictions of the suitability of NV in lonsdaleite for quantum sensing applications.

\section{Crystallographic notation for NV centres}

The usual form of diamond is packed in a cubic (ABC) stacking configuration (Figure~\ref{imageDefectSchematic}(a)), whereas lonsdaleite displays hexagonal (AB) stacking (Figure~\ref{imageDefectSchematic}(b)). This AB stacking results in each plane of adjacent carbon atoms being shifted above or below the mean by between 22-36~pm. The neighbouring planes therefore alternate in cross-plane distance, either $155\pm 5$~pm at its shortest or $255\pm 5$~pm at its longest. This alternating cross-plane distance is not seen in cubic stacking. 

To permit the easy discussion of colour centres in lonsdaleite, it is necessary to propose a notation to conveniently describe the structure of colour centres within the crystal lattice. Derived from the standard notation used for colour centres in SiC allotropes, we propose to notate the stacking sites a colour centre inhabits using the form WX-YZ. In this notation, W and X are stacking sites (either AA or AB, with BB being identical to AA), and YZ being the colour centres constituent components. For more complex colour centres comprised of more than two components, the notation could be extended to more accurately denote the colour centre's lattice structure in lonsdaleite. 

The NV centre is defined by a substitutional nitrogen adjacent to a carbon vacancy in the lattice.  In cubic diamond this requires the nitrogen and vacancy to be nearest neighbours with all possible configurations of adjacent sites being equivalent. This leads to a separation between N and V of 154~pm. However, due to the cross-plane distance variation seen in lonsdaleite, this is not preserved with five possible configurations. Of these, there are three symmetry-equivalent forms with N and V being situated in neighbouring A and B stacking sites (AB-NV, Figure~\ref{imageDefectSchematic}(c)). The remaining two forms are formed of N and V located in neighbouring A stacking sites (AA-NV, Figure~\ref{imageDefectSchematic}(d)). One of the AA-NV forms, short AA-NV, is formed where the N and V are each adjacent to three carbon atoms and is similar in structure to the c-NV seen in diamond.  The final form, which we term the long AA-NV, is found when the adjacent N and V are each bonded to four neighbouring carbon atoms.  Despite this change in bonding, the negatively charged long AA-NV is also a two-hole complex, similar to conventional NV, hence we expect the properties of long AA-NV to be similar to those of the other NV forms.  However we stress that this form is a novel form that can only emerge with hexagonal packing and is not present in cubic diamond. 

To summarise the NV notation:
\begin{itemize}
    \item \textit{AB-NV} for the three equivalent defects, where the N and V occupy neighbouring A and B stacking sites.
    \item \textit{Short AA-NV} where the N and V are adjacent and occupy the same stacking site. N and V each have three adjacent carbon atoms.
    \item \textit{Long AA-NV} where the N and V are adjacent in the same stacking site, but are not nearest neighbours (that is, there is no shared bond location between them). N and V each have four neighbour carbon atoms.
\end{itemize}

Importantly, none of the above mentioned studies\cite{atobeid,faothd110sfnvbqs,aDFTsonvcil} noted the existence of the long AA-NV defect. Therefore, an important question in studying lonsdaleite is whether one NV species is more favourable than the other, and how each species could be applied to specific applications, where one conformation may be more beneficial than another for certain technologies.

\section{Computational Details}
To study these systems in detail and to determine whether any of configurations is more thermodynamically stable, we require an appropriate first principles method. Similar works in the literature adopt simple electronic calculations for identification or characterisation, such as Tran \textit{et al.}\cite{qefhbnm} who looked at boron nitride monolayers, or by Alkauskas \textit{et al.}\cite{fpcolslsfdis} on gallium nitride and zinc oxide defects. Importantly, the latter work incorporates important vibrational contributions to the photoluminescence lineshape, however is only accurate in cases where initial and final states are of different symmetry\cite{dftbthkt}. 

Study of the lonsdaleite NV defect was carried out using a similar methodology to Karim \textit{et al.}\cite{aaiesspbfcocc,baipoNV+id}, 
in which cubic diamond nanocrystals containing NV$^+$ and NV$^-$ centres were defined in a purely quantum chemical nature to show quantum emission in both species. This methodology was recently employed on the NVN centre in diamond\cite{tnvncc} yielding very positive results. Here, an \textit{ab initio} model of an atomic lonsdaleite nanocrystal is generated to probe the chemical conformation of 3 species: the AB defect, and both AA defects. For each defect, a lonsdaleite nanocrystal is designed such that the defect is at its centre (see ESI for further details). We ensure that there are at least 2 carbon rings in the same plane as the NV defect, and 1 plane above and below. This means the typical "in-plane" lonsdaleite structure has less atoms than that of the "cross-plane" species. The largest crystal has 220 carbons in it, and can therefore be likened to simulations within the cluster regime\cite{sqditcr}. With such a (relatively) small crystal design, one could argue that surface effects would interfere with important NV physical properties, however the system size remains a rigid computational constraint in any \textit{ab initio} model. However, steps can be taken to minimise these artefacts. 

\begin{figure*}[tb]
    \centering
    \includegraphics[width=\linewidth]{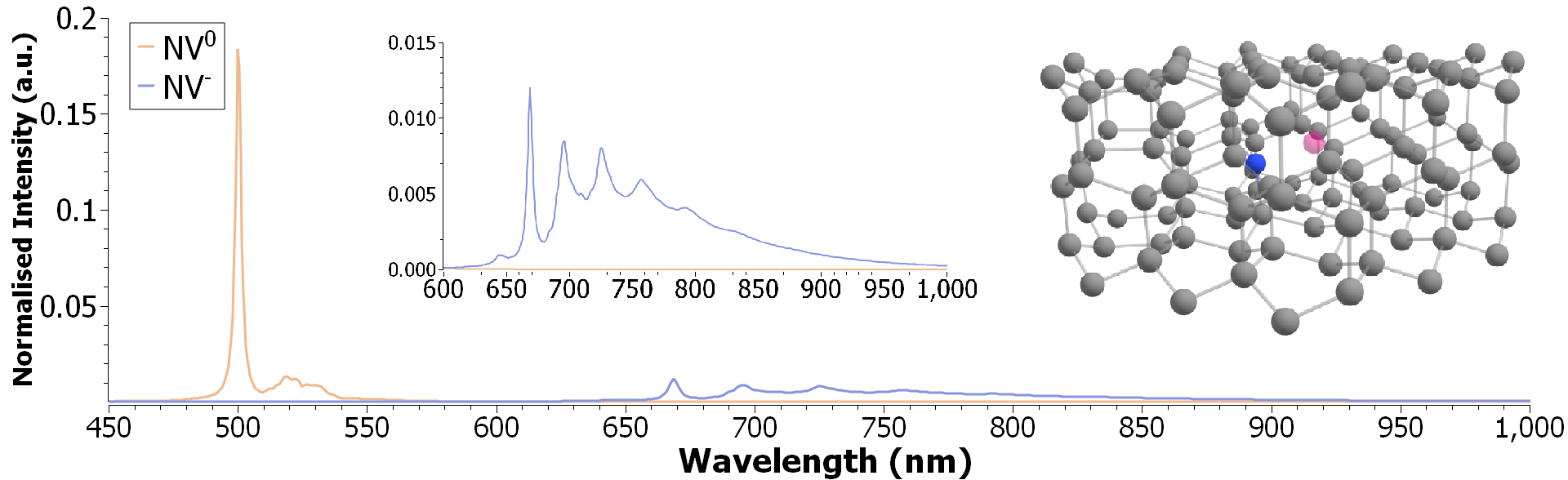}
    \caption{Calculated fluorescence spectra for the AB-NV$^0$ (orange) and AB-NV$^-$ (blue) defects in lonsdaleite nanocrystals. Insets show the NV$^-$ spectra in more detail, and the rendered nanocrystal; terminating hydrogens are not shown. Blue atoms are nitrogen, pink are vacancy. Our calculations show ZPLs at 500~nm and 669~nm for the AB-NV$^{^0}$ and AB-NV$^-$ respectively. These are comparable to the values for cubic diamond of 569-579~nm and 632-642~nm respectively\cite{iozplcienvca16300}.}
    \label{imageInPlane}
\end{figure*}

The three systems then undergo ground state optimisation; tested through frequency and electronic Hessian analysis. Following this, edge hydrogens are then frozen, before geometries and frequencies are recalculated; this is akin to the manual frequency damping performed by Karim \textit{et al.}\cite{aaiesspbfcocc,baipoNV+id} except now hydrogen contamination to the electronic Hessian are also minimised. The resulting thermochemistry is studied, as are the vertical excitation energies and natural transition orbitals (NTOs). 

Calculations are performed within the \textsc{Orca} software package\cite{ORCA,ORCA1}, using the hybrid rendition of the Perdew-Burke-Ernzerhof PBE0 density functional\cite{PBE,pbe01,pbe02} alongside the damped variant of the Becke-Johnson geometry dependent 3-parameter DFT-D3(BJ) dispersion correction\cite{dspbj1,dspbj}. Since the number of basis functions dictates the computational complexity of the problem, the AB species was calculated using the larger and more complete Karlsruhe variant of the single-$\zeta$ valence polarised except for hydrogen def2-SV(P) basis set\cite{def2TZVP}, while the AA species were calculated using the smaller but well respected double-$\zeta$ 6-primitive Slater-type orbital with 3-inner and 1-outer functions 6-31G basis set\cite{6-31G}. We acknowledge that the choice of basis set is relatively small and is not consistent due to computational constraints. However, a series of benchmarking calculations show there to be no observable basis set dependence, nor is there any major shift in values between results calculated using smaller vs. larger basis sets (ESI). 

Fluorescence spectra were generated using the vertical gradient approximation in \textsc{Orca}, which assumes the excited state electronic Hessian to be equivalent to the ground state electronic Hessian, and extrapolates the excited state geometry from the ground state using a displaced oscillator and augmentation method, resulting an expedited calculation\cite{rbaiaflom}. For the vertical excitation energies, 10 roots are examined, with the fluorescence focusing on the first as per the Franck-Condon approximation\cite{franck,condon,condon2}. 

An additional parameter likening our nanocrystal methodology to the bulk is the measure of orbital degeneracy. While typically associated with transition metal complexes, crystal field theory can be applied to yield qualitative datapoints on the symmetry effects and relative energy level splittings. We can approximate the crystal field splitting $\Delta$ using a simplified function respective of the absorption energy, cast as:
\begin{align}
    \Delta=\frac{2\pi\hbar c}{\lambda\Delta_c}
    \label{equationCFS}
\end{align}
where $\hbar$ is Planck's reduced constant, $c$ is the speed of light, $\lambda$ is the absorption energy of the first singlet excited state, and $\Delta_c$ is the splitting factor. As lonsdaleite is of tetrahedral character, we take $\Delta_c$ as $\sim\frac{4}{9}$, or 0.44. 

The fluorescence rate constant $k_r$ and lifetime $\tau_r$ can be calculated using Einstein's spontaneous emission function\cite{eetioledwtadfd,NRpaper,KMCstudy,InterIC,ICbenchmark}, cast as:
\begin{equation}
    k_r = \frac{1}{\tau_r} = \dfrac{4}{3\hbar  c^3}\left<\Psi_0\left|\hat{d}\right|\Psi_1\right>^2\int S_d(\omega)\omega^3d\omega
    \label{equationkr}
\end{equation}
Here, $S_d$ is the normalised emission bandshape with energy $\omega$, and $\left<\Psi_0\left|\hat{d}\right|\Psi_1\right>$ is the transition dipole moment between electronic states $S_1$ and $S_0$. No Herzberg-Teller or Duschinsky effects are accounted for in this simplified model.  

\section{\label{sec:NVg}NV in the graphene-like plane}
The AB-NV defect (Figure~\ref{imageInPlane}) without hydrogen contamination is observed to display C-C bond lengths of 1.54$\pm$0.02~\AA{} across both neutral and negatively charged defects. The range of values is large (between 1.48-1.58~\AA{}), as those in close proximity to the nitrogen centre are much lower than compared to the average C-C bond length. Average C-N bond lengths are calculated to be 1.48~\AA{} in AB-NV$^0$ and 1.47~\AA{} in AB-NV$^-$. $\alpha$ and $\beta$ band-gaps of 3.55~eV and 3.55~eV can be calculated for AB-NV$^0$, while the electron-paired bandgap for AB-NV$^-$ is 1.81~eV. Sun \textit{et al.} noted a much larger band gap of 4.45~eV for AB-NV$^0$ compared to this work\cite{faothd110sfnvbqs}, however this can easily be attributed to the difference in employed density functional. Specifically, Sun \textit{et al.} used PBE where we used PBE0; Song \textit{et al.} highlight that hybrid functionals yield more accurate band gap calculations compared to PBE\cite{csfhfwagaeeiapwb}. However, the lineshape matches up exceedingly well with that of Abdelghafar \textit{et al.}\cite{ACA2025}. Focusing on the defect centre, both nitrogen and vacancy were observed to break symmetry and sit slightly ajar from the plane, resulting in a longer nitrogen-to-vacancy spacing of approximately 1.85~\AA{} in both charge defects. 

\begin{figure*}[tb]
    \centering
    \includegraphics[width=\linewidth]{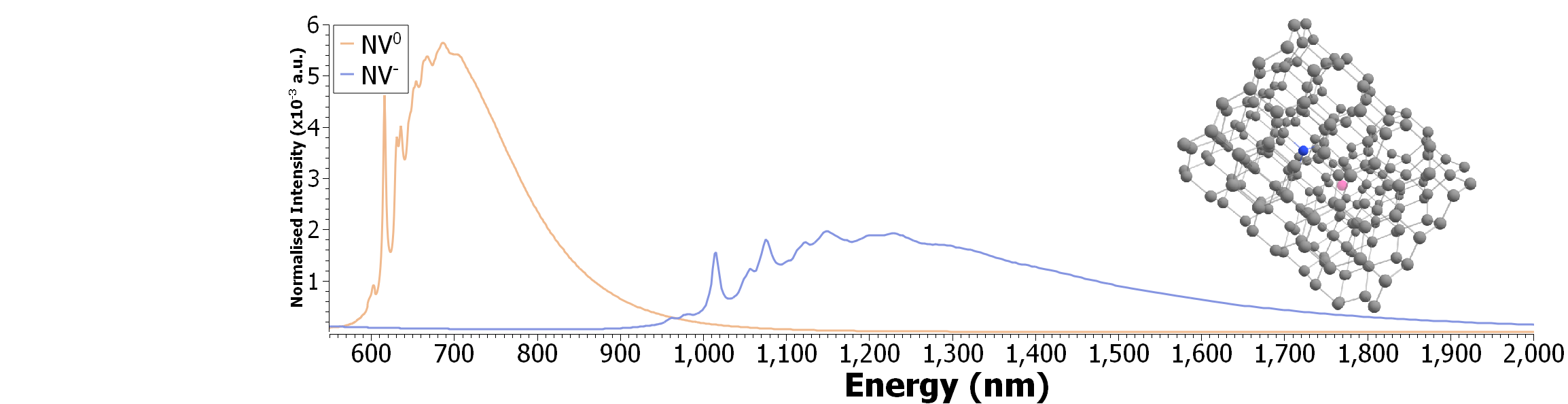}
    \includegraphics[width=\linewidth]{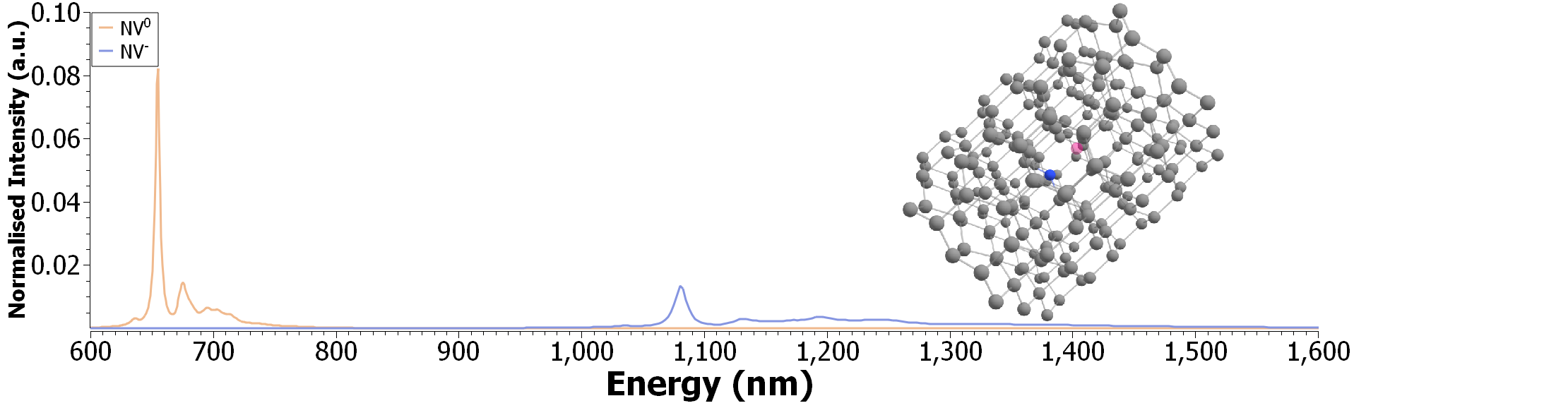}
    \caption{Visualised AA-NV$^0$ (orange) and AA-NV$^-$ (blue) defects in lonsdaleite nanocrystals; (top) shows the long subspecies with corresponding fluorescence spectra, and (bottom) shows the short subspecies with corresponding fluorescence spectra. Blue is nitrogen, pink is vacancy. Terminating hydrogen are not rendered.}
    \label{imageCrossPlane}
\end{figure*} 

Examination of the electronic structure from the vertical excitation spectrum (ESI), we see that for the AB-NV$^0$ system, the first singlet excited state is of mixed character between alpha highest occupied and lowest unoccupied molecular orbitals (H/L) $H_\alpha\rightarrow L_\alpha$ and $H_\beta\rightarrow L_\beta$ with an absorption energy of 2.24~eV, and weak transition and electric dipole moment of 0.31~au and 0.81~au, respectively. An estimated splitting corresponding to 541~THz is calculated (Equation~\ref{equationCFS}). Conversely, for the AB-NV$^-$ system, the excited states are significantly more photostable, with a first singlet excited state energy of 0.74~eV, a dominant $H\rightarrow L$ configuration state function, and while the transition dipole moment is weaker at 0.17~au, the electric dipole moment is much stronger at 1.92~au, alongside a weaker splitting corresponding to 407~THz. The corresponding natural transition orbitals for both charge species suggest electron activity is centered on the nitrogen vacancy. 

Looking at the fluorescence lineshape, a strong and sharp primary band at 500~nm is observed for AB-NV$^0$. A series of smaller bands at 520~nm and 530~nm are also noted, but they are significantly weaker than the primary peak. A fluorescence rate of \rate{2.41}{6} is calculated for this lineshape (Equation~\ref{equationkr}), corresponding to a lifetime of 416~ns. From this work, a zero-phonon line (called the adiabatic energy in the molecular phase) of 2.48~eV was observed. This value is slightly larger than those reported in similar studies\cite{aDFTsonvcil}, however this difference can be attributed to the employment of different functionals and different modelling techniques. For AB-NV$^-$, the spectra is significantly weaker; as the sidebands are more influential the normalised intensity is significantly lower compared to AB-NV$^0$. In addition to a primary peak at 670~nm, smaller bands can be seen at 700~nm, 710~nm, 740~nm, 760~nm, and 800~nm, with the first and third mentioned bands of similar intensity to the primary peak. A fluorescence rate of \rate{1.29}{5}, or a lifetime of 7.74~$\mu$s, is calculated using the corresponding lineshape, with a zero-phonon line of 1.85~eV. This compares well to similar studies\cite{aDFTsonvcil}. We note that the calculated adiabatic energies here are larger than the absorption energies, however this is due to the technique in which \textsc{Orca} approximates the excited state potential from the ground state electronic Hessian.

\section{\label{sec:NVc}AA-NV centres}
Identical to the AB-NV defect, the AA-NV defects (Figure~\ref{imageCrossPlane}) without hydrogen contamination are also observed with an average C-C bond length of 1.54$\pm$0.02~\AA{}. Importantly, while the range of bond values for the short AA-NV defect is similar to the AB-NV subspecies (between 1.46-1.60~\AA{}), the range for the long AA-NV defect is larger (between 1.34-1.61~\AA{}). Closer inspection of each respective geometry reveals the lower bond lengths to be parallel to the NV center in both charge states, while the longer to those surrounding the vacancy. Since the defect in the long AA-NV species displays a significantly larger nitrogen-to-vacancy spacing compared to other species (1.70~\AA{} in short AA-NV compared to 2.93~\AA{} in long AA-NV), it is likely the lower electronic potential would result in the lone-pair electron interacting with nearby bonding parameters, or increase the repulsive potential in the case of those bonds adjacent to the vacancy. The average C-N bond length for both defect species was 1.48~\AA{} across both subspecies and charge species'. In the case of NV$^0$, $\alpha$ and $\beta$ bandgaps of 3.22~eV and 3.77~eV for short AA-NV, and 3.09~eV and 3.22~eV for long AA-NV, respectively, were calculated, while for NV$^-$, electron-paired bandgaps of 1.55~eV and 2.24~eV for short and long AA subspecies, respectively. We again note the difference between these values and those reported by Sun \textit{et al.}\cite{faothd110sfnvbqs}, however also recall the discrepancy is likely due to the different employed functional as per Song \textit{et al.}\cite{csfhfwagaeeiapwb}. It is interesting that for both charge species, one AA-NV energy diverges from the AB-NV energy; for NV$^0$ this is the short AA species, where the $\alpha$/$\beta$ bandgaps diverge by 0.55~eV and away from 3.55/3.55~eV, while for NV$^-$ this is observed in the long AA species, where the bandgap is larger than the AB-NV defect by 0.43~eV. 

\begin{table*}[t!]
    \caption{\label{TableThermoChemistry} Thermochemistry results for each NV sub-species defect in lonsdaleite nanocrystals for both neutral and negatively charge defects, following a frozen edge-hydrogen approximation (no hydrogen contamination). Zero point energy (ZPE), entropy correction ($S$), enthalpy ($H$), thermal correction ($\partial T$), non-thermal correction ($\partial \Bar{T}$), Gibbs free energy ($G^0$), and the average $G^0$ per atom, with and without hydrogens.}
    \begin{tabular}{r|ccccccc}
        \multirow{2}{*}{System} & ZPE & $S$ & $H$ & $\partial T$ & $\partial \Bar{T}$ & $G^0$ & $G^0$/atom [-H] \\
         & (eV) & (kcal/mol) & (meV) & (kcal/mol) & (kcal/mol) & (kcal/mol) & (kcal/mol) \\
        \hline
        AB-NV$^0$ & 23.76 & 505.41 & 25.69 & 171.20 & 547.90 & 214.28 & 1.00 [1.71]\\
        Short AA-NV$^0$ & 42.01 & 707.28 & 25.69 & 244.66 & 968.83 & 506.80 & 1.46 [2.29]\\
        Long AA-NV$^0$ & 41.68 & 718.44 & 25.69 & 248.71 & 961.14 & 492.01 & 1.41 [2.23]\\
        AB-NV$^-$ & 31.19 & 505.04 & 25.69 & 171.22 & 547.94 & 214.71 & 1.00 [1.72]\\
        Short AA-NV$^-$ & 52.59 & 706.86 & 25.69 & 244.65 & 968.08 & 506.46 & 1.46 [2.29]\\
        Long AA-NV$^-$ & 52.74 & 717.34 & 25.69 & 248.22 & 968.07 & 499.54 & 1.43 [2.26]
    \end{tabular}
    \centering
\end{table*}

Looking at the electronic structure for the AA-NV defects (ESI), we see that vertical excitation energies diverge from their respective AB-NV counterparts. For NV$^0$ the AA defects absorb at lower energies, with $S_1$ absorption energies of 2.01~eV and 1.44~eV for short and long AA defects respectively, with the former being of pure $H_\alpha\rightarrow L_\alpha$ character, while the latter is of mixed $H_\alpha\rightarrow L_\alpha$ and $H_\beta\rightarrow L_\beta$ character. The short AA-NV species has a slightly brighter emission than the AB-NV species, with an $S_1$ transition dipole moment of 0.40~au, while the long AA-NV defect is significantly darker than either species, with a transition dipole moment of 0.11~au. Interestingly, the electric dipole moment of long AA-NV was much larger than any other defect at 0.72~au, while for short AA-NV it was a similar magnitude to AB-NV at 0.30~au. Conversely, NV$^-$ shows the AA defects to absorb at higher energies, with $S_1$ absorption energies of 1.20~eV and 1.08~eV respectively, both of localised $H\rightarrow L$ character. Here, the short AA-NV species is significantly brighter than the AB-NV species, with an $S_1$ transition dipole moment of 1.44~au, while the long AA-NV defect is much darker than the AB-NV defect with a transition dipole moment of 0.14~au. Similar to NV$^0$, the negative charge species displays an overly strong long AA-NV$^-$ permanent dipole moment of 3.31~au, compared to the short AA-NV$^-$ dipole moment of 1.86~au.

Further examination of the electronic structure shows that like the AB-NV defect, the majority of the electron density on the lower excited states are focused on the vacancy, however this is much more pronounced for the long AA-NV subspecies. For the $\beta$-spin orbitals in short AA-NV, the density is heavily delocalised across the nanocrystal. We can also see that in the case of the electron-paired configurational NV$^-$, the low-lying states are highly localised states and optically active. We can again use Equation~\ref{equationCFS} to estimate the crystal field splitting, yielding 487~THz and 348~THz for short and long AA-NV$^0$ subspecies, respectively, and 290~THz and 261~THz for short and long AA-NV$^-$ subspecies, respectively. 

Now with all splitting components, we can gain some qualitative comparison between NV$^0$ and NV$^-$: as a general rule we can see that the frequencies in the former are larger than in the latter. While crystal field theory is typically discussed in terms of transition metal d-orbitals, it can still be used to gain some understanding into molecular orbital splitting due to symmetry and local fields in defect centres. NV$^-$ in diamond can be envisaged through the sp$^3$ bonds of the nitrogen and adjacent carbons; the lower symmetry of lonsdalite splits these orbitals. Therefore, insight can be gained into how symmetry breakings splits degenerate energy levels. In the case of our calculated splittings, this implies the electronic structure of NV$^0$ to be more sensitive to local spin interactions compared to NV$^-$. This provides evidence that orbital splitting greater in the neutral spin-species compared to the negative (5 electrons vs. 6 electrons), and that spin-spin/spin-orbit coupling is greater in the neutral lonsdalite compared to the charged lonsdalite, similar to diamond; both the orbital splitting\cite{cefcoosoannvc} and higher order spin-type couplings\cite{tfsotnnvcid} are well documented phenomena in diamond. 

Looking again at the projected fluorescence spectra, vastly different lineshapes are observed compared to the AB-NV defect. Starting with the short AA-NV subspecies, primary and secondary peaks at 650~nm and 675~nm are observed for the neutral charge species, while in the case of the negatively charged short AA-NV$^-$ center, a clear primary peak at 1080~nm is observed, with a weak vibronic progression between 1130-1280~nm consisting of multiple tertiary peaks. Conversely, the long AA-NV$^-$ is found to display generally poor vibronic progression and fine structure conditions. For the neutral defect, a primary band of lower intensity than the tertiary band at around 610~nm is observed, while the second and third sit at around 625~nm and 680~nm, respectively. The negatively charged subspecies is shown to have a much broader lineshape ranging from 900-2000~nm, with a primary peak displaced from the head of the spectra at 1020~nm, a secondary at 1075~nm, and numerous tertiaries at 1140~nm, 1160~nm, 1200~nm, and 1230~nm. In particular, the tail for long AA-NV$^-$ runs particularly long, from 1300-2400~nm. In both cases, weak vibronic contributions can be seen before the first primary peaks. Qualitatively, these profiles are similar to fluorescence of NV nanodiamonds\cite{scomamrosdc,p15mmncnvcfsmFRET}, if not slightly blueshifted and broader. The corresponding $k_r$ and $\tau_r$ values for each AA defect (Equation~\ref{equationkr}), For long AA-NV$^0$ the values are \rate{6.89}{4} and 14~$\mu$s respectively, and \rate{1.10}{6} and 907~ns respectively for short AA-NV$^0$. For AA-NV$^-$, the values are \rate{1.54}{4} and 64~$\mu$s for long AA-NV$^-$, and \rate{2.26}{6} and 442~ns for short-NV$^-$, respectively. Adiabatic energies of 1.89~eV and 2.01~eV were calculated for short and long AA-NV$^0$, respectively, and 1.15~eV and 1.22~eV for short and long AA-NV$^-$, respectively.

\section{Evidence of chemical stability}
To understand the degree of stability of each of the studied defect subspecies with respect to one another, we can use thermochemical analysis (Table~\ref{TableThermoChemistry}) as a first-order approximation. Comparing both charge species, we see that the only large differences between the two are in the zero-point energies, where NV$^0$ is consistently 10~eV lower than the same defect in NV$^-$, while for every other parameter they are almost the same. The degree of entropy is lowest for AB-NV lonsdaleite at 505~kcal.mol$^{-1}$. This increases significantly when the defect resides across two carbon planes; short AA-NV is the lower at 707~kcal.mol$^{-1}$ while long AA-NV is the largest at 718~kcal.mol$^{-1}$. The thermal and non-thermal corrections, as well as the Gibbs free energy $G^0$, follow similar patterns. For AB-NV lonsdaleite, the overall $G^0$ comes to 214~kcal.mol$^{-1}$, but more than doubles for the larger AA-NV defects to 507~kcal.mol$^{-1}$ and 492~kcal.mol$^{-1}$ for short and long AA-NV, respectively. We therefore instead choose to examine $G^0$ in terms of the average free energy per atom; specifically the average free energy not including hydrogen contamination. Here, the range of values is not as vast; with an average $G^0$ for AB-NV of 1.71~kcal.mol$^{-1}$, and 2.29~kcal.mol$^{-1}$ and 2.23~kcal.mol$^{-1}$ for short and long AA-NV, respectively. Across all three defect species, the enthalpy remains constant, emphasising the stability of the system with respect to each other subspecies.

If we look at the vertical excitation energies in more detail for all three defect species (ESI), there in minimal level inversion. In NV$^0$, level is inversion is not observed until $S_5$, while for NV$^-$ it is not observed until $S_7$. The dominant configuration state functions for the low-lying states are similar across all defect subspecies. In NV$^0$, the long AA-NV subspecies has the lowest energy vertical excitation energy and is therefore the most energetically stable, more than 0.5~eV lower than short AA-NV and almost 0.8~eV less than AB-NV lonsdaleite for $S_1$ absorption. The higher states appear to have some degree of Rydberg contamination (diffuse orbitals) beginning at the $S_3$ state, however this is due to the nanocrystal itself (terminating hydrogens) and is therefore not representative of a larger bulk material. In NV$^-$, AB-NV$^-$ lonsdaleite is the most energetically stable, with short AA-NV$^-$ being of the highest energy by more than 0.45~eV. Long AA-NV$^-$ is similarly contaminated with numerous diffuse configuration state functions, however the $S_1$-$S_2$ energy gaps are much larger, and therefore less influential to the optical properties. 

From these thermochemical results, we have strong evidence for stable sub-species of NV$^0$ and NV$^-$ lonsdaleite. Firstly, no change in enthalpy is observed, suggesting each of the three nanocrystals are in stable configurations, require no additional energy to stabilise, and are chemically equivalent. Further, the average Gibbs free energy for each derivative are similar in value, where we would expect a large difference for systems with drastically different physics. Comparing the three subspecies across both NV$^0$ and NV$^-$, we can see that AB-NV lonsdaleite is the most stable species of the three sub-species in terms of $G^0$ alone, which makes some intuitive sense as it would otherwise be uncommon to assume the defect to be spread across different carbon planes. Importantly, while chemically similar, each defect species has very different physical and optical properties, due primarily to the different packing configurations. The long AA-NV subspecies for example appears to be the most optically active of the three, in addition to always displaying a stronger permanent dipole moment. AB-NV optical properties are consistently well energetically separated from the AA-NV optical activity. As well, all appear to display distinct optical and crystallographic properties. Therefore, this may allow for specific derivatives to be more favourable based on the specific applications. 

\begin{table*}[t!]
    \caption{\label{TabledDvL} AB-NV lonsdaleite nanocrystals compared to AB-NV diamond nanocrystals. Zero point energy (ZPE), entropy correction ($S$), enthalpy ($H$), thermal correction ($\partial T$), non-thermal correction ($\partial \Bar{T}$), Gibbs free energy ($G^0$), and the average $G^0$ per atom, with and without hydrogens. Thermochemistry is performed without hydrogen contamination.}
    \begin{tabular}{r|ccccccc}
        \multirow{2}{*}{System} & ZPE & $S$ & $H$ & $\partial T$ & $\partial \Bar{T}$ & $G^0$ & $G^0$/atom [-H] \\
         & (eV) & (kcal/mol) & (meV) & (kcal/mol) & (kcal/mol) & (kcal/mol) & (kcal/mol) \\
        \hline
        c-NV$^0$ & 20.37 & 543.09 & 25.69 & 182.76 & 469.86 & 110.12 & 1.05 [2.00]\\
        AB-NV$^0$ & 23.76 & 505.41 & 25.69 & 171.20 & 547.90 & 214.28 & 1.00 [1.71] \\
        \hline
        c-NV$^-$ & 28.27 & 542.68 & 25.69 & 182.77 & 469.26 & 109.94 & 0.53 [1.03]\\
        AB-NV$^-$ & 31.19 & 505.04 & 25.69 & 171.22 & 547.94 & 214.71 & 1.00 [1.72]
    \end{tabular}
    \centering
\end{table*}

It is worth briefly commenting on the efficacy of the model used to predict the properties discussed in this work. Specifically, on the employed method used to counteract hydrogen contamination (ESI); thermochemical analysis is significantly more difficult to parse. While disorder is supposedly lower by a factor of around 9 when ignoring hydrogen effects, the Gibbs free energy is larger by a factor of at least 8 across all lonsdaleite species, while the average Gibbs potential per non-hydrogen atom is around 4 times larger. If we consider hydrogen contamination in terms of vibronic progression (ESI), we see that not only is the spectral intensity lower, so is the energy and detail in the fine structure. Indeed, hydrogen contributions lower the $S_1$ projected adiabatic energy by 0.47~eV, as well as dampen the fine structure along the secondary band. This has an overall result of perturbing the fluorescence properties by a factor of 2; we therefore reemphasise importance of a frozen-hydrogen approximation, noted by Karim \textit{et al.} also\cite{aaiesspbfcocc,baipoNV+id}. 

Because all three centres are stable, we expect them to all be present in samples of lonsdaleite with sufficient levels of nitrogen and vacancies.  Confirming the presence of the long-AA defect is of interest for quantum applications, as well as informing about the structure of lonsdaleite.  Nevertheless, we have no information about formation pathways.  

Nitrogen is a common impurity in cubic diamond, and so we would expect it to naturally incorporate into lonsdaleite as well.  Creation of the NV centres, however, depends on how hexagonal stacking and point-defect populations are generated. If vacancies are injected into otherwise vacancy-free lonsdaleite, one would expect under annealing (or appropriate heating) that all of the NV orientations should be found with some probability.  We would therefore expect that fabrication routes should follow those known for NV in cubic diamond \cite{WON+2007,eonvccihpdbiiaa,DAK+2014,wiNVckvraastotoiNVc,hnvddnfbiiaoa}.

We would also expect that both nitrogen and vacancies could be trapped under extreme growth conditions. For example, nitrogen-vacancy centres are observed in detonation nanodiamonds \cite{CBD+2016,RCL+2017}, which are formed under highly non-equilibrium conditions.


\section{\label{sec:Comparison}Comparison between NV in hexagonal diamond to cubic diamond}
It is worth noting that since lonsdaleite is known to display $C_{3v}$ symmetry, we would typically expect similar properties between lonsdaleite and diamond, albeit with a bond length shift. However, explicit comparison of AB lonsdaleite with in-plane diamond shows almost identical bond lengths. Akin to lonsdaleite, the average C-C bond length in both c-NV$^0$ and c-NV$^-$ diamond is 1.54$\pm$0.02~\AA{}. The average C-N bond is also 1.48~\AA{}, while the $\alpha$ and $\beta$ band-gaps are larger at 4.80~eV and 4.84~eV, respectively. Similar values are observed in NV$^-$, but with a lower band gap of 1.88~eV. Just as in lonsdaleite, the NV centre breaks symmetry due to the unpaired electron and sits out of plane, resulting in a nitrogen-to-vacancy spacing of approximately 1.84~\AA{}. The estimated crystal field splittings (Equation~\ref{equationCFS}) are 901~THz for NV$^0$ and 726~THZ for NV$^-$, respectively. Comparing the calculated crystal field splitting in c-NV$^0$ with lonsdaleite (541~THz, 487~THz and 348~THz), we see that c-NV$^0$ has larger orbital degeneracy, and likely more spin-affinity compared to lonsdaleite. However, this is only qualitative, and requires further investigation.

\begin{figure}[tb]
    \centering
    \includegraphics[width=\linewidth]{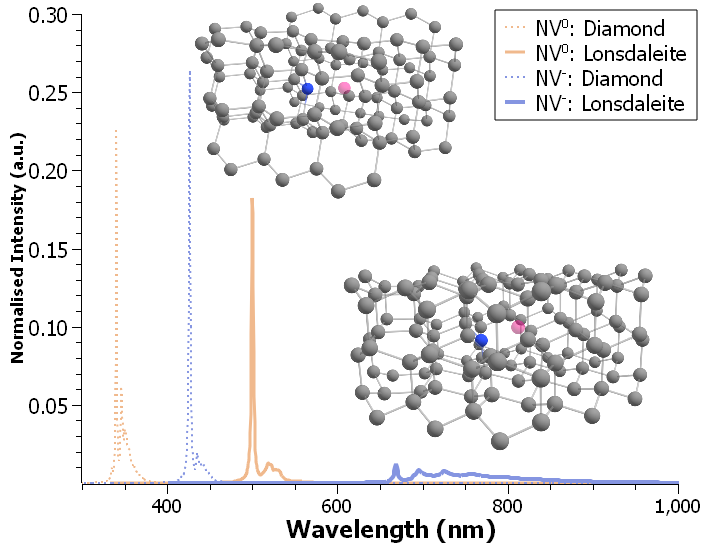}
    \caption{AB-NV (Lonsdaleite) compared to c-NV (diamond). Insets show differences in lattice configuration for the same defect species. }
    \label{imageSpectraComparison}
\end{figure}

Looking at the fluorescent profile for lonsdaleite compared to diamond (Figure~\ref{imageSpectraComparison}), we can see a number of similarities; the diamond emission profiles have distinct primary bands at 340~nm and 500~nm for neutral and negatively charged c-NV, respectively. This does differ to what is seen experimentally (575~nm and 637~nm for c-NV$^0$ and c-NV$^-$, respectively), however it is worth highlighting again that there will be some differences between the periodic system and nanocrystal structure. 
Unlike lonsdaleite however, the fine structure is much more detailed, with a clear secondary peak at 345~nm in c-NV$^0$, and tertiary peak at 350~nm. Part of the 0-0 band can also be seen at 335~nm. Interestingly, c-NV$^-$ appears almost identical in lineshape to c-NV$^0$ albeit with a lower energy and a lower intensity secondary peak. Importantly, lonsdaleite's emission energy is more than 1~eV lower than in diamond for both species; we suspect this is due to the higher degree of symmetry in lonsdaleite than in diamond. Corresponding adiabatic energies of 3.64~eV and 2.91~eV are noted for both neutral and negatively charged c-NV nanodiamonds. 

Similar to lonsdaleite's fluorescence rate and lifetime, we can also calculate these properties (Equation~\ref{equationkr}) for our diamond nanocrystal, yielding \rate{7.05}{6} and 142~ns for NV$^0$, and \rate{9.25}{5} and 1.08~$\mu$s for NV$^-$. 

Atomically, the structures of lonsdaleite and diamond nanoscrystals are similar, in both average bond lengths and NV behaviour. However, at the thermochemical level, they differ significantly. An important difference in the thermochemical properties of lonsdaleite compared to diamond is that of the Gibbs free energy of lonsdaleite is almost twice that of diamond (Table~\ref{TabledDvL}). This is also the case in terms of the potential per atom. Further, entropy is lower in lonsdaleite than in diamond. Overall, this suggests that lonsdaleite is significantly more chemically stable than diamond. We acknowledge the surprising nature of this result; due to the difficulty associated in fabricating lonsdaleite, it would typically be assumed to be \textit{less} thermodynamically stable than cubic diamond in spite of its metastability. This does infer there exists some fabrication methodology in which lonsdaleite is ultimately easier to fabricate compared to diamond, however this is purely speculation based on our findings. 

While diamond is the more efficient emitter, lonsdaleite emits more than 1~eV lower in energy, and has a smaller band gap than diamond. As lonsdaleite offers other advantageous properties that diamond does not; both remain useful for differing applications. However, the shift in packing has a significant effect on the chemical stabilisation of the carbon material, and therefore proves lonsdaleite to be a very important material to investigate with respect to properties and applications. Lonsdaleite also has the added benefit of subspecies of NV defects, which we have now shown for the first time to not only be photostable, but also to offer unique properties and characteristics important to a myriad of quantum applications. 

By comparing  our calculations to experiment, we do observe some differences. Specifically, the secondary peaks modeled in this work are not observed in experiment. We attribute these to computational artifacts caused by the employed nanocrystal methodology. Those peaks are likely due to surface effects that are more pronounced here due to the very small system size. Though we have taken efforts to minimise their influence, the C-H bonds would have some influence on the lineshape. In a larger (macroscale) system, these effects would likely be well separated from the NV defect; one could argue in the case of direct NV centre excitation, there would be minimal to no influence from the surface. That is not the case here, as the surface here is only a few carbons away. Despite this, we believe in the validity of our model, and were a similar nanocrystal measured in experiment the same peaks would be visible (where at the macroscale they would have negligible intensities).

\section{Prospects for ODMR in NV in lonsdaleite}

As discussed, there are good reasons to suspect that NV in lonsdaleite might show similar quantum properties to NV in diamond. To summarise, these are: rigid sp$^3$ bonded carbon lattice isolating ground state spins from phonon decoherence, two-hole structure for NV$^-$ systems providing triplet ground state and singlet transition for optical spin initialisation, ideal C$_{3V}$ symmetry for the long and short AA-NV configurations and close to C$_{3V}$ symmetry for the AB-NV orientations. While NV$^-$ is of electron-paired (singlet) character, NV$^0$ is of doublet spin nature (s=2) due to a unpaired electron, and is therefore paramagnetic\cite{eprsotnnvid}. Taken together, this implies that all NV$^-$ centres in lonsdaleite are good candidates to explore for quantum applications.

\begin{table*}
    \centering
    \begin{tabular}{c|c|c|c|c|c}
          & Calculated & Predicted & & G.S. fine &  \\
          N-V && N-V & Effective & structure & $|0\rangle$ to $|\pm 1\rangle$ \\
          & separation & separation & strain & shift & splitting\\ \hline
          c-NV$^-$ & 184 pm & 154~pm & 0\% & 0~GHz & 2.88~GHz \\
         AB-NV$^-$ & 185 pm & 155~pm & -0.7\% & -0.139~GHz & 2.74~GHz\\
         Short AA-NV$^-$ & 170~pm & 142~pm & 7.8\% & 1.68~GHz & 4.56~GHz\\
         Long AA-NV$^-$ & 293~pm & 245~pm & -37.4\%  & - & - \\ \hline
    \end{tabular}
    \caption{Table showing separation between N and V in lonsdaleite compared with effective strain (relative to cubic diamond), and predicted ground state (g.s.) shift from the value for cubic diamond (2.87~GHz) for the AB-NV$^-$ and Short AA-NV$^-$. We do not show the hypothesised shifts for Long AA-NV$^-$ as the effective strain is too large to warrant a linear approximation to the fine structure shift.}
    \label{tab:splitting}
\end{table*}

Although our modelling does not show the ground state properties of the NV$^-$ in lonsdaleite, we may draw inspiration from NV$^-$ in diamond to provide a first estimate to guide experimental studies. By assuming that the \emph{only} difference between c-NV$^-$ in diamond and NV$^-$ in lonsdaleite is the lattice spacing, we may treat NV$^-$ in lonsdaleite as a form of strained diamond, with the strain accounting for the change of lattice spacing between the two allotropes. Again, we stress that this is unlikely to be only distinction between the properties of the centres, but is likely to be a guide for future experimental studies.

For the purposes of predicting the ground state properties, and in particular the ground state crystal field splitting of the NV$^-$, we follow Knauer \textit{et al.} \cite{KHR2020}.  Due to the two-hole configuration of the NV$^-$ centres, we expect a ground state triplet. The degeneracy between the spin 0 and spin $\pm 1$ states at zero magnetic field is lifted by the ground state fine structure, which is set, ultimately, by spin-spin interactions \cite{LR1996,Mar1999,MHS2006}.  Knauer \textit{et al.} report that the ground state fine structure shift for NV$^-$ in cubic diamond has a susceptibility of 
\begin{align}
    k^{\perp}_{\sigma gs}=21.5\pm 1.2 \textrm{GHz (per unit strain)}. \label{eq:strainshift}
\end{align}
The lattice spacing for NV$^-$ in cubic diamond is 154~pm and ground state fine structure splitting is 2.88~GHZ. Our modelling of cubic diamond gives a nitrogen to vacancy separation of 184~pm, likely due to the small cell size that we have modelled.  Considering this, to make a hypothesis for the ground state fine structure splitting that might be observed in NV$^-$ in lonsdaleite, we rescale the calculated nitrogen to vacancy separations for each configuration by $83.7~\%$, and then treat the change in separation as an effective strain; applying Equation \ref{eq:strainshift} allows us to predict changes in the fine structure splitting for the NV$^-$ in lonsdaleite. 

The results in Table~\ref{tab:splitting} highlight that the AB-NV and short AA-NV are relatively close to the values seen for cubic diamond.  By comparison, we note that the diamond lattice retains integrity until a strain of around 16\% is achieved \cite{FRL+2012}. Hence the results for AB-NV and Short AA-NV should be approximately accurate.  We have less confidence in using this method to predict ground state properties for Long AA-NV, as N to V separation is considerably further from what is seen for c-NV in cubic diamond.

Nevertheless, these results suggest that it is possible to search for optically detected magnetic resonance for NV$^-$ centres in lonsdaleite using equipment substantially similar to that used currently for NV$^-$ in cubic diamond.

The fact that the splittings and emission spectra for the AB- and AA-NV$^-$ are distinct is a potential advantage for vector magnetometry. Vector magnetometry using c-NV$^-$ in cubic diamond requires a small bias magnetic field to lift the degeneracy between the centres  \cite{SDN+2010}, and the centres are not optically distinguishable, being only distinguishable by the ODMR. Although the three AB-NV$^-$ are degenerate at zero field, the two AA-NV which are in a predominantly perpendicular orientations should have a distinct ODMR resonance and spectral resonance.  This should aid distinguishability, and hence provide new opportunities for vector magnetometry compared to c-NV$^-$ in cubic diamond.

\section{Summary and comparison to established work}

The possible existence of nitrogen-vacancy type defects in hexagonal diamond has been explored briefly, and here we briefly summarise our results and compare to the existing literature.  In addition to the present work, we are aware of three other manuscripts.  Two of which are by Abdelghafar \textit{et al.} \cite{aDFTsonvcil,ACA2025} which we discuss together due to their methodological similarities.  The other is Sun \text{et al.} \cite{faothd110sfnvbqs}.

Although differing in scope and approach, these studies all converge on the plausibility of NV defects in lonsdaleite. Importantly however, this work fills in the gaps of knowledge that cannot be easily gained through a purely plane-wave approach, like optical properties. As well, this work highlights 3 unique defect species, where Abdelgafar \textit{et al.} did not differentiate between two of them. Despite this, there is critical overlap and strong agreement between these works and the present one. Taken together, these independent analyses provide increased certainty for the existence of nitrogen-vacancy defects in lonsdaleite, and increase prospects for identifying them in natural and synthetic samples. We summarise the findings, methodologies and goals of the work in Table~\ref{tab:comparison}.

\begin{table*}[t]
    \centering
    \begin{tabular}{p{3cm}|p{4cm}p{4cm}p{4cm}}
       Feature  & Present work & Abdelgafar \textit{et al.} [\citenum{aDFTsonvcil,ACA2025}] & Sun \textit{et al.} [\citenum{faothd110sfnvbqs}]\\ \hline
       Defect Model  & NV in lonsdaleite nanocrystal & NV in solid-state lonsdaleite and diamond/lonsdaleite heterostructure & NV in fluorine and oxygen terminated hexagonal diamond surfaces\\ \hline
        Focus & Exploratory: NV-related derivative species in molecular analogues & Energetics, stability, electronic levels of NV center in bulk lonsdaleite & Luminescence lineshapes, strain, and dual diamond/lonsdaleite interfaces, surface effects\\ \hline
        Discussed Applications & Fluorescence and ODMR & Optical properties and energetics & Luminescence and sensing\\ \hline
        Unique Defects & 3 & 2 & 1 \\ \hline
        Computational Method & Molecular quantum chemistry (DFT, ORCA) & Solid-state quantum chemistry (DFT, Quantum Espresso) & Solid-state quantum chemistry (DFT, VASP) \\ \hline
        Exchange-Correlation & PBE0/D3(BJ) & PBE, HSE06 & PBE \\ \hline
        Basis set/Pseudopotential & def2-SV(P), 6-31G & ONCVPSP norm-conserving pseudopotential & PAW \\ \hline
        System Size & $\sim 220$ atoms & $\sim 980$ atoms & Slab model \\ \hline
        Zero phonon line & AB-NV$^0$ - 500~nm; AB-NV$^-$ - 670~nm; AA-NV$^0$ - 610-680~nm; AA-NV$^-$ - 1020-1080~nm & NV$^-$ 2.04-2.29~eV  & NV$^0$ - 2.32~eV \\ \hline
        Band Gap & NV$^0$ - 3.22-3.77~eV; NV$^-$ - 1.08-2.24~eV & not reported & 4.45 eV \\ \hline
        Fluorescence Lifetimes &
        AB-NV$^0$ - 416~ns; AB-NV$^-$ - 7.74~$\mu$s; AA-NV$^0$ - 907~ns / 14~$\mu$s; AA-NV$^-$ - 442~ns / 64~$\mu$s & not reported & not reported \\  \hline
        Symmetry Considerations & Symmetry resulting in 3 distinct defect species &  Symmetry evolution from C3v to C1h for off-axis defects &
        Site/lattice symmetry breaking and their effects \\ \hline
        Strain Analysis & 
        Highlight finestructure splitting with strain effects & 
        Notes bond elongation for off-axis defects & 
        Strain at heterostructure interface 

    \end{tabular}
    \caption{Comparison of the current work with previous theoretical analyses of nitrogen-vacancy defects in hexagonal diamond}
    \label{tab:comparison}
\end{table*}

\section{Conclusion}
In this work, we examined hexagonal diamond, which we referred to as lonsdaleite for the sake of brevity, as an alternative candidate quantum sensing applications from a theoretical perspective. A schematic representation of a standard nitrogen-vacancy center yields 3 unique sub-species defect configurations that could exist. Two of these configurations are similar to those found with c-NV in cubic diamond, with the third (the long AA-NV) comprising a new centre that cannot form in cubic diamond.

Adopting a solid-state to nanocrystal methodology, exploration of each species' thermochemical stability across both the NV$^0$ and NV$^-$ charge defects highlighted all to be highly stable, each with distinct properties important for various quantum applications. AB was observed to emit cyan light, while short and long AA emit red and broad red-infrared light, respectively. Individually, long AA-NV was shown to uniquely possess a permanent dipole moment more than twice as large as the other two species alongside a very bright emission, while short AA-NV was observed with a dark emission profile. 

\begin{figure}[tb]
    \centering
    \includegraphics[width=\linewidth]{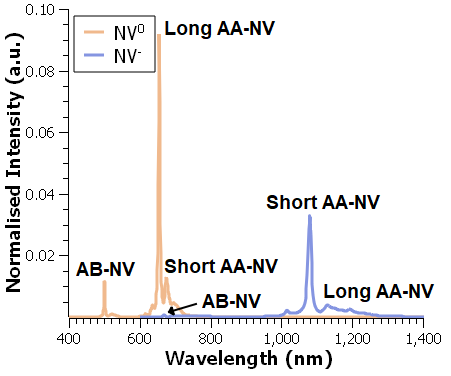}
    \caption{Fluorescence spectra superimposed across all three subspecies, for both NV$^0$ and NV$^-$. Labels correspond to expected contribution from which subspecies.}
    \label{imageSpectraMultiplied}
\end{figure}

If we consider how each sub-species would exist in a bulk structure, without any special fabrication pathway, the configurations would likely be found in some thermodynamically weighted average; whereby the most favourable structure is the dominant species, and the other two exist with some lower probability. In such a multi-defect system, spectral properties would be combined; we can begin to understand what this combined spectral profile would look like by assuming an even  weighting of each species (detailed analysis is not in the scope of this work). The resulting lineshapes (Figure~\ref{imageSpectraMultiplied}) show the long AA-NV profiles to be the most dominant contributors to the overall lineshape. AB-NV$^0$ can be seen at 500~nm, with the short AA contribution at 660~nm being amplified by the cross-section of the long AA-NV vibronic progression between 600-700~nm. AB-NV$^-$ is almost negligible in its contribution to the overall lineshape, and while the short AA-NV peak at around 1080~nm appears dominant, this is again due to constructive interference from the long AA-NV vibronic progression.

Despite a slight disagreement between our model and what is observed in the solid state due to many-body chemical stabilisation effects, we fully expect these results to drive future investigation into hexagonal diamond. Compared to conventional NV in cubic diamond, lonsdaleite manifests in multiple subspecies, each with their own characteristics, which will allow for the material to facilitate numerous quantum technological applications, particularly optically detected magnetic resonance and magnetometry devices.


We also note that the identification of the new long AA-NV centre may have implications for the search for colour centres in silicon carbide \cite{Castelletto2020}, especially the 2H polymorph which has so far not been observed to host colour centres.

\section*{Data Availability Statement}
The data supporting this article has been included as part of the Electronic Supplementary Information.

\section*{Acknowledgments}
AM and SPR acknowledge support from the Australian Government through the Australian Research Council (ARC) under the Centre of Excellence scheme (project number CE170100026), and support of computational resources provided by the Australian Government through the National Computational Infrastructure National Facility and the Pawsay Supercomputer Centre. AM acknowledges that parts of this research were supported by the ARC Centre of Excellence in Quantum Biotechnology through project number CE230100021. MOdV acknowledges part of this research was supported through RMIT's Research Stipend Scholarship (RRSS-SC).

AM would like to thank Rika Kobayashi for her help in working through some computational hiccups in the early development of this work.  The team also acknowledges Giannis Thalassinos and Dougal Mcculloch for useful conversations. Figure 1 generated using \textsc{VESTA 3}\cite{Momma2011}.

\section*{Declarations}
The authors declare no conflicts of interest.



\balance


\bibliography{Lonsdaleite,NVLons} 
\bibliographystyle{rsc} 

\end{document}


\preprint{APS/123-QED}

\title{Nitrogen-vacancy centre in lonsdaleite: a novel nanoscale sensor?}

\author{Anjay Manian}
\email{amanian@uow.edu.au}
\affiliation{ARC Centre of Excellence in Exciton Science, School of Science, RMIT, Australia, 3001.}
\affiliation{School of Science and Molecular Horizons, University of Wollongong, Australia, 2522.}
\affiliation{ARC Centre of Excellence in Quantum Biotechnology, University of Wollongong, Australia, 2522.}
\author{Mitchell O. de Vries}
\affiliation{ARC Centre of Nanoscale Biophotonics, School of Science, RMIT, Australia, 3001}
\affiliation{Quantum Machines Unit, Okinawa Institute of Science and Technology Graduate University, Onna, Okinawa 904-0495, Japan.}
\author{Daniel Stavrevski}
\affiliation{ARC Centre of Nanoscale Biophotonics, School of Science, RMIT, Australia, 3001.}
\author{Qiang Sun}
\affiliation{ARC Centre of Nanoscale Biophotonics, School of Science, RMIT, Australia, 3001.}
\author{Salvy P. Russo}
\affiliation{ARC Centre of Excellence in Exciton Science, School of Science, RMIT, Australia, 3001.}
\author{Andrew D. Greentree}
\email{andrew.greentree@rmit.edu.au} \affiliation{ARC Centre of Nanoscale Biophotonics, School of Science, RMIT, Australia, 3001.}

\date{\today}

\begin{abstract}
Electronic supplementary information contains further details as to how the nanocrystals are defined, a brief exploration of any possible basis set dependence, examination into the effects of hydrogen contamination, tabulated vertical excitation energies and their corresponding configuration state functions, virtualised $\alpha$ and $\beta$ natural transition orbitals, and the optimised coordinates for diamond, and each of the three lonsdaleite derivatives studied in this work, within the frozen hydrogen approximation.
\end{abstract}

\maketitle

\tableofcontents

\newpage
\section{Nanocrystal setup}
\begin{figure}[tb]
    \centering
    \includegraphics[width=\linewidth]{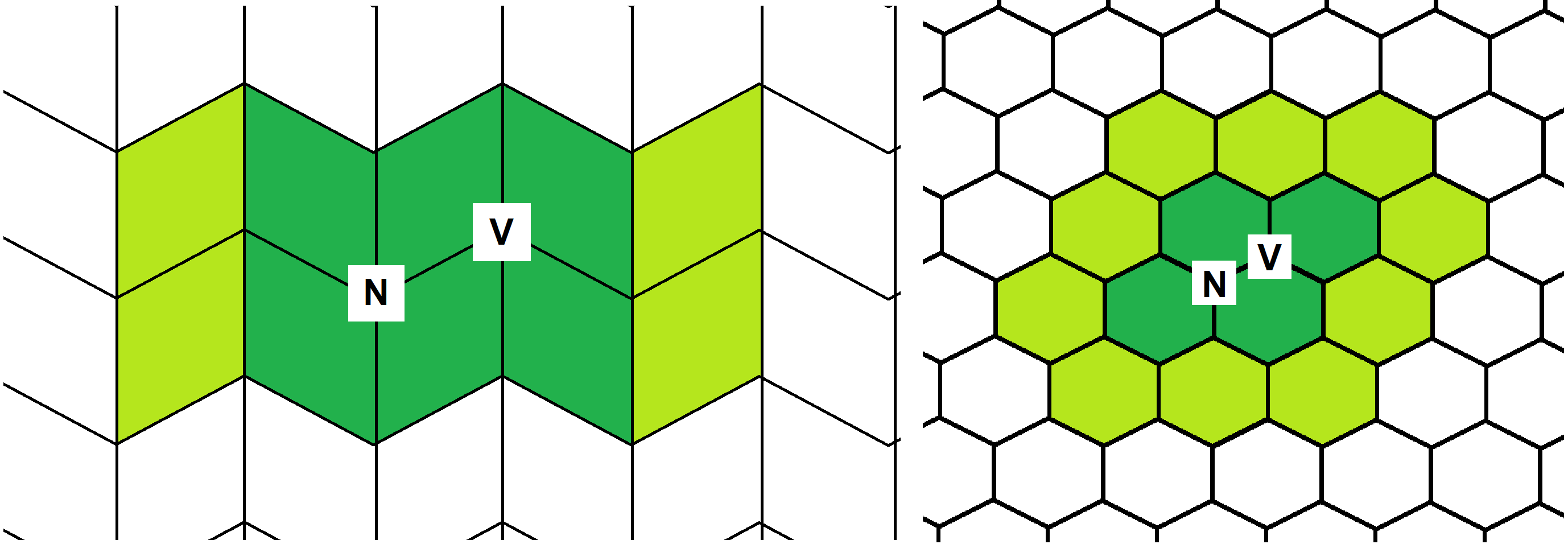}
    \caption{Schematic representation of the 2:1 carbon "shell" scheme. }
    \label{imageCrystalSetup}
\end{figure}

Because quantum chemical analysis is limited by the many body problem, a needle must be threaded that balances computational cost while maximising the system size. We need an atomic system that includes enough carbon to capture the physics of lonsdaleite; however too much and the problem becomes intractable. To satisfy this condition, a 2:1 carbon "shell" is defined, such there are least two carbons in surrounding the defect on the same carbon plane, and one carbon plane above/below the defect (Figure~\ref{imageCrystalSetup}). Note this means the cross-plane lonsdaleite systems therefore contain more atoms than the AB species, due to the phase of the nanocrystal. The edge sites are them terminated through hydrogenation. Each of these systems has an electron multiplicity of 2, due to the electronic configuration of NV$^0$ lonsdaleite. 

\section{Basis set dependence}
\begin{table}
    \caption{\label{TableBasisSet} Basis set comparison for two basis sets for lonsdaleite nanocrystals with hydrogen contamination; zero point energy (ZPE), entropy correction ($S$), enthalpy ($H$), thermal correction ($\partial T$), non-thermal correction ($\partial \Bar{T}$), Gibbs free energy ($G^0$), and the average $G^0$ per atom, with and without hydrogens.}
    \begin{tabular}{r|ccccccc}
        \multirow{2}{*}{Basis Set} & ZPE & $S$ & $H$ & $\partial T$ & $\partial \Bar{T}$ & $G^0$ & $G^0$/atom [-H] \\
         & (eV) & (kcal/mol) & (meV) & (kcal/mol) & (kcal/mol) & (kcal/mol) & (kcal/mol) \\
        \hline
        6-31G & 83.67 & 87.30 & 25.69 & 43.40 & 1929.53 & 1886.22 & 5.40 [8.53] \\
        3-21G & 83.30 & 88.71 & 25.69 & 44.37 & 1921.64 & 1877.90 & 5.38 [8.50]
    \end{tabular}
\end{table}

To understand whether a basis set dependence exists, we take the simplified case of long AA-NV$^0$ lonsdaleite at two different basis sets. 

Even with the smaller 6-31G basis set, calculations of the cross-plane defects take weeks to conclude. Of course the easiest method to cut down on computational costs is to reduce the number of basis functions, however the question of accuracy arises. As a case study, we can examine any possibility of basis set dependence by calculating the thermochemical properties using 3-21G. Thankfully, minimal difference is observed (Figure~\ref{TableBasisSet}), with properties differing by less than 1\%. We can therefore say with some certainty that due to the relative chemical simplicity of the system, the small and ill-used 3-21G basis set can work in the case of significantly larger systems. 

\section{Hydrogen contamination}
\begin{table}[t]
    \caption{\label{TableThermoChemistry} Thermochemistry results for lonsdaleite NV$^0$ nanocrystals with and without hydrogen contamination; zero point energy (ZPE), entropy correction ($S$), enthalpy ($H$), thermal correction ($\partial T$), non-thermal correction ($\partial \Bar{T}$), Gibbs free energy ($G^0$), and the average $G^0$ per atom, with and without hydrogens.}
    \begin{tabular}{r|ccccccc}
        \multirow{2}{*}{System} & ZPE & $S$ & $H$ & $\partial T$ & $\partial \Bar{T}$ & $G^0$ & $G^0$/atom [-H] \\
         & (eV) & (kcal/mol) & (meV) & (kcal/mol) & (kcal/mol) & (kcal/mol) & (kcal/mol) \\
        \hline
        \multicolumn{8}{c}{Free-Hydrogen Systems}\\
        \hline
        AB & 53.25 & 62.20 & 25.69 & 26.57 & 1227.99 & 1292.96 & 3.66 [5.05]\\
        short AA & 80.98 & 92.91 & 25.69 & 46.93 & 1867.42 & 1822.03 & 5.25 [8.24]\\
        long AA & 83.67 & 87.30 & 25.69 & 43.40 & 1929.53 & 1886.22 & 5.40 [8.53]\\
        \hline
        \multicolumn{8}{c}{Frozen-Hydrogen Systems}\\
        \hline
        AB & 23.76 & 505.41 & 25.69 & 171.20 & 547.90 & 214.28 & 1.00 [1.71]\\
        short AA & 42.01 & 707.28 & 25.69 & 244.66 & 968.83 & 506.80 & 1.46 [2.29]\\
        long AA & 41.68 & 718.44 & 25.69 & 248.71 & 961.14 & 492.01 & 1.41 [2.23]
    \end{tabular}
\end{table}

\begin{figure}[tb]
    \centering
    \includegraphics[width=\linewidth]{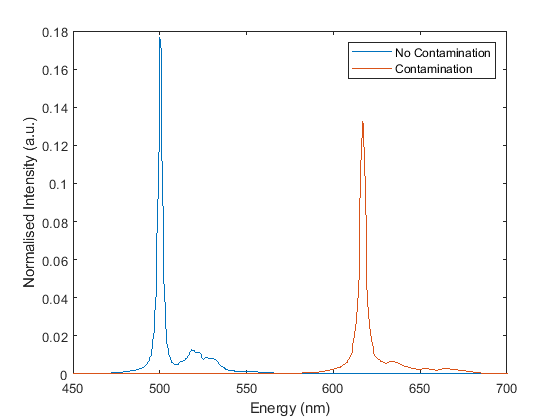}
    \caption{Spectral features for AB-NV$^0$ lonsdaleite with and without hydrogen contamination.}
    \label{imageContamination}
\end{figure}

To probe the efficacy of our modelling method, we need to further understand the effect of hydrogen contamination and its impact on the calculated physics of the studied nanoscystals. One such way we can gain some insight is to calculate the thermochemistry with and without a frozen hydrogen approximation for each defect species (Table~\ref{TableThermoChemistry}). Simplified to only consider NV$^0$, its effect is instantly noticeable, but not immediately obvious. Of course the zero-point energy will change, as that is dependent on the system size. What is surprising is that disorder is supposedly lower by a factor of around 9 when ignoring hydrogen effects (or by a factor of around 4 in terms of an average free energy). This results in a much larger Gibbs free energy in hydrogen contaminated systems by a factor of at least 8 across all lonsdaleite species. We know that since enthalpy remains constant across all calculations, the systems are chemically invariant and can be considered stable configurations. However the free energy lowers significantly upon accounting for hydrogenation. Therefore hydrogen has a destabilising effect on the thermochemistry, likely due to the high energy normal modes now incorporated into calculating the electronic Hessian. 

Alternatively, we may also consider hydrogen contamination in terms of the spectral properties and lineshape. For this, we will simplify the problem and only consider the emission of AB lonsdaleite (Figure~\ref{imageContamination}). In doing so, we see that the spectral intensity for hydrogen contaminated lonsdaleite is lower in both energy and intensity. Where with a frozen hydrogen approximation emission is observed at 500~nm, a hydrogen contaminated emission profile shows a redshifted fluorescence of 620~nm, corresponding to a decrease in the adiabatic energy of 0.47~eV. In addition, the fine structure can be observed to lose a significant degree of detail around the second band. The overall result of this can be observed clearly when calculating the fluorescence properties in more detail: whereby a contaminated structure yields a rate and lifetime of \rate{1.30}{6} and 769~ns, a frozen hydrogen system is faster at \rate{2.41}{6} and 416~ns, an overall damping of the fluorescence by a factor of almost 2. 

From these results, we have a clear idea as to the effect of hydrogen contamination, and clear reasons as to the importance of a frozen-hydrogen approximation. Not only are the thermochemical properties larger than what would be expected, the spectral properties are dampened due to the larger number of high energy normal modes otherwise not considered in the solid state. 

\section{Vertical excitation energies}
\begin{table}
    \caption{\label{TableTDDFT} Vertical excitation energies and character from the $S_0$ Franck-Condon point for each of the 3 lonsdaleite NV$^0$ species. Energies given in eV, while character is in terms of highest occupied (H) and lowest unoccupied (L) molecular orbitals with half $\alpha$/$\beta$ and full spin; weightings are given in parenthesis. }
    {\tiny \begin{tabular}{r|cc|cc|cc|cc|cc|cc}
        \multirow{3}{*}{Root} & \multicolumn{6}{c|}{NV$^0$} & \multicolumn{6}{c}{NV$^-$}\\
        & \multicolumn{2}{c|}{AB} & \multicolumn{2}{c|}{short AA} & \multicolumn{2}{c|}{long AA} & \multicolumn{2}{c|}{AB} & \multicolumn{2}{c|}{short AA} & \multicolumn{2}{c}{long AA} \\
         & Energy & Character & Energy & Character & Energy & Character & Energy & Character & Energy & Character & Energy & Character\\
        \hline
        1 & 2.236 & $H_\alpha\rightarrow L_\alpha$ (0.62) & 2.014 & $H_\alpha\rightarrow L_\alpha$ (0.96) & 1.439 & $H_\alpha\rightarrow L_\alpha$ (0.48) & 0.735 & $H\rightarrow L$ (0.99) & 1.201 & $H\rightarrow L$ (0.96) & 1.082 & $H\rightarrow L$ (0.98)\\
         & & $H_\beta\rightarrow L_\beta$ (0.31) & & & & $H_\beta\rightarrow L_\beta$ (0.30) & & & & & & \\
        2 & 2.509 & $H_\beta\rightarrow L_\beta$ (0.62) & 2.490 & $H_\beta\rightarrow L_\beta$ (0.72) & 2.055 & $H_\alpha\rightarrow L_\alpha$ (0.46) & 1.722 & $H\rightarrow L^1$ (0.98) & 1.931 & $H\rightarrow L^1$ (0.86) & 2.142 & $H\rightarrow L^1$ (0.98)\\
         & & $H_\alpha\rightarrow L_\alpha$ (0.32) & & $H^1_\alpha\rightarrow L_\alpha$ (0.21) & & $H_\beta\rightarrow L_\beta$ (0.30) & & & & & & \\
         & & & & & & $H^1_\alpha\rightarrow L_\alpha$ (0.15) & & & & & & \\
        3 & 2.981 & $H^1_\alpha\rightarrow L_\alpha$ (0.95) & 3.060 & $H^1_\alpha\rightarrow L_\alpha$ (0.73) & 2.288 & $H^1_\alpha\rightarrow L_\alpha$ (0.63) & 1.950 & $H\rightarrow L^2$ (0.93) & 2.059 & $H\rightarrow L^2$ (0.83) & 2.330 & $H\rightarrow L^2$ (0.91)\\
         & & & & $H_\beta\rightarrow L_\beta$ (0.22) & & $H_\beta\rightarrow L_\beta$ (0.30) & & & & $H\rightarrow L^1$ (0.12) & & \\
        4 & 3.128 & $H_\beta\rightarrow L^1_\beta$ (0.95) & 3.380 & $H_\beta\rightarrow L^1_\beta$ (0.96) & 2.374 & $H_\beta\rightarrow L^1_\beta$ (0.97) & 2.021 & $H\rightarrow L^3$ (0.94) & 2.127 & $H\rightarrow L^3$ (0.86) & 2.485 & $H\rightarrow L^3$ (0.91)\\
        5 & 3.946 & $H^1_\beta\rightarrow L_\beta$ (0.79) & 4.048 & $H^1_\beta\rightarrow L_\beta$ (0.59) & 3.333 & $H_\alpha\rightarrow L^8_\alpha$ (0.20) & 2.158 & $H\rightarrow L^4$ (0.98) & 2.179 & $H\rightarrow L^4$ (0.79) & 2.511 & $H\rightarrow L^4$ (0.75)\\
         & & & & $H^2_\beta\rightarrow L_\beta$ (0.18) & & $H_\beta\rightarrow L^{11}_\beta$ (0.13) & & & & $H\rightarrow L^3$ (0.10) & & $H\rightarrow L^5$ (0.18)\\
        6 & 4.165 & $H_\alpha\rightarrow L^1_\alpha$ (0.95) & 4.100 & $H^3_\beta\rightarrow L_\beta$ (0.27) & 3.452 & $H^2_\alpha\rightarrow L_\alpha$ (0.21) & 2.258 & $H\rightarrow L^5$ (0.96) & 2.331 & $H\rightarrow L^6$ (0.48) & 2.645 & $H\rightarrow L^5$ (0.65)\\
         & & & & $H^1_\beta\rightarrow L_\beta$ (0.27) & & $H^7_\beta\rightarrow L_\beta$ (0.17) & & & & $H\rightarrow L^5$ (0.46) & & $H\rightarrow L^4$ (0.16)\\
         & & & & $H^2_\beta\rightarrow L_\beta$ (0.25) & & & & & & & & \\
        7 & 4.366 & $H_\alpha\rightarrow L^3_\alpha$ (0.70) & 4.184 & $H_\alpha\rightarrow L^1_\alpha$ (0.32) & 3.820 & $H^2_\alpha\rightarrow L_\alpha$ (0.75) & 2.323 & $H\rightarrow L^6$ (0.95) & 2.342 & $H\rightarrow L^6$ (0.49) & 2.727 & $H\rightarrow L^7$ (0.81)\\
         & & $H_\alpha\rightarrow L^2_\alpha$ (0.22) & & $H^3_\beta\rightarrow L_\beta$ (0.24) & & & & & & $H\rightarrow L^5$ (0.44) & & \\
        8 & 4.358 & $H^2_\beta\rightarrow L_\beta$ (0.70) & 4.193 & $H_\alpha\rightarrow L^1_\alpha$ (0.56) & 3.880 & $H^1_\beta\rightarrow L_\alpha$ (0.85) & 2.452 & $H\rightarrow L^7$ (0.96) & 2.423 & $H\rightarrow L^7$ (0.84) & 2.741 & $H\rightarrow L^6$ (0.71)\\
         & & & & $H^2_\beta\rightarrow L_\beta$ (0.16) & & & & & & & & \\
        9 & 4.374 & $H_\alpha\rightarrow L^2_\alpha$ (0.54) & 4.272 & $H^1_\alpha\rightarrow L^1_\alpha$ (0.38) & 3.951 & $H^3_\alpha\rightarrow L_\alpha$ (0.72) & 2.592 & $H\rightarrow L^8$ (0.91) & 2.454 & $H\rightarrow L^7$ (0.58) & 2.843 & $H\rightarrow L^8$ (0.59)\\
         & & $H_\alpha\rightarrow L^3_\alpha$ (0.21) & & $H^2_\alpha\rightarrow L^1_\alpha$ (0.33) & & $H^2_\beta\rightarrow L_\beta$ (0.20) & & & & $H\rightarrow L^{11}$ (0.13) & & $H\rightarrow L^9$ (0.27)\\
        10 & 4.423 & $H_\beta\rightarrow L^2_\beta$ (0.93) & 4.367 & $H_\alpha\rightarrow L^3_\alpha$ (0.88) & 4.146 & $H^2_\beta\rightarrow L_\beta$ (0.61) & 2.626 & $H\rightarrow L^9$ (0.93) & 2.544 & $H\rightarrow L^9$ (0.75) & 2.851 & $H\rightarrow L^9$ (0.59)\\
         & & & & & & $H^3_\alpha\rightarrow L_\alpha$ (0.16) & & & & $H\rightarrow L^8$ (0.14) & & $H\rightarrow L^8$ (0.22)
    \end{tabular}}
\end{table}

\begin{figure}[tb]
    \centering
    \includegraphics[width=\linewidth]{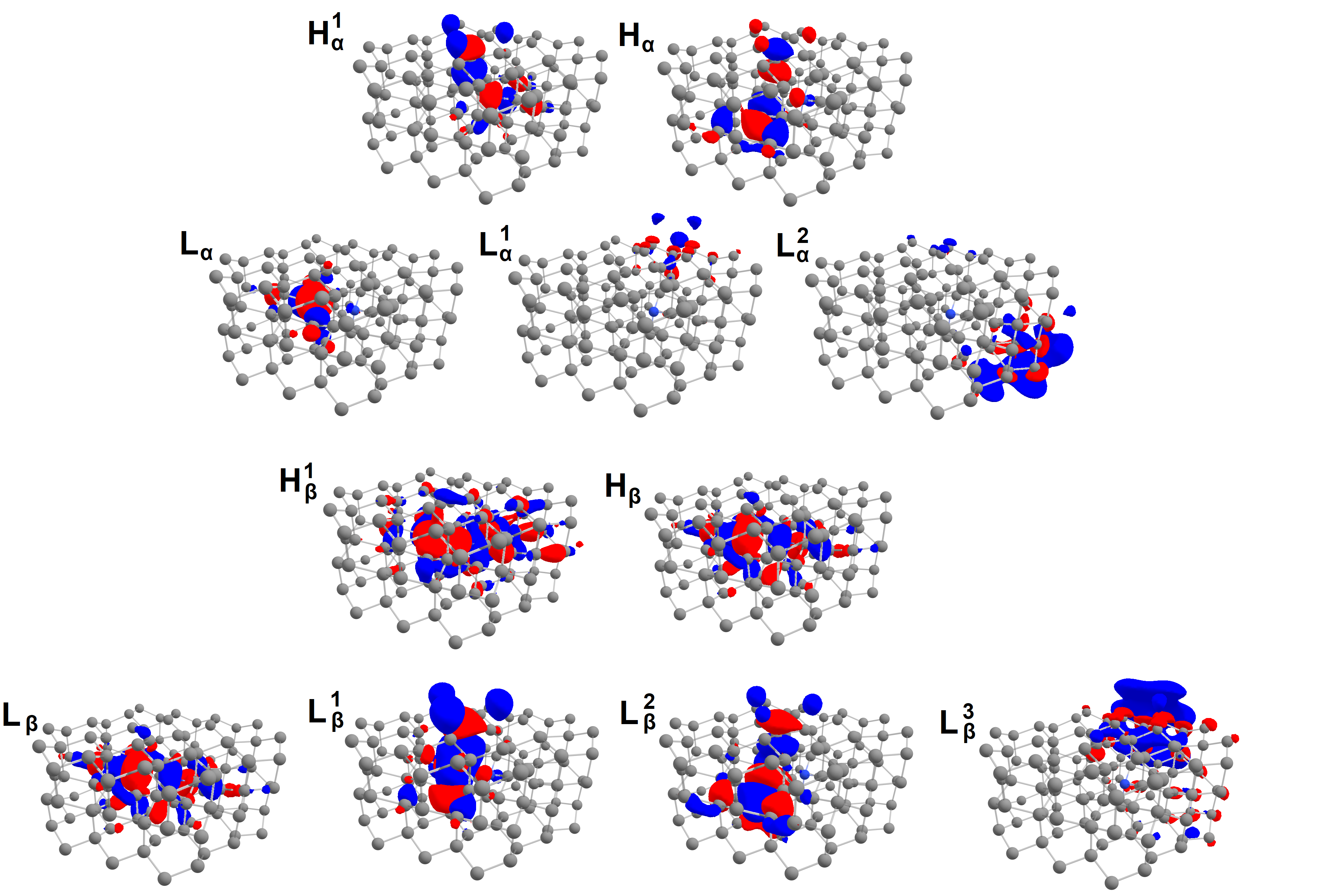}
    \caption{Highest occupied and lowest unoccupied $\alpha$ and $\beta$ natural transition molecular orbitals for AB-NV$^0$ lonsdaleite nanocrystals. Hydrogens have been removed for the sake of brevity. }
    \label{imageOrbitalsInPlane0}
\end{figure}

\begin{figure}[tb]
    \centering
    \includegraphics[width=\linewidth]{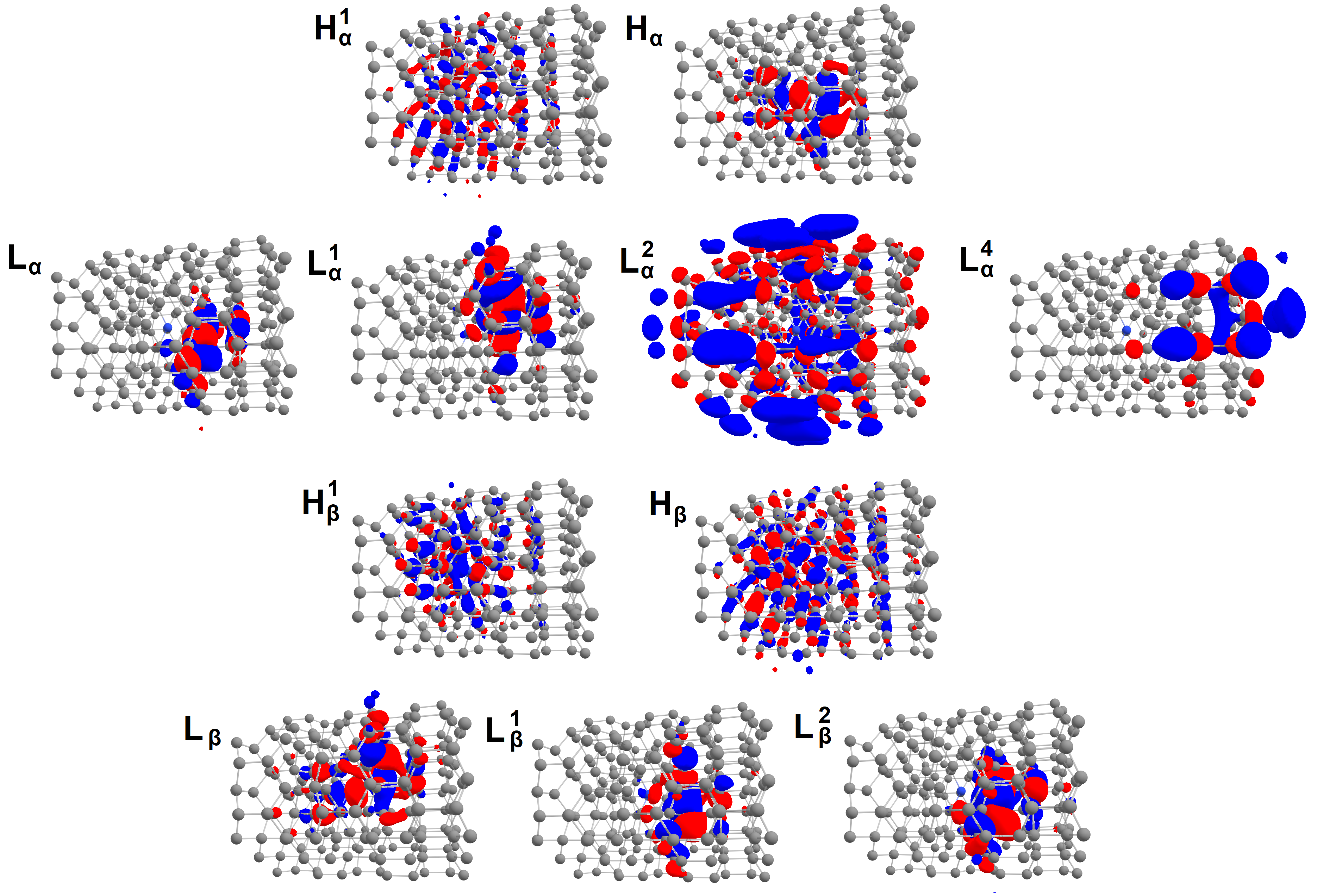}
    \caption{Highest occupied and lowest unoccupied $\alpha$ and $\beta$ natural transition molecular orbitals for short AA-NV$^0$ lonsdaleite nanocrystals. Hydrogens have been removed for the sake of brevity. }
    \label{imageOrbitalsShortCross0}
\end{figure}

\begin{figure}[tb]
    \centering
    \includegraphics[width=\linewidth]{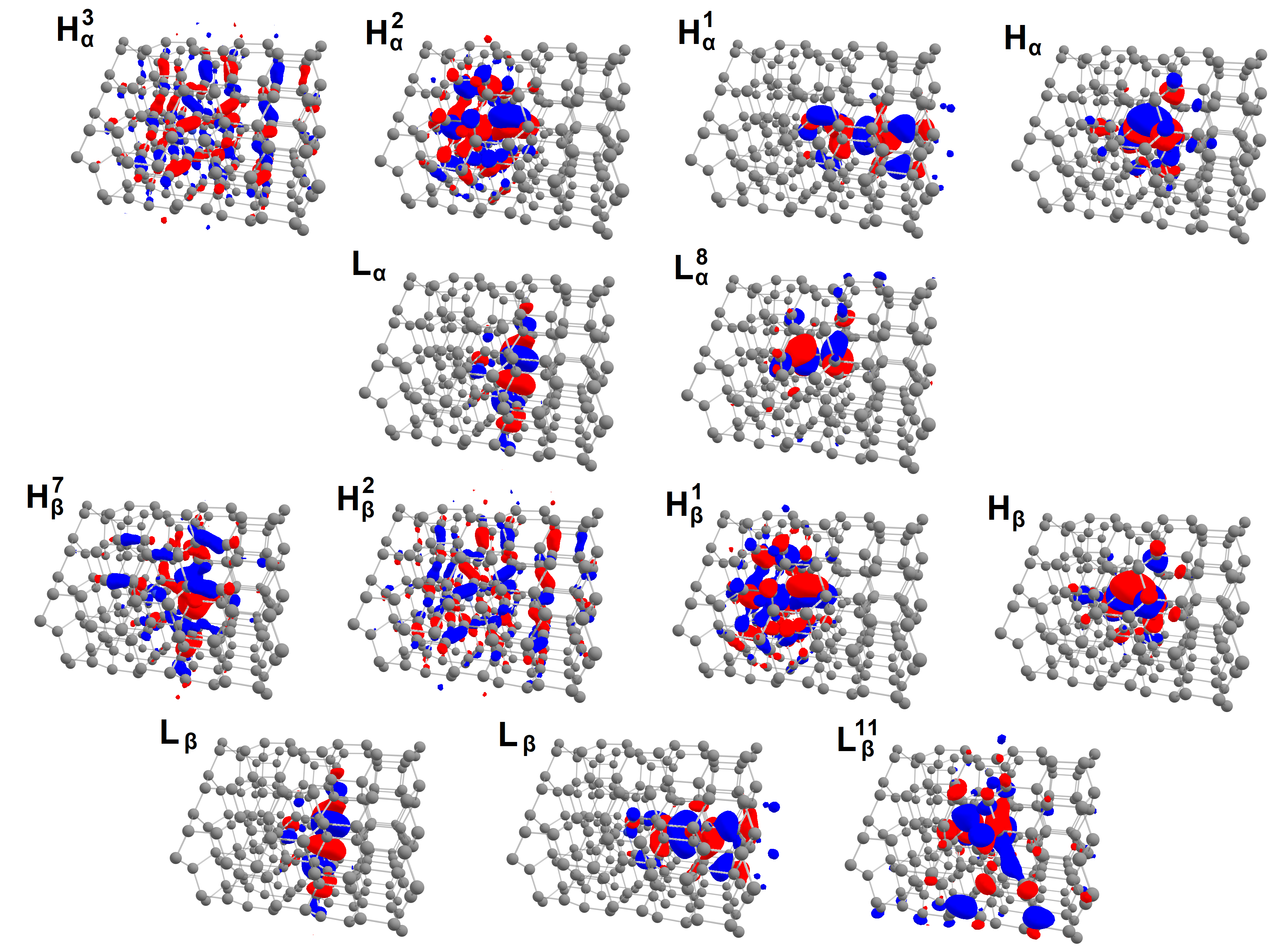}
    \caption{Highest occupied and lowest unoccupied $\alpha$ and $\beta$ natural transition molecular orbitals for long AA-NV$^0$ lonsdaleite nanocrystals. Hydrogens have been removed for the sake of brevity. }
    \label{imageOrbitalsLongCross0}
\end{figure}

\begin{figure}[tb]
    \centering
    \includegraphics[width=\linewidth]{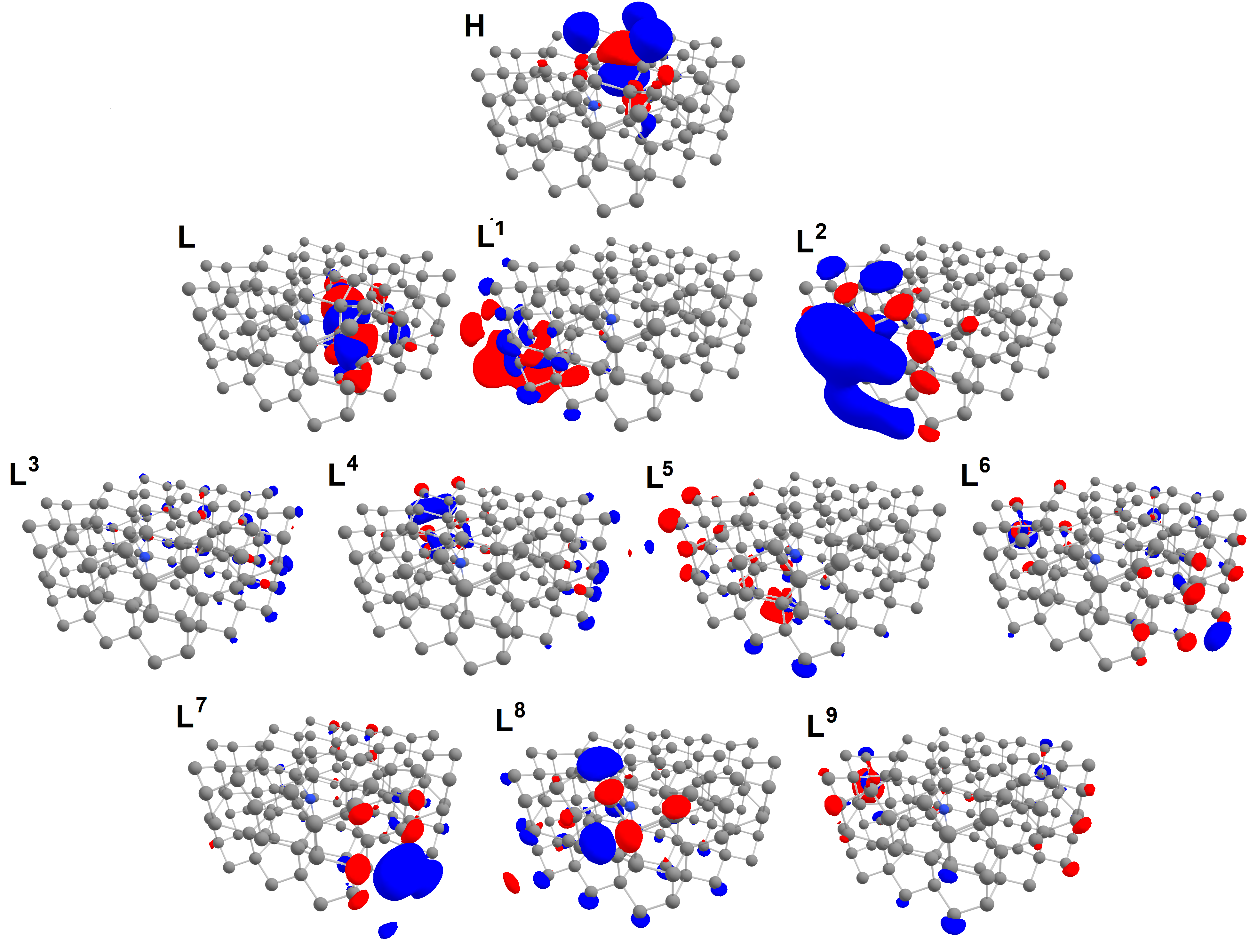}
    \caption{Highest occupied and lowest unoccupied natural transition molecular orbitals for AB-NV$^-$ lonsdaleite nanocrystals. Hydrogens have been removed for the sake of brevity. }
    \label{imageOrbitalsInPlane-}
\end{figure}

\begin{figure}[tb]
    \centering
    \includegraphics[width=\linewidth]{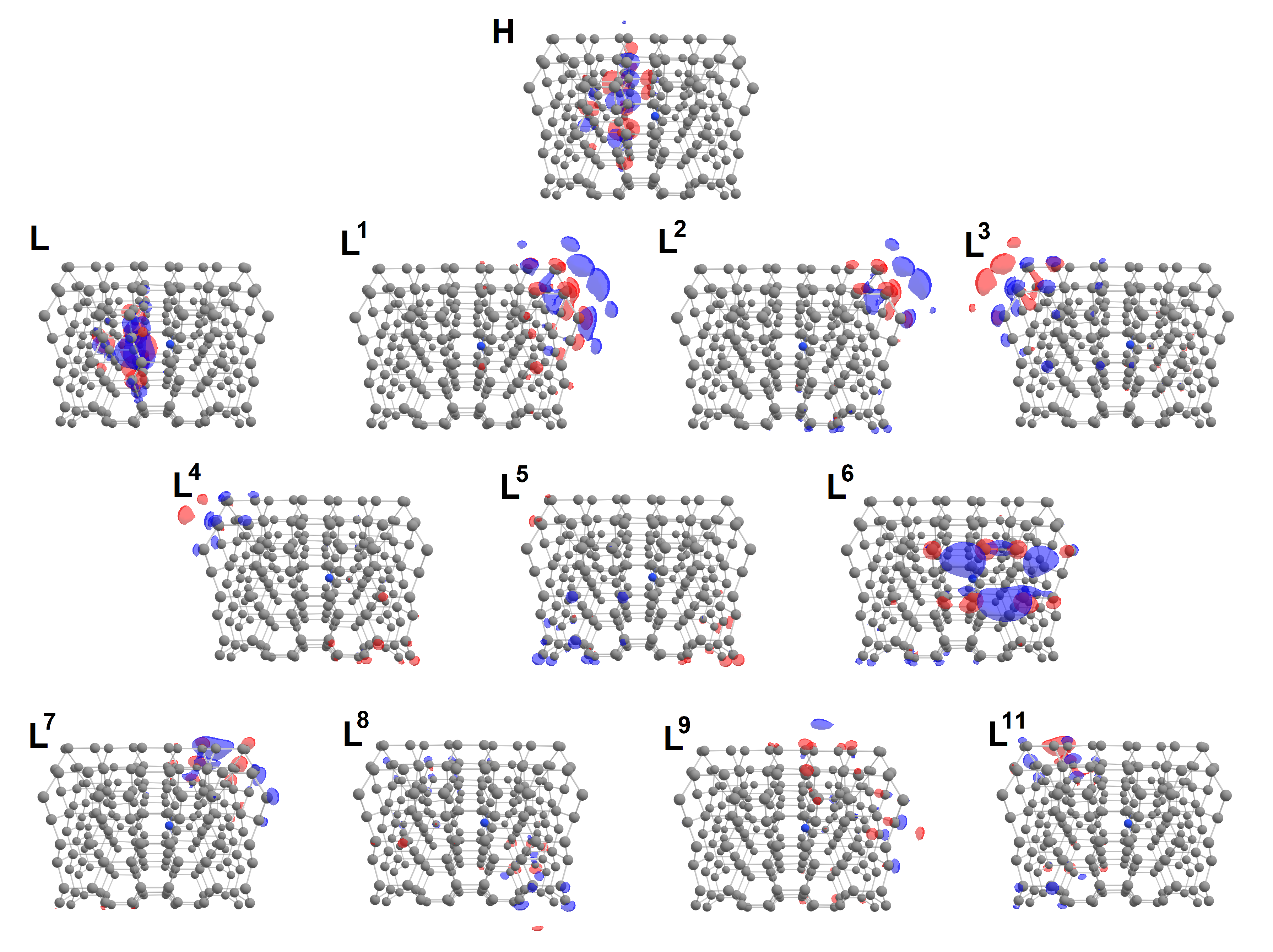}
    \caption{Highest occupied and lowest unoccupied natural transition molecular orbitals for short AA-NV$^-$ lonsdaleite nanocrystals. Hydrogens have been removed for the sake of brevity. }
    \label{imageOrbitalsShortCross-}
\end{figure}

\begin{figure}[tb]
    \centering
    \includegraphics[width=\linewidth]{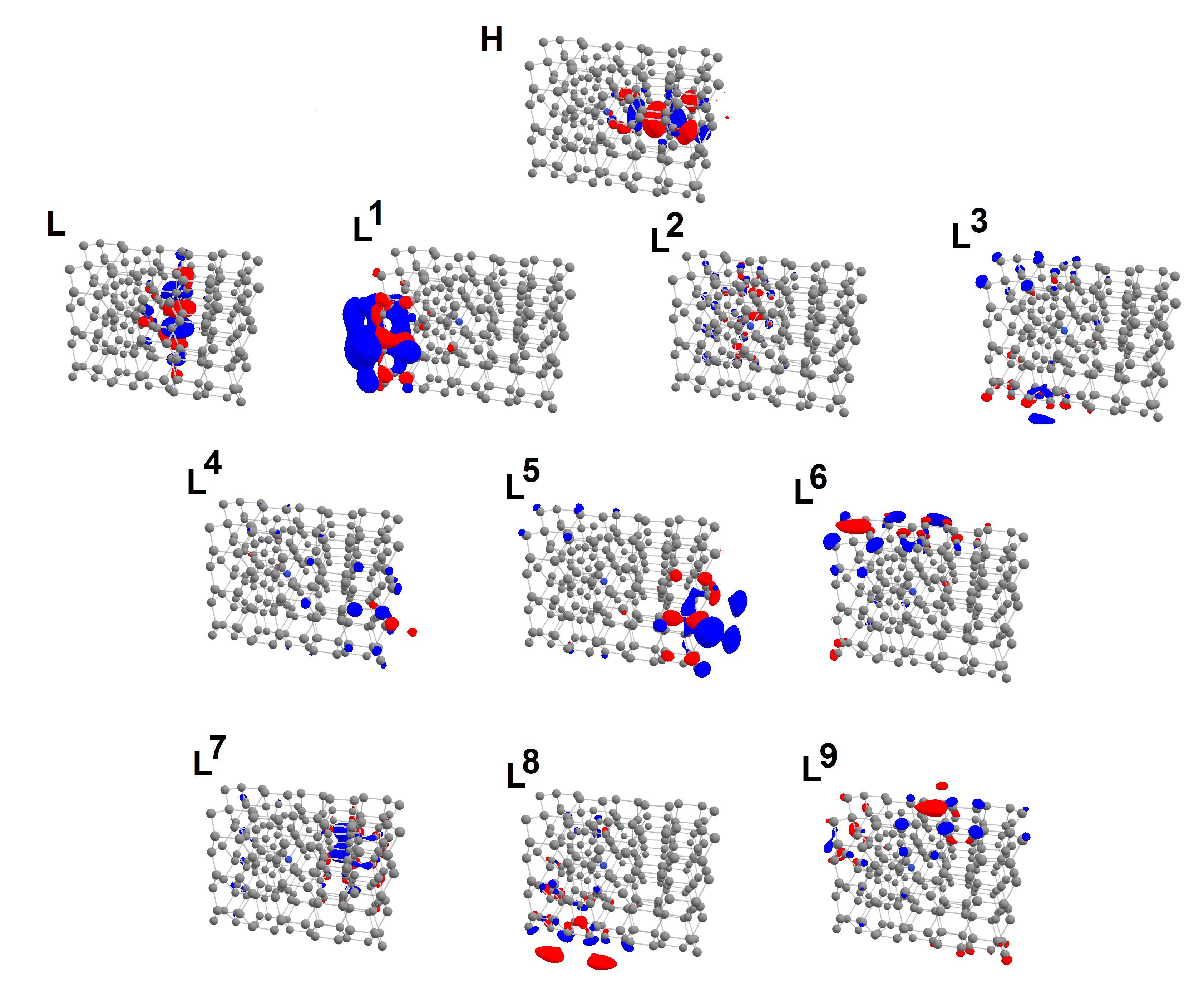}
    \caption{Highest occupied and lowest unoccupied natural transition molecular orbitals for long AA-NV$^-$ lonsdaleite nanocrystals. Hydrogens have been removed for the sake of brevity. }
    \label{imageOrbitalsLongCross-}
\end{figure}

The full vertical excitation energies for both NV$^0$ and NV$^-$ are tabulated in Table~\ref{TableTDDFT}; natural transition orbitals for each species are presented in Figures~\ref{imageOrbitalsInPlane0}, \ref{imageOrbitalsShortCross0}, and \ref{imageOrbitalsLongCross0}, for NV$^0$ AB, short AA, and long AA defects, respectively, and Figures~\ref{imageOrbitalsInPlane-}, \ref{imageOrbitalsShortCross-}, and \ref{imageOrbitalsLongCross-}, for NV$^-$ AB, short AA, and long AA defects, respectively. 

No level inversion is observed until the fifth excited state. While only the short AA displays a pure $S_1$ state, each excited state wavefunction is of mixed character. However, the dominant contributions for the low-lying states are similar across all defect species. Long AA has the lowest energy vertical excitations and is therefore the most energetically stable, more than 0.5~eV lower than short AA and almost 0.8~eV less than AB lonsdaleite for $S_1$ absorption. This large difference is there across all excited states, with respect to long AA, however short AA begin to converge energetically towards that of AB by the $S_3$ state. The higher states appear to have some degree of rydberg contamination (diffuse orbitals) beginning at the $S_3$ state, however this is due to the nanocrystal itself (terminating hydrogens) and is therefore not representative of a larger bulk material. 

\newpage {\color{white} INTENTIONAL WHITE SPACE}
\newpage {\color{white} INTENTIONAL WHITE SPACE}
\newpage {\color{white} INTENTIONAL WHITE SPACE}
\newpage {\color{white} INTENTIONAL WHITE SPACE}
\newpage {\color{white} INTENTIONAL WHITE SPACE}
\newpage {\color{white} INTENTIONAL WHITE SPACE}
\newpage
\section{Optimised coordinates}
\subsection{NV$^0$}
\subsubsection{AB diamond}
\begin{verbatim}
1       -3.254786000     10.322834000     -6.963026000
1       -2.201795000     10.547836000     -5.583647000
1       -3.563957000      9.193528000     -4.206799000
1       -4.750567000      9.758661000     -5.351829000
1       -4.820666000      7.819974000     -6.558316000
1       -6.107490000      7.664842000     -4.685887000
1       -4.819137000      7.161000000     -3.624053000
1       -6.150565000      5.763596000     -5.933784000
1       -6.130797000      5.201571000     -2.998561000
1       -7.432648000      5.597189000     -4.083773000
1       -7.723660000      3.492136000     -4.872070000
1       -8.860523000      0.648730000     -7.768714000
1       -8.285865000      0.934824000     -9.528681000
1       -7.643553000     -0.333360000    -10.549430000
1       -6.649670000      1.237283000    -12.019662000
1       -7.970363000      2.185537000    -11.394167000
1       -6.710254000      3.388882000     -9.942631000
1       -6.611187000      4.240482000    -12.046703000
1       -5.305172000      3.224563000    -12.597063000
1       -5.382733000      5.420042000    -10.544352000
1       -4.034467000      5.229082000    -13.204862000
1       -5.264755000      6.313544000    -12.620380000
1       -3.848300000      7.717275000    -11.558402000
1       -2.725765000      7.155516000    -12.775845000
1       -3.438444000      8.913859000    -10.320943000
1       -2.769808000     10.623137000     -8.746307000
1       -1.473617000     10.385449000     -9.918571000
1       -4.164398000      8.489057000     -8.078653000
1       -4.783111000      6.832732000     -9.655915000
1       -5.483127000      6.239546000     -7.496355000
1       -5.790143000      4.596387000     -9.099005000
1       -6.714027000      4.429569000     -6.970846000
1       -7.442463000      2.774284000     -8.443842000
1       -8.153213000      2.358993000     -6.211076000
1       -0.536159000     10.480664000     -7.600593000
1        0.555129000      9.204535000     -9.407904000
1        1.591630000      9.223404000     -7.168779000
1       -1.131816000      8.250531000    -11.051129000
1        0.964748000      7.010575000    -10.628456000
1        3.032229000      7.205702000     -7.575967000
1        2.594813000      7.982622000     -9.097842000
1        2.335213000      5.511954000     -9.152089000
1       -0.687450000      6.132800000    -12.216582000
1        1.244574000      4.808740000    -11.968665000
1        1.856507000      4.346966000    -10.397440000
1        1.596158000      2.260309000    -11.219467000
1        0.308871000      2.662413000    -12.321506000
1        0.276392000      2.074287000     -9.391822000
1        0.254096000      0.181970000    -10.644288000
1       -1.017420000      0.686702000    -11.725550000
1       -1.050287000      0.043116000     -8.788841000
1       -1.089606000     -1.879843000     -9.987614000
1       -2.283449000     -1.340744000    -11.135751000
1       -3.628099000     -2.702678000     -9.746025000
1       -2.571776000     -2.457231000     -8.372783000
1        1.635107000      5.064468000     -6.904992000
1       -1.039530000      3.694902000    -12.557419000
1       -2.499732000      4.557044000    -12.856773000
1       -3.889301000      2.400485000    -12.256397000
1       -2.436027000      1.541525000    -11.950185000
1       -3.839523000     -0.602589000    -11.284557000
1       -5.298864000      0.261508000    -11.587049000
1       -5.697147000     -1.553676000     -9.934463000
1       -5.297760000     -2.636064000     -7.734476000
1        0.960988000      3.442238000     -8.492855000
1        0.132205000      2.918307000     -6.360825000
1       -0.663403000      1.568729000     -8.061599000
1       -1.022085000      1.061173000     -5.822129000
1       -1.685678000     -0.631888000     -7.280722000
1       -2.373936000     -1.051979000     -5.042360000
1       -3.059726000     -2.772749000     -6.600199000
1       -1.985616000      0.107322000     -3.716882000
1       -0.561409000      1.537380000     -2.682883000
1       -1.787960000      2.635048000     -2.119349000
1       -0.440093000      2.381391000     -4.775919000
1        0.795192000      3.583498000     -3.287559000
1       -0.510714000      4.607509000     -2.752150000
1        0.916579000      4.436549000     -5.394927000
1        0.825133000      6.598232000     -3.317547000
1        2.168589000      5.674047000     -3.934627000
1        2.464805000      6.928251000     -5.804864000
1        1.804428000      8.184544000     -4.780838000
1       -0.158715000      9.370097000     -5.383037000
1       -2.039077000      8.364565000     -4.094921000
1       -0.575722000      7.503769000     -3.788889000
1       -3.390617000      6.283456000     -3.458350000
1       -1.930741000      5.422427000     -3.158127000
1       -3.279292000      3.357771000     -2.542991000
1       -4.737036000      4.218644000     -2.844827000
1       -8.434043000     -0.127461000     -6.243381000
1       -4.697569000     -0.446897000     -4.277468000
1       -3.083675000      0.685206000     -2.478981000
1       -5.125952000      1.765508000     -3.041794000
1       -6.838260000      0.819505000     -4.709169000
1       -6.400982000     -1.379493000     -5.931472000
1       -4.350832000     -2.542290000     -5.421252000
1       -7.429206000     -1.375728000     -8.165514000
1       -7.103724000      3.065880000     -3.290060000
6       -5.724156000      2.668457000     -4.908334000
6       -2.220919000      3.464639000     -5.118329000
6       -3.031482000      6.809840000     -6.151327000
6       -5.622341000      1.567574000     -8.442448000
6       -2.407781000      2.456436000     -8.780407000
6       -2.949534000      5.653079000     -9.639684000
6       -4.015702000      2.430380000     -6.597410000
6       -4.590018000      5.707095000     -7.844483000
6       -5.933275000      3.702163000     -7.243646000
7       -2.529050000      4.706158000     -7.284547000
6       -4.351275000      4.723756000     -5.548708000
6       -4.062236000      3.433606000     -9.115541000
6       -5.013448000      3.409733000     -6.074068000
6       -5.065980000      4.341340000     -8.319220000
6       -3.615114000      5.492518000     -6.700632000
6       -6.818285000      3.534597000     -4.250232000
6       -4.637922000      2.308700000     -3.874704000
6       -6.337991000      1.384326000     -5.521198000
6       -5.525927000      5.635856000     -5.044180000
6       -1.553801000      4.274730000     -6.284173000
6       -3.302963000      4.364683000     -4.460881000
6       -1.079813000      3.005895000     -4.144589000
6       -2.895944000      2.158732000     -5.647809000
6       -2.344619000      7.540658000     -7.336499000
6       -4.203103000      7.710994000     -5.659674000
6       -1.956941000      6.431807000     -5.073086000
6       -4.925915000      2.280046000     -9.651849000
6       -6.653857000      2.458279000     -7.745132000
6       -4.580578000      1.221575000     -7.339338000
6       -6.297638000      0.281908000     -8.979505000
6       -1.513089000      3.129399000     -9.837980000
6       -1.426096000      2.191012000     -7.582996000
6       -2.856429000      1.056815000     -9.233567000
6       -2.274732000      6.430948000    -10.799592000
6       -3.983842000      6.541091000     -8.959189000
6       -1.912505000      5.329237000     -8.511806000
6       -3.572602000      4.346366000    -10.255147000
6        0.472326000      7.555223000     -6.358568000
6       -4.821161000     -1.711034000     -7.356989000
6        0.487718000      6.457470000     -9.795037000
6       -3.993124000      3.588218000     -3.336339000
6       -5.170291000      7.074653000     -4.664831000
6       -0.122964000      4.089954000     -3.643905000
6       -1.301105000      7.718872000     -4.577775000
6       -7.358984000      1.707492000     -6.609077000
6       -3.160484000     -0.773626000     -5.761477000
6       -0.610131000      3.356255000     -7.039673000
6       -1.250215000      6.633504000     -7.983319000
6       -2.008854000      9.899971000     -9.082096000
6       -7.370812000      0.610934000    -10.042875000
6       -3.090634000     -1.920322000     -9.178389000
6       -3.892388000      1.357816000    -10.341181000
6       -4.739593000      4.815160000    -11.191425000
6       -0.327805000      2.207732000    -10.294629000
6       -1.194206000      5.543987000    -11.427764000
6       -6.490877000      5.020202000     -4.023223000
6       -1.481084000      2.059578000     -3.006528000
6       -2.646698000      5.653012000     -3.949937000
6       -3.834354000      9.144084000     -5.273468000
6       -5.271107000      0.459800000     -6.152362000
6       -1.903749000      1.320485000     -6.430372000
6       -3.353697000      7.848277000     -8.455221000
6        0.845420000      5.369300000     -7.607332000
6        0.063847000      8.655583000     -8.580936000
6       -5.211002000     -0.616599000     -9.598757000
6       -6.078720000      2.770404000    -10.589301000
6       -1.658432000      0.157872000     -9.691339000
6       -2.505558000      3.485176000    -10.966108000
6       -3.300065000      6.826753000    -11.890247000
6        0.956474000      4.311453000    -11.024448000
6       -3.555269000      1.387335000     -4.471135000
6       -0.850199000      5.520460000     -5.711122000
6       -1.695329000      8.837192000     -6.796222000
6       -6.962333000     -0.445355000     -7.789040000
6       -3.509932000      0.318212000     -8.016189000
6       -0.853870000      4.414992000     -9.215759000
6       -1.634556000      7.710972000    -10.223686000
6       -4.200028000      0.120040000     -5.089844000
6       -4.149858000     -0.988567000     -8.546542000
6       -2.531133000      1.001009000     -3.383721000
6        0.285074000      5.059429000     -4.750812000
6       -0.635159000      8.439176000     -5.748466000
6       -5.904548000     -0.817945000     -6.747443000
6       -2.501631000      0.004034000     -6.907545000
6        0.158111000      4.095302000     -8.121804000
6       -0.574213000      7.373869000     -9.161497000
6       -7.032658000      1.684035000    -11.089299000
6       -2.015504000     -1.280149000    -10.069561000
6       -3.180020000      2.171486000    -11.458237000
6       -4.346062000      5.783603000    -12.306418000
6        0.651059000      2.831023000    -11.288957000
6       -2.745931000      9.775126000     -6.157866000
6       -8.010993000      0.422460000     -7.103773000
6       -2.673060000      8.624800000     -9.584575000
6       -4.558813000      0.067290000    -10.806492000
6       -5.688060000      3.735945000    -11.700395000
6       -0.680128000      0.776199000    -10.680995000
6       -1.807039000      4.288974000    -12.055915000
6       -0.186009000      6.264534000     -6.901483000
6       -1.015357000      9.557434000     -7.979101000
6       -0.149188000      5.175214000    -10.369579000
6        1.222881000      6.156619000     -4.244706000
6        1.544862000      7.238094000     -5.290059000
6        1.129179000      8.290884000     -7.545345000
6       -3.821090000     -2.052699000     -6.258810000
6        1.534785000      6.141796000     -8.734210000
6        2.178371000      7.429604000     -8.236187000
\end{verbatim}

\subsubsection{AB lonsdaleite}
\begin{verbatim}
1       -3.254786000     10.322834000     -6.963026000
1       -2.201795000     10.547836000     -5.583647000
1       -3.563957000      9.193528000     -4.206799000
1       -4.750567000      9.758661000     -5.351829000
1       -4.820666000      7.819974000     -6.558316000
1       -6.107490000      7.664842000     -4.685887000
1       -4.819137000      7.161000000     -3.624053000
1       -6.150565000      5.763596000     -5.933784000
1       -6.130797000      5.201571000     -2.998561000
1       -7.432648000      5.597189000     -4.083773000
1       -7.723660000      3.492136000     -4.872070000
1       -8.860523000      0.648730000     -7.768714000
1       -8.285865000      0.934824000     -9.528681000
1       -7.643553000     -0.333360000    -10.549430000
1       -6.649670000      1.237283000    -12.019662000
1       -7.970363000      2.185537000    -11.394167000
1       -6.710254000      3.388882000     -9.942631000
1       -6.611187000      4.240482000    -12.046703000
1       -5.305172000      3.224563000    -12.597063000
1       -5.382733000      5.420042000    -10.544352000
1       -4.034467000      5.229082000    -13.204862000
1       -5.264755000      6.313544000    -12.620380000
1       -3.848300000      7.717275000    -11.558402000
1       -2.725765000      7.155516000    -12.775845000
1       -3.438444000      8.913859000    -10.320943000
1       -2.769808000     10.623137000     -8.746307000
1       -1.473617000     10.385449000     -9.918571000
1       -4.164398000      8.489057000     -8.078653000
1       -4.783111000      6.832732000     -9.655915000
1       -5.483127000      6.239546000     -7.496355000
1       -5.790143000      4.596387000     -9.099005000
1       -6.714027000      4.429569000     -6.970846000
1       -7.442463000      2.774284000     -8.443842000
1       -8.153213000      2.358993000     -6.211076000
1       -0.536159000     10.480664000     -7.600593000
1        0.555129000      9.204535000     -9.407904000
1        1.591630000      9.223404000     -7.168779000
1       -1.131816000      8.250531000    -11.051129000
1        0.964748000      7.010575000    -10.628456000
1        3.032229000      7.205702000     -7.575967000
1        2.594813000      7.982622000     -9.097842000
1        2.335213000      5.511954000     -9.152089000
1       -0.687450000      6.132800000    -12.216582000
1        1.244574000      4.808740000    -11.968665000
1        1.856507000      4.346966000    -10.397440000
1        1.596158000      2.260309000    -11.219467000
1        0.308871000      2.662413000    -12.321506000
1        0.276392000      2.074287000     -9.391822000
1        0.254096000      0.181970000    -10.644288000
1       -1.017420000      0.686702000    -11.725550000
1       -1.050287000      0.043116000     -8.788841000
1       -1.089606000     -1.879843000     -9.987614000
1       -2.283449000     -1.340744000    -11.135751000
1       -3.628099000     -2.702678000     -9.746025000
1       -2.571776000     -2.457231000     -8.372783000
1        1.635107000      5.064468000     -6.904992000
1       -1.039530000      3.694902000    -12.557419000
1       -2.499732000      4.557044000    -12.856773000
1       -3.889301000      2.400485000    -12.256397000
1       -2.436027000      1.541525000    -11.950185000
1       -3.839523000     -0.602589000    -11.284557000
1       -5.298864000      0.261508000    -11.587049000
1       -5.697147000     -1.553676000     -9.934463000
1       -5.297760000     -2.636064000     -7.734476000
1        0.960988000      3.442238000     -8.492855000
1        0.132205000      2.918307000     -6.360825000
1       -0.663403000      1.568729000     -8.061599000
1       -1.022085000      1.061173000     -5.822129000
1       -1.685678000     -0.631888000     -7.280722000
1       -2.373936000     -1.051979000     -5.042360000
1       -3.059726000     -2.772749000     -6.600199000
1       -1.985616000      0.107322000     -3.716882000
1       -0.561409000      1.537380000     -2.682883000
1       -1.787960000      2.635048000     -2.119349000
1       -0.440093000      2.381391000     -4.775919000
1        0.795192000      3.583498000     -3.287559000
1       -0.510714000      4.607509000     -2.752150000
1        0.916579000      4.436549000     -5.394927000
1        0.825133000      6.598232000     -3.317547000
1        2.168589000      5.674047000     -3.934627000
1        2.464805000      6.928251000     -5.804864000
1        1.804428000      8.184544000     -4.780838000
1       -0.158715000      9.370097000     -5.383037000
1       -2.039077000      8.364565000     -4.094921000
1       -0.575722000      7.503769000     -3.788889000
1       -3.390617000      6.283456000     -3.458350000
1       -1.930741000      5.422427000     -3.158127000
1       -3.279292000      3.357771000     -2.542991000
1       -4.737036000      4.218644000     -2.844827000
1       -8.434043000     -0.127461000     -6.243381000
1       -4.697569000     -0.446897000     -4.277468000
1       -3.083675000      0.685206000     -2.478981000
1       -5.125952000      1.765508000     -3.041794000
1       -6.838260000      0.819505000     -4.709169000
1       -6.400982000     -1.379493000     -5.931472000
1       -4.350832000     -2.542290000     -5.421252000
1       -7.429206000     -1.375728000     -8.165514000
1       -7.103724000      3.065880000     -3.290060000
6       -5.724156000      2.668457000     -4.908334000
6       -2.220919000      3.464639000     -5.118329000
6       -3.031482000      6.809840000     -6.151327000
6       -5.622341000      1.567574000     -8.442448000
6       -2.407781000      2.456436000     -8.780407000
6       -2.949534000      5.653079000     -9.639684000
6       -4.015702000      2.430380000     -6.597410000
6       -4.590018000      5.707095000     -7.844483000
6       -5.933275000      3.702163000     -7.243646000
7       -2.529050000      4.706158000     -7.284547000
6       -4.351275000      4.723756000     -5.548708000
6       -4.062236000      3.433606000     -9.115541000
6       -5.013448000      3.409733000     -6.074068000
6       -5.065980000      4.341340000     -8.319220000
6       -3.615114000      5.492518000     -6.700632000
6       -6.818285000      3.534597000     -4.250232000
6       -4.637922000      2.308700000     -3.874704000
6       -6.337991000      1.384326000     -5.521198000
6       -5.525927000      5.635856000     -5.044180000
6       -1.553801000      4.274730000     -6.284173000
6       -3.302963000      4.364683000     -4.460881000
6       -1.079813000      3.005895000     -4.144589000
6       -2.895944000      2.158732000     -5.647809000
6       -2.344619000      7.540658000     -7.336499000
6       -4.203103000      7.710994000     -5.659674000
6       -1.956941000      6.431807000     -5.073086000
6       -4.925915000      2.280046000     -9.651849000
6       -6.653857000      2.458279000     -7.745132000
6       -4.580578000      1.221575000     -7.339338000
6       -6.297638000      0.281908000     -8.979505000
6       -1.513089000      3.129399000     -9.837980000
6       -1.426096000      2.191012000     -7.582996000
6       -2.856429000      1.056815000     -9.233567000
6       -2.274732000      6.430948000    -10.799592000
6       -3.983842000      6.541091000     -8.959189000
6       -1.912505000      5.329237000     -8.511806000
6       -3.572602000      4.346366000    -10.255147000
6        0.472326000      7.555223000     -6.358568000
6       -4.821161000     -1.711034000     -7.356989000
6        0.487718000      6.457470000     -9.795037000
6       -3.993124000      3.588218000     -3.336339000
6       -5.170291000      7.074653000     -4.664831000
6       -0.122964000      4.089954000     -3.643905000
6       -1.301105000      7.718872000     -4.577775000
6       -7.358984000      1.707492000     -6.609077000
6       -3.160484000     -0.773626000     -5.761477000
6       -0.610131000      3.356255000     -7.039673000
6       -1.250215000      6.633504000     -7.983319000
6       -2.008854000      9.899971000     -9.082096000
6       -7.370812000      0.610934000    -10.042875000
6       -3.090634000     -1.920322000     -9.178389000
6       -3.892388000      1.357816000    -10.341181000
6       -4.739593000      4.815160000    -11.191425000
6       -0.327805000      2.207732000    -10.294629000
6       -1.194206000      5.543987000    -11.427764000
6       -6.490877000      5.020202000     -4.023223000
6       -1.481084000      2.059578000     -3.006528000
6       -2.646698000      5.653012000     -3.949937000
6       -3.834354000      9.144084000     -5.273468000
6       -5.271107000      0.459800000     -6.152362000
6       -1.903749000      1.320485000     -6.430372000
6       -3.353697000      7.848277000     -8.455221000
6        0.845420000      5.369300000     -7.607332000
6        0.063847000      8.655583000     -8.580936000
6       -5.211002000     -0.616599000     -9.598757000
6       -6.078720000      2.770404000    -10.589301000
6       -1.658432000      0.157872000     -9.691339000
6       -2.505558000      3.485176000    -10.966108000
6       -3.300065000      6.826753000    -11.890247000
6        0.956474000      4.311453000    -11.024448000
6       -3.555269000      1.387335000     -4.471135000
6       -0.850199000      5.520460000     -5.711122000
6       -1.695329000      8.837192000     -6.796222000
6       -6.962333000     -0.445355000     -7.789040000
6       -3.509932000      0.318212000     -8.016189000
6       -0.853870000      4.414992000     -9.215759000
6       -1.634556000      7.710972000    -10.223686000
6       -4.200028000      0.120040000     -5.089844000
6       -4.149858000     -0.988567000     -8.546542000
6       -2.531133000      1.001009000     -3.383721000
6        0.285074000      5.059429000     -4.750812000
6       -0.635159000      8.439176000     -5.748466000
6       -5.904548000     -0.817945000     -6.747443000
6       -2.501631000      0.004034000     -6.907545000
6        0.158111000      4.095302000     -8.121804000
6       -0.574213000      7.373869000     -9.161497000
6       -7.032658000      1.684035000    -11.089299000
6       -2.015504000     -1.280149000    -10.069561000
6       -3.180020000      2.171486000    -11.458237000
6       -4.346062000      5.783603000    -12.306418000
6        0.651059000      2.831023000    -11.288957000
6       -2.745931000      9.775126000     -6.157866000
6       -8.010993000      0.422460000     -7.103773000
6       -2.673060000      8.624800000     -9.584575000
6       -4.558813000      0.067290000    -10.806492000
6       -5.688060000      3.735945000    -11.700395000
6       -0.680128000      0.776199000    -10.680995000
6       -1.807039000      4.288974000    -12.055915000
6       -0.186009000      6.264534000     -6.901483000
6       -1.015357000      9.557434000     -7.979101000
6       -0.149188000      5.175214000    -10.369579000
6        1.222881000      6.156619000     -4.244706000
6        1.544862000      7.238094000     -5.290059000
6        1.129179000      8.290884000     -7.545345000
6       -3.821090000     -2.052699000     -6.258810000
6        1.534785000      6.141796000     -8.734210000
6        2.178371000      7.429604000     -8.236187000
\end{verbatim}

\subsubsection{Cross-plane (short) lonsdaleite}
\begin{verbatim}
1        1.490652000      5.355814000      2.523229000
1       -0.998782000      5.327771000      2.707621000
1       -3.145585000      4.068234000      2.872606000
1       -5.245297000      2.837701000      2.882547000
1       -6.533854000      0.686475000      3.140243000
1       -6.537094000     -1.851409000      3.216486000
1       -5.319731000     -3.988571000      3.251981000
1       -3.102149000     -5.230558000      3.324234000
1       -0.924602000     -6.463708000      3.292692000
1        1.565924000     -6.432944000      3.096734000
1        3.696399000     -5.219103000      2.787971000
1        3.661930000      4.108832000      2.468271000
1        4.900558000      1.945527000      2.536196000
1        4.908277000     -0.546268000      2.595930000
1        4.930401000     -3.039439000      2.686462000
1       -7.007159000      0.389560000     -7.705318000
1       -5.707219000      2.550516000     -7.681925000
1       -3.616078000      3.772944000     -7.916433000
1       -1.463813000      5.033239000     -8.008152000
1        1.032569000      5.060217000     -8.044515000
1        3.199822000      3.812555000     -8.110936000
1        3.229615000     -5.515075000     -7.919640000
1        1.087453000     -6.736065000     -7.981690000
1       -1.415053000     -6.771306000     -7.959405000
1       -3.590959000     -5.533178000     -7.875198000
1       -5.795161000     -4.284841000     -7.681977000
1       -7.010261000     -2.146571000     -7.648548000
1       -6.869485000      0.467684000     -3.384669000
1       -5.585982000      2.624374000     -3.354460000
1       -3.463170000      3.843134000     -3.642403000
1       -1.307912000      5.103713000     -3.766939000
1        1.171929000      5.137408000     -3.736964000
1        3.352920000      3.892800000     -3.798897000
1        3.369443000     -5.397014000     -3.559636000
1        1.241505000     -6.650497000     -3.619642000
1       -1.258915000     -6.668589000     -3.385042000
1       -3.422035000     -5.432465000     -3.324083000
1       -5.603394000     -4.168708000     -3.265611000
1       -6.849890000     -2.059288000     -3.326541000
1        1.097622000      5.087876000     -5.966160000
1       -1.373434000      5.089995000     -5.743607000
1       -3.557751000      3.807121000     -5.646813000
1       -5.690861000      2.578236000     -5.707293000
1       -6.943152000      0.422710000     -5.518397000
1       -6.932130000     -2.099773000     -5.427595000
1       -5.716533000     -4.218677000     -5.482103000
1       -3.528091000     -5.496949000     -5.534389000
1       -1.348152000     -6.754972000     -5.591456000
1        1.152781000     -6.734708000     -5.579316000
1        3.290933000     -5.467516000     -5.799056000
1        3.262708000      3.847584000     -6.028876000
1        1.263403000      5.197165000     -1.777274000
1       -1.203244000      5.183167000     -1.543038000
1       -3.369373000      3.908338000     -1.404409000
1       -5.517161000      2.678346000     -1.433100000
1       -6.775227000      0.523619000     -1.180865000
1       -6.759905000     -1.988362000     -1.114127000
1       -5.557479000     -4.112899000     -1.150534000
1       -3.340847000     -5.392830000     -1.215918000
1       -1.156913000     -6.636849000     -1.270875000
1        1.345709000     -6.585482000     -1.259374000
1        3.448561000     -5.345328000     -1.561019000
1        3.436700000      3.960567000     -1.837031000
1        4.703891000      1.833272000     -1.714720000
1        4.699975000     -0.697413000     -1.640709000
1        4.704747000     -3.196017000     -1.567434000
1        4.598540000     -3.237902000     -3.789847000
1        4.588058000     -0.754149000     -3.862311000
1        4.581506000      1.734815000     -3.934997000
1        4.540877000     -3.317815000     -5.778213000
1        4.529677000     -0.801950000     -5.819633000
1        4.522835000      1.710921000     -5.900322000
1        4.431413000      1.646288000     -8.161528000
1        4.462817000     -3.334941000     -8.039649000
1       -6.669264000      0.600561000      0.956305000
1       -5.411463000      2.759250000      0.917978000
1       -3.286888000      3.981952000      0.607862000
1       -1.104026000      5.264418000      0.443311000
1        1.374351000      5.272693000      0.449232000
1        3.547678000      4.031463000      0.390507000
1        3.570776000     -5.286405000      0.675985000
1        1.429511000     -6.554669000      0.700858000
1       -1.066532000     -6.567994000      0.936568000
1       -3.241657000     -5.309239000      0.992884000
1       -5.426251000     -4.043293000      1.062489000
1       -6.658512000     -1.928137000      0.996501000
1        4.813810000     -3.146881000      0.428115000
1        4.803115000     -0.630875000      0.330775000
1        4.798221000      1.890251000      0.273874000
1        4.440165000     -0.842571000     -8.086788000
1       -5.251986000     -3.292082000     -9.031438000
1       -3.007704000     -4.437463000     -9.125513000
1       -0.831638000     -5.666327000     -9.201056000
1        2.632658000     -4.547378000     -9.267976000
1        2.648532000     -2.119947000     -9.233160000
1        2.624651000      0.357087000     -9.285811000
1        2.572303000      2.792086000     -9.401586000
1        0.413843000      4.043082000     -9.342279000
1       -1.681040000      2.857128000     -9.183595000
1       -3.483436000      1.595903000     -9.184559000
1       -5.668382000      1.610275000     -9.151253000
1       -5.673731000     -0.915095000     -9.055526000
1       -3.524241000     -0.886541000     -9.135277000
1       -1.660223000     -2.126923000     -9.099539000
1        0.495874000     -3.354604000     -9.159023000
1        0.483056000     -0.877285000     -9.163015000
1        0.462720000      1.605221000     -9.219775000
1       -1.665493000      0.360282000     -9.112504000
1        0.994753000      4.412771000      3.925178000
1        3.155626000      3.165384000      3.865907000
1        3.201976000      0.724877000      3.879603000
1        3.224476000     -1.751371000      3.961362000
1        3.216480000     -4.181045000      4.130444000
1       -0.242011000     -5.289369000      4.415126000
1       -2.419636000     -4.068353000      4.460201000
1       -4.664929000     -2.929006000      4.499071000
1       -5.091493000     -0.550255000      4.432841000
1       -5.078818000      1.979372000      4.391872000
1       -2.899123000      1.963949000      4.239900000
1       -1.104577000      3.218857000      4.012475000
1        1.037831000      1.972073000      3.931598000
1        1.051756000     -0.509587000      4.005700000
1        1.076302000     -2.986519000      4.141169000
1       -1.078418000     -1.753465000      4.197929000
1       -2.943206000     -0.512362000      4.332441000
1       -1.087170000      0.718302000      4.071691000
6       -1.165873000      0.684390000      2.970936000
6       -1.179047000      3.199970000      2.913529000
6       -1.159029000     -1.819168000      3.099736000
6        0.990006000      1.948179000      2.828882000
6        0.968615000      4.439552000      2.826190000
6        1.015025000     -3.054831000      3.041927000
6        3.183342000     -1.799644000      2.861317000
6        3.167198000      0.700070000      2.778634000
6        3.145926000      3.187575000      2.766466000
6        3.186576000     -4.279887000      3.035680000
6       -3.278117000      1.938475000      3.204963000
6       -3.288395000     -0.570589000      3.286930000
6       -0.468998000     -0.583835000      2.481160000
6       -0.479152000      1.929032000      2.403501000
6       -0.480055000      4.430194000      2.344134000
6       -0.436014000     -3.050456000      2.591540000
6        1.708578000     -1.823372000      2.430858000
6        1.696279000      0.685019000      2.336854000
6        1.691692000      3.203223000      2.284115000
6        1.748021000     -4.291669000      2.544605000
6        3.875433000     -3.077257000      2.382618000
6        3.855576000     -0.564444000      2.278019000
6        3.848569000      1.947895000      2.219863000
6       -2.660716000      0.663026000      2.602697000
6       -2.651503000      3.154386000      2.512232000
6       -2.626920000     -1.803816000      2.679998000
6       -0.379223000     -5.510247000      3.346154000
6       -2.562465000     -4.273123000      3.387772000
6       -4.819594000      1.931092000      3.325231000
6       -4.846080000     -0.572824000      3.359794000
6       -4.764989000     -3.053904000      3.410611000
6       -1.209454000     -4.401607000      2.711316000
6       -3.392215000     -3.163078000      2.771207000
6       -5.500862000      0.678333000      2.761691000
6       -5.509217000     -1.854540000      2.827364000
6        0.966813000     -5.627370000      2.656697000
6        1.001744000     -0.553870000      2.903557000
6       -1.567237000      0.441522000     -5.437728000
6       -1.577069000      2.956502000     -5.484816000
6       -1.563807000     -2.088701000     -5.371945000
6        0.608939000      1.703839000     -5.528800000
6        0.586436000      4.195651000     -5.582868000
6        0.617365000     -3.328527000     -5.410785000
6        2.800847000     -2.058971000     -5.529309000
6        2.793529000      0.455683000     -5.588293000
6        2.768524000      2.947096000     -5.642580000
6        2.788771000     -4.552979000     -5.459648000
6       -3.686522000      1.682773000     -5.370560000
6       -3.684207000     -0.825447000     -5.304280000
6       -0.906497000     -0.857485000     -7.518321000
6       -0.915756000      1.654469000     -7.568492000
6       -0.752215000      1.721292000     -3.367833000
6       -0.915383000      4.154259000     -7.642917000
6       -0.755467000      4.222028000     -3.420462000
6       -0.874953000     -3.327274000     -7.483927000
6       -0.717953000     -3.206053000     -3.222964000
6        1.271269000     -2.101043000     -7.577972000
6        1.404257000     -1.995909000     -3.360993000
6        1.259666000      0.407412000     -7.620718000
6        1.402689000      0.496035000     -3.424268000
6        1.254313000      2.923972000     -7.704573000
6        1.413753000      3.013966000     -3.492636000
6        1.308400000     -4.570240000     -7.558731000
6        1.457019000     -4.466055000     -3.284761000
6        3.437918000     -3.354532000     -7.645173000
6        3.572004000     -3.250995000     -3.403940000
6        3.419053000     -0.841255000     -7.678603000
6        3.562205000     -0.745358000     -3.471990000
6        3.410382000      1.668222000     -7.757119000
6        3.559116000      1.761140000     -3.539667000
6       -3.103368000      0.386644000     -7.507165000
6       -2.908233000      0.465347000     -3.234657000
6       -3.090873000      2.878189000     -7.552424000
6       -2.918994000      2.957015000     -3.291177000
6       -3.066561000     -2.081183000     -7.448744000
6       -2.876882000     -1.975125000     -3.158601000
6       -0.851007000     -0.829399000     -5.947067000
6       -0.860365000      1.687609000     -5.999824000
6       -0.859498000      4.184588000     -6.088097000
6       -0.826277000     -3.308539000     -5.920942000
6        1.327192000     -2.076506000     -5.998375000
6        1.319975000      0.438751000     -6.051859000
6        1.311703000      2.955307000     -6.125265000
6        1.356149000     -4.553244000     -5.983821000
6        3.487346000     -3.331281000     -6.082302000
6        3.475741000     -0.811591000     -6.125521000
6        3.468521000      1.704486000     -6.201087000
6       -3.043454000      0.418439000     -5.928886000
6       -3.037774000      2.905216000     -5.994722000
6       -3.008893000     -2.051513000     -5.885755000
6       -1.484484000      0.446691000     -3.856159000
6       -1.646794000      0.384455000     -8.008996000
6       -1.511448000      2.973231000     -3.903199000
6       -1.659383000      2.898071000     -8.083007000
6       -1.482564000     -2.015092000     -3.804573000
6       -1.643019000     -2.124927000     -7.996583000
6        0.648456000     -0.768203000     -3.909527000
6        0.523618000     -0.857520000     -8.060023000
6        0.669936000      1.738213000     -3.952813000
6        0.511155000      1.644550000     -8.117427000
6        0.643567000      4.225208000     -4.039091000
6        0.485504000      4.131381000     -8.248502000
6        0.677323000     -3.275595000     -3.837900000
6        0.531248000     -3.361152000     -8.056334000
6        2.843941000     -2.010281000     -3.952508000
6        2.703191000     -2.104686000     -8.132869000
6        2.838654000      0.488094000     -4.017769000
6        2.686746000      0.394909000     -8.186545000
6        2.825711000      2.979495000     -4.099786000
6        2.660927000      2.877652000     -8.308833000
6        2.843902000     -4.502528000     -3.914170000
6        2.698736000     -4.588706000     -8.171159000
6       -3.613401000      1.707822000     -3.801450000
6       -3.773392000      1.629130000     -8.121463000
6       -3.619871000     -0.788750000     -3.739920000
6       -3.782238000     -0.881232000     -8.063178000
6       -0.804757000     -5.836419000     -5.315980000
6       -2.986016000     -4.578145000     -5.252962000
6       -5.236392000      1.670139000     -5.301581000
6       -5.245211000     -0.820353000     -5.260371000
6       -5.164869000     -3.309970000     -5.199775000
6       -1.651095000     -4.681206000     -7.470688000
6       -1.488339000     -4.575968000     -3.172818000
6       -3.833856000     -3.439872000     -7.408612000
6       -3.662552000     -3.336029000     -3.113018000
6       -5.944222000      0.401371000     -7.421584000
6       -5.805609000      0.482563000     -3.099911000
6       -5.951980000     -2.130195000     -7.352469000
6       -5.787752000     -2.034694000     -3.043986000
6        0.524399000     -5.908425000     -7.535458000
6        0.677609000     -5.808401000     -3.210725000
6       -1.604114000     -4.669923000     -5.891883000
6       -3.779618000     -3.412823000     -5.832779000
6       -5.897802000      0.426890000     -5.866501000
6       -5.894724000     -2.094816000     -5.791924000
6        0.566837000     -5.901648000     -5.976382000
6       -0.738030000     -5.775413000     -3.764277000
6       -0.876084000     -5.825045000     -8.113455000
6       -2.917209000     -4.528501000     -3.701361000
6       -3.057745000     -4.583314000     -8.035426000
6       -5.172605000      1.696252000     -3.755771000
6       -5.319269000      1.620272000     -8.109709000
6       -5.186548000     -0.786088000     -3.681138000
6       -5.340479000     -0.881127000     -8.006958000
6       -5.094962000     -3.268894000     -3.640264000
6       -5.257118000     -3.360536000     -7.933384000
6       -1.408915000      0.612810000     -1.126466000
6       -1.395916000      3.063811000     -1.257612000
6       -1.470566000     -2.089961000     -1.012811000
6        0.791301000      1.817658000     -1.327388000
6        0.762284000      4.305458000     -1.382042000
6        0.787347000     -3.193919000     -1.163300000
6        2.975908000     -1.941482000     -1.313005000
6        2.984950000      0.579309000     -1.387823000
6        2.947288000      3.061664000     -1.443309000
6        2.958882000     -4.427870000     -1.213406000
6       -3.505773000      1.788413000     -1.107151000
6       -3.503926000     -0.717223000     -1.024546000
6       -0.677897000      1.763254000     -1.785503000
6       -0.686242000      4.275027000     -1.878039000
6       -0.657414000     -3.162645000     -1.655622000
6        1.468651000     -1.944263000     -1.775940000
6        1.488874000      0.549987000     -1.848223000
6        1.484763000      3.061503000     -1.923601000
6        1.520677000     -4.419485000     -1.720402000
6        3.646064000     -3.213320000     -1.856003000
6        3.639765000     -0.700595000     -1.927174000
6        3.644675000      1.817463000     -2.001129000
6       -2.833011000      0.501672000     -1.660885000
6       -2.852622000      3.002266000     -1.747320000
6       -2.824626000     -1.945259000     -1.594780000
6       -0.620911000     -5.712950000     -0.998436000
6       -2.807605000     -4.464830000     -0.949905000
6       -5.062903000      1.774341000     -1.019972000
6       -5.068341000     -0.711928000     -0.960181000
6       -4.995699000     -3.204393000     -0.886523000
6       -1.426471000     -4.555634000     -1.597319000
6       -3.622098000     -3.311580000     -1.539707000
6       -5.738818000      0.524489000     -1.554113000
6       -5.723627000     -1.990583000     -1.481429000
6        0.743111000     -5.767541000     -1.662575000
6        0.904640000     -0.664214000     -1.200384000
6        0.623898000     -0.809286000     -5.480179000
6       -0.558493000      1.879493000      0.838557000
6       -0.559718000      4.377923000      0.790813000
6       -0.521214000     -3.112807000      1.036529000
6        1.627731000     -1.883319000      0.855041000
6        1.621877000      0.635355000      0.771948000
6        1.612670000      3.150606000      0.706323000
6        1.658594000     -4.356519000      0.973501000
6        3.790302000     -3.139049000      0.822631000
6        3.779071000     -0.618796000      0.726489000
6        3.773714000      1.900868000      0.665198000
6       -2.744075000      0.610491000      1.026886000
6       -2.736512000      3.098347000      0.956598000
6       -2.708648000     -1.853468000      1.124873000
6       -1.311366000      0.605399000      0.394119000
6       -1.324099000      3.121498000      0.319480000
6       -1.318211000     -1.929472000      0.476295000
6        0.881936000     -0.638754000      0.314435000
6        0.865315000      1.870035000      0.243716000
6        0.837231000      4.359661000      0.161540000
6        0.871101000     -3.161149000      0.403596000
6        3.058056000     -1.897107000      0.260406000
6        3.054167000      0.622659000      0.181021000
6        3.023111000      3.111997000      0.100198000
6        3.040669000     -4.389958000      0.330794000
6       -3.438367000      1.845790000      0.458803000
6       -3.435020000     -0.663153000      0.533464000
6       -1.296936000     -4.470665000      1.132451000
6       -3.473951000     -3.215143000      1.193096000
6       -5.597218000      0.619461000      1.210326000
6       -5.593054000     -1.905186000      1.267598000
6        0.875107000     -5.701669000      1.100784000
6       -0.550236000     -5.671712000      0.556348000
6       -2.730399000     -4.411222000      0.606187000
6       -4.990490000      1.829951000      0.524459000
6       -4.996410000     -0.660466000      0.618209000
6       -4.911790000     -3.151100000      0.676929000
6       -0.551415000     -0.639877000      0.907145000
7       -0.754412000     -0.768809000     -3.426595000
\end{verbatim}

\subsubsection{Cross-plane (long) lonsdaleite}
\begin{verbatim}
1       -7.022155000      0.377851000     -7.729793000
1       -5.772597000      2.549011000     -7.676150000
1       -3.611524000      3.784213000     -7.873246000
1       -1.432307000      5.070805000     -7.966476000
1        1.043361000      5.089899000     -7.906985000
1        3.238645000      3.831395000     -7.942640000
1        3.218134000     -5.487648000     -7.811405000
1        1.098487000     -6.740386000     -7.926782000
1       -1.393462000     -6.781754000     -7.788770000
1       -3.588372000     -5.538757000     -7.735750000
1       -5.785070000     -4.255383000     -7.673956000
1       -7.004150000     -2.140733000     -7.716959000
1       -6.866938000      0.458076000     -3.453204000
1       -5.526808000      2.583899000     -3.366213000
1       -3.506109000      3.777629000     -3.674400000
1       -1.330627000      5.112571000     -3.740651000
1        1.146322000      5.159588000     -3.686027000
1        3.353956000      3.893283000     -3.725238000
1        3.321541000     -5.397448000     -3.546572000
1        1.217598000     -6.682245000     -3.606699000
1       -1.267196000     -6.800222000     -3.420152000
1       -3.511685000     -5.527523000     -3.378516000
1       -5.754264000     -4.204789000     -3.334871000
1       -6.895083000     -2.044950000     -3.383815000
1        1.098009000      5.120602000     -5.915241000
1       -1.370506000      5.109578000     -5.730759000
1       -3.561838000      3.781168000     -5.632617000
1       -5.648168000      2.560335000     -5.712355000
1       -6.931633000      0.422341000     -5.523394000
1       -6.952035000     -2.080918000     -5.500627000
1       -5.796999000     -4.237217000     -5.548490000
1       -3.563607000     -5.547898000     -5.587167000
1       -1.333560000     -6.818524000     -5.624092000
1        1.162651000     -6.713749000     -5.568751000
1        3.255646000     -5.449152000     -5.785947000
1        3.286795000      3.863743000     -5.954313000
1        1.201716000      5.181723000     -1.707668000
1       -1.270175000      5.152514000     -1.508245000
1       -3.434359000      3.873415000     -1.437429000
1       -5.605460000      2.639989000     -1.445484000
1       -6.849114000      0.473966000     -1.243920000
1       -6.830092000     -2.039163000     -1.168284000
1       -5.624873000     -4.158422000     -1.193918000
1       -3.418220000     -5.448908000     -1.205438000
1       -1.207336000     -6.691164000     -1.238235000
1        1.287383000     -6.630199000     -1.251042000
1        3.389704000     -5.376443000     -1.537520000
1        3.396232000      3.924621000     -1.749747000
1        4.624905000      1.765449000     -1.621308000
1        4.611563000     -0.732796000     -1.599276000
1        4.619617000     -3.209414000     -1.555691000
1        4.500283000     -3.211854000     -3.769405000
1        4.513475000     -0.805893000     -3.823168000
1        4.561265000      1.729012000     -3.847553000
1        4.481507000     -3.282295000     -5.721826000
1        4.491132000     -0.828486000     -5.762221000
1        4.527592000      1.721704000     -5.826874000
1        4.470206000      1.678546000     -8.057252000
1        4.462168000     -3.333807000     -7.942021000
1       -6.740986000      0.557252000      1.121870000
1       -5.469124000      2.705013000      0.946205000
1       -3.372646000      3.945187000      0.666498000
1       -1.201979000      5.224262000      0.569410000
1        1.291950000      5.213128000      0.556919000
1        3.451989000      3.973239000      0.516610000
1        3.501223000     -5.318291000      0.725897000
1        1.402987000     -6.534475000      0.706244000
1       -1.108242000     -6.599544000      0.949633000
1       -3.308437000     -5.354259000      0.984232000
1       -5.520282000     -4.093513000      1.026912000
1       -6.764457000     -1.970803000      1.038871000
1        4.749072000     -3.161972000      0.472521000
1        4.719830000     -0.670436000      0.464810000
1        4.717473000      1.825258000      0.446605000
1       -3.683716000     -3.492430000      2.264255000
1       -2.611947000     -1.631493000      2.185433000
1       -5.406683000      0.572798000      2.273242000
1       -2.732272000      3.015032000      2.020946000
1       -0.593973000      4.264237000      1.914221000
1        1.528548000      3.080411000      1.818064000
1        3.623452000      1.834578000      1.826136000
1        3.636378000     -0.650843000      1.852825000
1        3.730433000     -3.109934000      1.906317000
1        0.731141000     -5.970235000      2.213926000
1       -1.486323000     -4.722463000      2.229464000
1       -0.405725000     -2.869002000      2.119826000
1        1.761949000     -4.055178000      2.069726000
1        1.577722000     -1.892419000      1.912693000
1        1.547300000      0.612857000      1.853744000
1       -0.597514000      1.855743000      1.904759000
1       -2.772904000      0.588751000      2.052038000
1       -0.588544000     -0.646421000      1.980006000
1       -5.537898000     -1.930088000      2.303664000
1        4.459283000     -0.839852000     -7.991219000
1        1.004441000      5.036503000    -10.172390000
1       -1.486348000      5.050070000    -10.053621000
1       -3.662555000      3.771534000     -9.981665000
1       -5.777152000      2.522558000    -10.078212000
1       -7.050422000      0.367065000    -10.092477000
1       -7.060250000     -2.152805000     -9.917103000
1       -5.808391000     -4.271605000     -9.899246000
1       -3.606386000     -5.527820000     -9.932991000
1       -1.420420000     -6.770029000     -9.981909000
1        1.102558000     -6.709018000     -9.884894000
1        3.194530000     -5.505577000    -10.078382000
1        3.167062000      3.793615000    -10.207552000
1        4.461804000     -3.347306000     -9.986837000
1        4.438180000     -0.851772000    -10.069254000
1        4.434555000      1.648828000    -10.140157000
1       -5.789021000      0.338272000    -11.322385000
1       -3.118774000      0.352461000    -11.244672000
1       -3.102399000      2.784336000    -11.331414000
1       -0.950473000      4.039402000    -11.392347000
1        1.173911000      2.861220000    -11.368494000
1        3.261109000      1.612311000    -11.451678000
1        3.272355000     -0.884801000    -11.387757000
1        3.352811000     -3.347264000    -11.353381000
1        1.380590000     -4.279638000    -11.357451000
1        0.344157000     -6.200228000    -11.370965000
1       -1.855712000     -4.928637000    -11.300803000
1       -4.042571000     -3.702338000    -11.255632000
1       -5.907856000     -2.157699000    -11.250420000
1       -2.943586000     -1.861428000    -11.282064000
1       -0.958723000     -0.878865000    -11.225445000
1       -0.972622000      1.607471000    -11.278232000
1        1.177152000      0.365227000    -11.303050000
1        1.197711000     -2.107970000    -11.261667000
1       -0.795654000     -3.064122000    -11.296300000
6       -1.513585000      0.450224000     -5.477431000
6       -1.568498000      2.976099000     -5.482864000
6       -1.747086000     -2.397741000     -5.236764000
6        0.618253000      1.739383000     -5.527219000
6        0.589534000      4.221460000     -5.547139000
6        0.565250000     -3.326172000     -5.413907000
6        2.732743000     -2.042828000     -5.504991000
6        2.788004000      0.473819000     -5.553925000
6        2.777418000      2.966090000     -5.583926000
6        2.744746000     -4.542343000     -5.440534000
6       -3.649893000      1.652229000     -5.386535000
6       -3.671523000     -0.859378000     -5.330235000
6       -0.878711000     -0.810124000     -7.554457000
6       -0.894522000      1.691962000     -7.590008000
6       -0.757532000      1.724753000     -3.387947000
6       -0.894894000      4.178738000     -7.621296000
6       -0.780417000      4.227489000     -3.397252000
6       -0.885292000     -3.327847000     -7.498803000
6       -0.800018000     -3.320635000     -3.229754000
6        1.264728000     -2.089577000     -7.583379000
6        1.258832000     -1.988153000     -3.361032000
6        1.288596000      0.441012000     -7.619589000
6        1.355690000      0.510687000     -3.429735000
6        1.285131000      2.959851000     -7.669870000
6        1.398155000      3.032335000     -3.454851000
6        1.300350000     -4.561279000     -7.550387000
6        1.391431000     -4.492877000     -3.282941000
6        3.421258000     -3.341951000     -7.596503000
6        3.473904000     -3.252463000     -3.382668000
6        3.421751000     -0.826597000     -7.631573000
6        3.484804000     -0.760983000     -3.441322000
6        3.437514000      1.693921000     -7.687211000
6        3.528586000      1.754888000     -3.477851000
6       -3.075419000      0.393499000     -7.511808000
6       -2.862258000      0.378788000     -3.268525000
6       -3.069938000      2.891447000     -7.534319000
6       -2.930570000      2.916286000     -3.311205000
6       -3.047595000     -2.087797000     -7.457772000
6       -2.966452000     -2.074642000     -3.184124000
7       -0.772253000     -0.669459000     -6.083078000
6       -0.845596000      1.728430000     -6.019464000
6       -0.852473000      4.204660000     -6.071209000
6       -0.861930000     -3.353348000     -5.938892000
6        1.268797000     -2.059077000     -6.002911000
6        1.335618000      0.477048000     -6.054143000
6        1.327081000      2.987956000     -6.095955000
6        1.323821000     -4.539185000     -5.979683000
6        3.433602000     -3.311869000     -6.043444000
6        3.442617000     -0.804889000     -6.085156000
6        3.476673000      1.720282000     -6.140409000
6       -3.001333000      0.389304000     -5.945514000
6       -3.016342000      2.899152000     -5.990357000
6       -3.037023000     -2.102226000     -5.906309000
6       -1.396943000      0.448158000     -3.921053000
6       -1.632192000      0.421006000     -8.078032000
6       -1.520842000      2.980947000     -3.907703000
6       -1.646884000      2.936314000     -8.125574000
6       -1.708388000     -2.373188000     -3.895140000
6       -1.636542000     -2.119210000     -8.041375000
6        0.567050000     -0.680138000     -3.961906000
6        0.551915000     -0.828627000     -8.104459000
6        0.665837000      1.773984000     -3.947280000
6        0.544741000      1.685404000     -8.144674000
6        0.628985000      4.245262000     -3.999574000
6        0.518746000      4.180133000     -8.222493000
6        0.592906000     -3.302781000     -3.844280000
6        0.526073000     -3.365425000     -8.095554000
6        2.738790000     -2.003475000     -3.935694000
6        2.717559000     -2.090302000     -8.138412000
6        2.808440000      0.494041000     -3.988576000
6        2.728255000      0.426711000     -8.186631000
6        2.817977000      2.989881000     -4.039047000
6        2.712903000      2.921823000     -8.256609000
6        2.780151000     -4.511944000     -3.900867000
6        2.699507000     -4.585861000     -8.157184000
6       -3.574991000      1.648269000     -3.832306000
6       -3.771337000      1.650192000     -8.070305000
6       -3.622966000     -0.847962000     -3.777933000
6       -3.763348000     -0.876778000     -8.040354000
6       -0.827646000     -5.886367000     -5.324790000
6       -3.029430000     -4.629883000     -5.287882000
6       -5.203522000      1.641927000     -5.320464000
6       -5.241736000     -0.842220000     -5.299187000
6       -5.223321000     -3.346492000     -5.252633000
6       -1.654940000     -4.700341000     -7.499017000
6       -1.559824000     -4.700798000     -3.157839000
6       -3.836274000     -3.449778000     -7.457454000
6       -3.771446000     -3.429283000     -3.119540000
6       -5.959469000      0.398706000     -7.438459000
6       -5.811356000      0.449895000     -3.137626000
6       -5.947079000     -2.124078000     -7.413897000
6       -5.834965000     -2.064188000     -3.091609000
6        0.523175000     -5.906816000     -7.515523000
6        0.627275000     -5.853816000     -3.206083000
6       -1.627221000     -4.729314000     -5.920117000
6       -3.836748000     -3.458392000     -5.879602000
6       -5.892055000      0.413292000     -5.887601000
6       -5.909099000     -2.104410000     -5.848147000
6        0.553780000     -5.896786000     -5.964298000
6       -0.787692000     -5.870306000     -3.766006000
6       -0.873629000     -5.865589000     -8.111757000
6       -2.993237000     -4.615679000     -3.719500000
6       -3.068310000     -4.622760000     -8.061105000
6       -5.138927000      1.647863000     -3.775200000
6       -5.333653000      1.632187000     -8.077853000
6       -5.194618000     -0.821176000     -3.716899000
6       -5.332249000     -0.863972000     -8.020969000
6       -5.187366000     -3.327802000     -3.681307000
6       -5.252080000     -3.357000000     -8.016735000
6       -1.448165000      0.502995000     -1.204131000
6       -1.460872000      3.025150000     -1.265412000
6       -1.463186000     -2.029849000     -1.105821000
6        0.718495000      1.784422000     -1.325354000
6        0.698966000      4.284028000     -1.328810000
6        0.705211000     -3.255638000     -1.164643000
6        2.879933000     -1.972705000     -1.308358000
6        2.887504000      0.532963000     -1.357571000
6        2.889588000      3.029124000     -1.370996000
6        2.886179000     -4.466001000     -1.191391000
6       -3.581433000      1.746185000     -1.149602000
6       -3.576375000     -0.774939000     -1.067049000
6       -0.755252000     -0.769638000     -1.568461000
6       -0.742390000      1.752343000     -1.800347000
6       -0.743213000      4.253046000     -1.850804000
6       -0.736655000     -3.250276000     -1.668609000
6        1.391017000     -1.989793000     -1.778399000
6        1.412412000      0.517555000     -1.848543000
6        1.433545000      3.045627000     -1.880747000
6        1.454615000     -4.465690000     -1.713382000
6        3.559788000     -3.237662000     -1.838014000
6        3.554793000     -0.732495000     -1.899864000
6        3.573507000      1.779013000     -1.935129000
6       -2.905480000      0.465174000     -1.695911000
6       -2.911918000      2.966488000     -1.770178000
6       -2.901210000     -2.011750000     -1.631744000
6       -0.677154000     -5.760107000     -0.979563000
6       -2.880479000     -4.521337000     -0.946612000
6       -5.142459000      1.737452000     -1.038308000
6       -5.142868000     -0.756937000     -0.997379000
6       -5.067559000     -3.251682000     -0.916515000
6       -1.500005000     -4.620694000     -1.584109000
6       -3.685917000     -3.369880000     -1.547433000
6       -5.802599000      0.483851000     -1.589434000
6       -5.789891000     -2.035553000     -1.525579000
6        0.683580000     -5.813570000     -1.656183000
6        0.705794000     -0.729351000     -1.258173000
6        0.585208000     -0.754505000     -5.511930000
6       -0.653620000      1.822114000      0.802156000
6       -0.652548000      4.307485000      0.817056000
6       -0.606303000     -3.155978000      1.072783000
6        1.529829000     -1.929748000      0.809974000
6        1.514757000      0.576309000      0.750405000
6        1.507875000      3.081277000      0.716453000
6        1.580071000     -4.378828000      1.031926000
6        3.705087000     -3.165864000      0.808827000
6        3.660091000     -0.670512000      0.752801000
6        3.657008000      1.831531000      0.727082000
6       -2.823449000      0.554760000      0.950071000
6       -2.812194000      3.036634000      0.923638000
6       -2.796105000     -1.913486000      1.134933000
6       -1.395229000      0.555177000      0.386721000
6       -1.407432000      3.071532000      0.318698000
6       -1.401434000     -1.986460000      0.490008000
6        0.785823000     -0.692882000      0.322699000
6        0.771982000      1.815384000      0.248278000
6        0.752631000      4.315354000      0.225360000
6        0.779915000     -3.214210000      0.412997000
6        2.964418000     -1.937952000      0.266975000
6        2.954119000      0.568659000      0.219152000
6        2.938609000      3.061255000      0.182689000
6        2.966363000     -4.428683000      0.367296000
6       -3.517990000      1.794471000      0.424189000
6       -3.520461000     -0.726722000      0.501744000
6       -1.374485000     -4.506666000      1.155474000
6       -3.569494000     -3.265031000      1.193008000
6       -5.643819000      0.558610000      1.199233000
6       -5.679795000     -1.957289000      1.212855000
6        0.819998000     -5.719883000      1.147962000
6       -0.600871000     -5.699099000      0.573225000
6       -2.804790000     -4.452789000      0.601789000
6       -5.063380000      1.782525000      0.515313000
6       -5.080170000     -0.713870000      0.578992000
6       -5.005223000     -3.201898000      0.642838000
6       -0.645481000     -0.688746000      0.878157000
6       -1.663415000      0.390030000     -9.645422000
6       -1.679948000      2.911651000     -9.709546000
6       -1.668506000     -2.136790000     -9.625854000
6        0.506312000      1.654329000     -9.713673000
6        0.484794000      4.152249000     -9.779426000
6        0.504171000     -3.380318000     -9.671839000
6        2.698554000     -2.107940000     -9.714335000
6        2.692096000      0.401825000     -9.763126000
6        2.672702000      2.895606000     -9.812918000
6        2.685327000     -4.601302000     -9.719566000
6       -3.794225000      1.633662000     -9.640875000
6       -3.795322000     -0.889334000     -9.606883000
6       -0.943061000     -0.868393000    -10.122141000
6       -0.952195000      1.637695000    -10.174864000
6       -0.950909000      4.123770000    -10.295998000
6       -0.919869000     -3.331130000    -10.232830000
6        1.228586000     -2.111528000    -10.158416000
6        1.218412000      0.390808000    -10.200211000
6        1.213850000      2.900500000    -10.268122000
6        1.261597000     -4.567923000    -10.300355000
6        3.398646000     -3.359343000    -10.255159000
6        3.362824000     -0.860259000    -10.291177000
6        3.359253000      1.646085000    -10.356819000
6       -3.121705000      0.370249000    -10.141972000
6       -3.117685000      2.852145000    -10.233244000
6       -3.094927000     -2.096813000    -10.215814000
6       -0.887669000     -5.860754000     -9.666504000
6       -3.082933000     -4.613995000     -9.611850000
6       -5.344330000      1.617370000     -9.637468000
6       -5.356553000     -0.878571000     -9.600309000
6       -5.277550000     -3.365033000     -9.576751000
6       -1.688520000     -4.684367000    -10.240249000
6       -3.874597000     -3.445583000    -10.198269000
6       -5.959899000      0.367457000    -10.236162000
6       -5.987584000     -2.144479000    -10.153197000
6        0.495398000     -5.910049000    -10.322300000
6        0.514572000     -0.855933000     -9.661416000
\end{verbatim}

\subsection{NV$^-$}
\subsubsection{AB diamond}
\begin{verbatim}
1       -3.254283000     10.323378000     -6.962778000
1       -2.204233000     10.543672000     -5.578443000
1       -3.572559000      9.186530000     -4.211371000
1       -4.756843000      9.753523000     -5.360062000
1       -4.821510000      7.810295000     -6.566378000
1       -6.113982000      7.657340000     -4.697980000
1       -4.824942000      7.155125000     -3.632664000
1       -6.153931000      5.744323000     -5.938369000
1       -6.134553000      5.211559000     -3.000489000
1       -7.436339000      5.595950000     -4.092857000
1       -7.739319000      3.471803000     -4.831134000
1       -8.861028000      0.652742000     -7.769578000
1       -8.281548000      0.936234000     -9.523528000
1       -7.638182000     -0.328830000    -10.551433000
1       -6.645394000      1.249000000    -12.012098000
1       -7.964488000      2.198738000    -11.380269000
1       -6.697042000      3.392429000     -9.928055000
1       -6.613287000      4.243166000    -12.036932000
1       -5.307676000      3.225554000    -12.592333000
1       -5.373427000      5.422579000    -10.542950000
1       -4.033078000      5.234077000    -13.210445000
1       -5.264412000      6.319907000    -12.624241000
1       -3.844434000      7.722855000    -11.565051000
1       -2.721944000      7.160246000    -12.784980000
1       -3.436313000      8.917070000    -10.323677000
1       -2.769674000     10.624882000     -8.744285000
1       -1.471276000     10.390218000     -9.918025000
1       -4.163887000      8.487853000     -8.081911000
1       -4.783281000      6.832508000     -9.657150000
1       -5.482936000      6.221428000     -7.503848000
1       -5.762987000      4.578640000     -9.098279000
1       -6.705213000      4.428919000     -6.986853000
1       -7.439884000      2.772947000     -8.438757000
1       -8.150368000      2.365302000     -6.211321000
1       -0.536509000     10.484464000     -7.596182000
1        0.562558000      9.217355000     -9.404760000
1        1.595336000      9.224391000     -7.163721000
1       -1.128823000      8.261543000    -11.056710000
1        0.973821000      7.017732000    -10.633007000
1        3.032537000      7.205039000     -7.573734000
1        2.598096000      7.986471000     -9.096647000
1        2.337328000      5.514285000     -9.155392000
1       -0.680144000      6.143556000    -12.223762000
1        1.248212000      4.812428000    -11.978467000
1        1.859975000      4.350576000    -10.404727000
1        1.600774000      2.262437000    -11.223009000
1        0.312575000      2.664019000    -12.326967000
1        0.272815000      2.085175000     -9.393546000
1        0.254067000      0.184096000    -10.633946000
1       -1.018027000      0.684733000    -11.719359000
1       -1.055739000      0.060359000     -8.777316000
1       -1.083176000     -1.865893000     -9.969590000
1       -2.275356000     -1.332134000    -11.124971000
1       -3.624085000     -2.697044000     -9.746975000
1       -2.573693000     -2.452671000     -8.366195000
1        1.635241000      5.062991000     -6.909805000
1       -1.031298000      3.707060000    -12.574702000
1       -2.492912000      4.570783000    -12.871809000
1       -3.890732000      2.397850000    -12.262826000
1       -2.435818000      1.538607000    -11.956997000
1       -3.839643000     -0.607469000    -11.280984000
1       -5.300812000      0.256755000    -11.584923000
1       -5.698987000     -1.555624000     -9.929662000
1       -5.292898000     -2.643012000     -7.735272000
1        0.960233000      3.441510000     -8.494780000
1        0.117638000      2.908764000     -6.369839000
1       -0.689756000      1.584384000     -8.072237000
1       -1.022342000      1.065409000     -5.839460000
1       -1.690098000     -0.630351000     -7.277526000
1       -2.369910000     -1.046576000     -5.041834000
1       -3.057935000     -2.773033000     -6.597764000
1       -2.007200000      0.089767000     -3.676615000
1       -0.558134000      1.532201000     -2.689630000
1       -1.778598000      2.636537000     -2.119015000
1       -0.453411000      2.366287000     -4.782694000
1        0.796622000      3.570900000     -3.299888000
1       -0.509447000      4.596526000     -2.759614000
1        0.912411000      4.430789000     -5.404142000
1        0.822048000      6.587826000     -3.320143000
1        2.168367000      5.666492000     -3.939284000
1        2.464455000      6.929251000     -5.803821000
1        1.799406000      8.182466000     -4.776106000
1       -0.158109000      9.369993000     -5.383966000
1       -2.040132000      8.361578000     -4.096037000
1       -0.575806000      7.499786000     -3.791252000
1       -3.391609000      6.280224000     -3.461144000
1       -1.930270000      5.418215000     -3.162344000
1       -3.278234000      3.359918000     -2.542441000
1       -4.736555000      4.221069000     -2.845367000
1       -8.436021000     -0.125677000     -6.241349000
1       -4.697829000     -0.464015000     -4.264606000
1       -3.093207000      0.701483000     -2.438374000
1       -5.129775000      1.765307000     -3.022316000
1       -6.857238000      0.812345000     -4.700522000
1       -6.422134000     -1.413893000     -5.934863000
1       -4.350871000     -2.541465000     -5.416363000
1       -7.437719000     -1.373890000     -8.170261000
1       -7.096357000      3.073173000     -3.245468000
6       -5.732796000      2.664499000     -4.896427000
6       -2.231148000      3.454505000     -5.125736000
6       -3.032253000      6.805409000     -6.155337000
6       -5.614065000      1.564125000     -8.430662000
6       -2.436224000      2.488381000     -8.787870000
6       -2.944502000      5.657054000     -9.648651000
6       -3.993475000      2.454240000     -6.638566000
6       -4.582560000      5.697889000     -7.848608000
6       -5.917533000      3.694518000     -7.239449000
7       -2.512620000      4.729360000     -7.298490000
6       -4.352764000      4.708696000     -5.554285000
6       -4.020119000      3.424296000     -9.108757000
6       -5.002858000      3.381857000     -6.071739000
6       -5.038082000      4.324265000     -8.316967000
6       -3.611391000      5.483857000     -6.707413000
6       -6.823564000      3.523124000     -4.223313000
6       -4.643009000      2.302250000     -3.866558000
6       -6.344372000      1.377457000     -5.510452000
6       -5.525716000      5.623370000     -5.049617000
6       -1.560927000      4.272585000     -6.293199000
6       -3.305631000      4.357995000     -4.465778000
6       -1.088345000      2.997991000     -4.152345000
6       -2.924729000      2.154276000     -5.653813000
6       -2.343344000      7.540647000     -7.339676000
6       -4.205776000      7.704016000     -5.666140000
6       -1.958498000      6.428032000     -5.076030000
6       -4.914452000      2.279305000     -9.643480000
6       -6.646159000      2.456675000     -7.741925000
6       -4.583322000      1.228130000     -7.306759000
6       -6.294641000      0.282997000     -8.972933000
6       -1.511761000      3.134710000     -9.851557000
6       -1.453529000      2.206561000     -7.591948000
6       -2.864390000      1.067191000     -9.228407000
6       -2.270068000      6.437215000    -10.806325000
6       -3.980812000      6.539113000     -8.963166000
6       -1.906291000      5.338655000     -8.516396000
6       -3.565686000      4.348493000    -10.267391000
6        0.472229000      7.554353000     -6.359587000
6       -4.823837000     -1.713928000     -7.352541000
6        0.492763000      6.463591000     -9.800668000
6       -3.995075000      3.585149000     -3.336129000
6       -5.175845000      7.065782000     -4.674141000
6       -0.124368000      4.078059000     -3.653087000
6       -1.302237000      7.715554000     -4.579864000
6       -7.356482000      1.707819000     -6.604615000
6       -3.162876000     -0.770938000     -5.757534000
6       -0.620506000      3.357309000     -7.047219000
6       -1.244430000      6.640804000     -7.991136000
6       -2.007366000      9.903005000     -9.081910000
6       -7.366130000      0.613536000    -10.038060000
6       -3.090916000     -1.914973000     -9.172536000
6       -3.892228000      1.357076000    -10.341589000
6       -4.733960000      4.819001000    -11.194250000
6       -0.327715000      2.214424000    -10.298869000
6       -1.188898000      5.551326000    -11.436385000
6       -6.494915000      5.015644000     -4.023616000
6       -1.483075000      2.053489000     -3.006491000
6       -2.648661000      5.647503000     -3.953699000
6       -3.840133000      9.138259000     -5.279178000
6       -5.281692000      0.445051000     -6.143783000
6       -1.918319000      1.330280000     -6.432560000
6       -3.351871000      7.847812000     -8.458884000
6        0.845077000      5.369618000     -7.611913000
6        0.068389000      8.662521000     -8.581804000
6       -5.211610000     -0.617110000     -9.593591000
6       -6.067860000      2.775358000    -10.577448000
6       -1.661999000      0.171517000     -9.681016000
6       -2.500248000      3.490354000    -10.981551000
6       -3.296282000      6.832135000    -11.897709000
6        0.960152000      4.315617000    -11.032621000
6       -3.557725000      1.379972000     -4.458456000
6       -0.852289000      5.516708000     -5.715329000
6       -1.695627000      8.836757000     -6.797253000
6       -6.964750000     -0.447660000     -7.785507000
6       -3.520094000      0.325222000     -8.007555000
6       -0.850885000      4.419854000     -9.225163000
6       -1.630132000      7.717987000    -10.229059000
6       -4.206062000      0.113239000     -5.079173000
6       -4.151998000     -0.985021000     -8.540119000
6       -2.545942000      0.994810000     -3.359279000
6        0.282569000      5.052553000     -4.757215000
6       -0.635777000      8.437909000     -5.749194000
6       -5.915260000     -0.831467000     -6.736079000
6       -2.507408000      0.009657000     -6.906709000
6        0.155676000      4.095960000     -8.126937000
6       -0.568188000      7.382383000     -9.166641000
6       -7.026584000      1.692662000    -11.078500000
6       -2.012469000     -1.269658000    -10.056712000
6       -3.178907000      2.171239000    -11.465027000
6       -4.344940000      5.788412000    -12.311196000
6        0.654398000      2.833226000    -11.293867000
6       -2.747766000      9.772805000     -6.157545000
6       -8.011794000      0.425161000     -7.102113000
6       -2.670688000      8.627606000     -9.586604000
6       -4.559081000      0.064919000    -10.804084000
6       -5.686469000      3.739679000    -11.694369000
6       -0.680249000      0.780433000    -10.674631000
6       -1.801358000      4.296963000    -12.070672000
6       -0.185546000      6.264806000     -6.904908000
6       -1.014171000      9.560425000     -7.977863000
6       -0.144858000      5.181038000    -10.376668000
6        1.221100000      6.148671000     -4.248464000
6        1.542993000      7.235930000     -5.289078000
6        1.131140000      8.292919000     -7.543978000
6       -3.820228000     -2.052123000     -6.255149000
6        1.536268000      6.144279000     -8.737048000
6        2.180085000      7.431310000     -8.236010000
\end{verbatim}

\subsubsection{AB lonsdaleite}
\begin{verbatim}
1       -2.804266000      0.000156000      3.580921000
1       -0.663705000      1.195802000      3.487124000
1        1.657877000      0.210234000      3.556754000
1       -2.640285000     -2.213311000      3.671748000
1       -0.488489000     -3.395963000      3.636817000
1        1.675046000     -2.653035000      3.606652000
1        0.583001000     -1.330845000     -3.313947000
1        0.676474000     -3.763135000     -3.327003000
1       -1.500680000     -2.553471000     -3.294901000
1       -1.502647000     -0.126748000     -3.297546000
1        0.654287000      1.108686000     -3.362591000
1       -3.383636000     -1.345002000     -3.356701000
1       -6.868435000     -0.057127000     -1.955139000
1       -5.554398000      2.098917000     -1.879139000
1       -5.521243000      1.167615000     -3.362096000
1       -3.337762000      1.135806000     -3.401093000
1       -3.433065000      3.304811000     -2.129504000
1       -1.490889000      2.370358000     -3.359893000
1       -1.286066000      4.536715000     -2.156796000
1        1.204719000      4.551462000     -2.143359000
1        0.610956000      3.537497000     -3.465737000
1        3.347822000      3.328764000     -2.176736000
1        2.760872000      2.299525000     -3.491037000
1        2.860626000     -0.110945000     -3.405049000
1        2.873005000     -2.544028000     -3.368263000
1        3.374886000     -5.945097000     -2.051414000
1        2.805303000     -4.956252000     -3.408396000
1        1.269534000     -7.142724000     -2.101449000
1       -0.632306000     -6.084246000     -3.373856000
1       -1.229724000     -7.192491000     -2.126929000
1       -3.400968000     -5.988126000     -2.096505000
1       -2.809400000     -4.875578000     -3.341843000
1       -5.080273000     -3.726968000     -3.293888000
1       -5.651295000     -4.741614000     -1.959596000
1       -6.869953000     -2.611181000     -1.929645000
1       -5.528361000     -1.347480000     -3.286780000
1       -6.788005000     -0.009433000      2.627047000
1       -5.451628000     -0.018158000      3.793579000
1       -5.505458000      2.136869000      2.501287000
1       -2.762439000      2.403326000      3.604615000
1       -3.376211000      3.376451000      2.254681000
1       -0.605953000      3.635858000      3.525071000
1       -1.207367000      4.621272000      2.182998000
1        1.279006000      4.597363000      2.165331000
1        1.500358000      2.459545000      3.438495000
1        3.419433000      3.377361000      2.151231000
1        3.681268000      1.260463000      3.422860000
1        3.701386000     -1.224805000      3.456106000
1        3.718033000     -3.704853000      3.496727000
1        3.466358000     -5.870565000      2.302290000
1        1.704383000     -4.625991000      3.619447000
1        1.357737000     -7.091879000      2.236169000
1        0.642300000     -6.516526000      3.727754000
1       -1.154030000     -7.132993000      2.386822000
1       -1.530954000     -5.273655000      3.712026000
1       -3.325717000     -5.908469000      2.412700000
1       -3.717183000     -4.067475000      3.727552000
1       -5.537346000     -4.666022000      2.465027000
1       -6.801560000     -2.557264000      2.497992000
1       -5.577477000     -2.527734000      3.778891000
1        1.246428000      4.509796000     -0.092048000
1       -1.230673000      4.558230000      0.111066000
1       -3.399540000      3.342789000      0.142817000
1       -5.637947000      2.118748000      0.110927000
1       -6.870683000     -0.047763000      0.235257000
1       -6.846210000     -2.596607000      0.292733000
1       -5.619964000     -4.698536000      0.243084000
1       -3.350724000     -5.954850000      0.212897000
1       -1.187244000     -7.148813000      0.190793000
1        1.318600000     -7.071510000      0.284450000
1        3.370247000     -5.904705000      0.032373000
1        3.352858000      3.342356000     -0.118344000
1       -0.897699000     -1.226786000      3.604098000
1        6.814341000     -2.547731000     -2.227052000
1        5.574591000     -4.707530000     -2.111490000
1        4.977200000     -3.725358000     -3.456665000
1        4.953618000      1.087994000     -3.526827000
1        5.541045000      2.112728000     -2.208818000
1        6.801675000     -0.037160000     -2.264971000
1        5.034499000     -1.319856000     -3.433966000
1        6.911452000     -2.493571000      2.141921000
1        5.779363000     -2.471512000      3.503974000
1        5.643625000     -4.634768000      2.244698000
1        5.611775000      2.161358000      2.147809000
1        6.900130000      0.030037000      2.106722000
1        5.767728000      0.035994000      3.468928000
1        5.613340000     -4.672876000     -0.019692000
1        6.880239000     -2.541883000      0.041646000
1        6.869979000      0.016988000      0.003430000
1        5.582239000      2.133930000     -0.117002000
6       -1.498233000     -0.083891000      0.346874000
6       -1.457331000      2.429794000      0.333946000
6       -1.486932000     -2.507993000      0.410891000
6        0.725376000      1.074370000      0.322030000
6        0.705653000      3.630910000      0.293775000
6        0.740451000     -3.649575000      0.419066000
6        2.946064000     -2.496621000      0.322818000
6        2.935138000     -0.054759000      0.286352000
6        2.878464000      2.422488000      0.262702000
6        2.897140000     -4.975648000      0.391513000
6       -3.618946000      1.220355000      0.416226000
6       -3.620614000     -1.301286000      0.461757000
6       -0.797198000     -1.318206000     -1.661789000
6       -0.667656000     -1.253696000      2.476871000
6       -0.761618000      1.154890000     -1.739511000
6       -0.683316000      1.208142000      2.377683000
6       -0.747947000      3.643903000     -1.790920000
6       -0.663611000      3.690350000      2.422483000
6       -0.719548000     -3.746646000     -1.669802000
6       -0.641033000     -3.686784000      2.577823000
6        1.449119000     -2.522508000     -1.732642000
6        1.519620000     -2.475877000      2.483036000
6        1.440154000     -0.092071000     -1.763639000
6        1.508244000     -0.015253000      2.442403000
6        1.411531000      2.407521000     -1.817478000
6        1.482465000      2.462983000      2.333490000
6        1.466433000     -4.971750000     -1.712192000
6        1.538369000     -4.927934000      2.570861000
6        3.625167000     -3.799886000     -1.779009000
6        3.697384000     -3.734659000      2.392339000
6        3.633476000     -1.300994000     -1.792453000
6        3.692066000     -1.242633000      2.346537000
6        3.600825000      1.194266000     -1.849443000
6        3.672161000      1.248238000      2.317747000
6       -2.975764000     -0.076876000     -1.718536000
6       -2.858035000     -0.019708000      2.476393000
6       -2.929624000      2.395058000     -1.754214000
6       -2.831151000      2.447797000      2.501579000
6       -2.938695000     -2.540893000     -1.663621000
6       -2.828855000     -2.474395000      2.616266000
7       -0.914300000     -1.298474000     -0.186746000
6       -0.736896000      1.142252000     -0.178094000
6       -0.719809000      3.648185000     -0.241653000
6       -0.692232000     -3.693784000     -0.120805000
6        1.537588000     -2.416721000     -0.200476000
6        1.529456000     -0.160615000     -0.240393000
6        1.438463000      2.383058000     -0.246198000
6        1.486854000     -4.907166000     -0.155785000
6        3.660099000     -3.779999000     -0.195917000
6        3.693315000     -1.279807000     -0.222085000
6        3.639566000      1.214442000     -0.273813000
6       -2.977152000     -0.045515000     -0.140780000
6       -2.911958000      2.415810000     -0.199519000
6       -2.931780000     -2.523438000     -0.110853000
6       -1.420562000     -0.043853000      1.921764000
6       -1.509848000     -0.087805000     -2.191583000
6       -1.422300000      2.469759000      1.913661000
6       -1.489485000      2.402686000     -2.253356000
6       -1.419559000     -2.509966000      1.989557000
6       -1.505908000     -2.569333000     -2.188434000
6        0.773930000     -1.255911000      2.194119000
6        0.664950000     -1.315935000     -2.207128000
6        0.759033000      1.187137000      1.877576000
6        0.680496000      1.142054000     -2.254160000
6        0.738988000      3.689403000      1.839967000
6        0.658632000      3.618329000     -2.363608000
6        0.770105000     -3.724580000      1.985386000
6        0.698420000     -3.772622000     -2.218248000
6        2.952214000     -2.489859000      1.896551000
6        2.886939000     -2.540970000     -2.258827000
6        2.941212000     -0.015537000      1.858736000
6        2.874965000     -0.080596000     -2.296076000
6        2.909264000      2.456665000      1.809817000
6        2.822223000      2.381292000     -2.389043000
6        2.931935000     -4.971202000      1.942588000
6        2.856022000     -5.004153000     -2.303595000
6       -3.551953000      1.230229000      1.975197000
6       -3.624780000      1.165259000     -2.333994000
6       -3.564842000     -1.283349000      2.015255000
6       -3.645305000     -1.338809000     -2.282763000
6       -0.659766000     -6.227032000      0.502705000
6       -2.852261000     -5.014542000      0.518188000
6       -5.168463000      1.205585000      0.505608000
6       -5.174403000     -1.293384000      0.506676000
6       -5.068916000     -3.789129000      0.541065000
6       -1.482678000     -5.097179000     -1.657517000
6       -1.402787000     -5.041993000      2.638265000
6       -3.693393000     -3.896577000     -1.641776000
6       -3.595253000     -3.822782000      2.656313000
6       -5.808510000     -0.052820000     -1.643498000
6       -5.684503000     -0.015748000      2.712149000
6       -5.813797000     -2.589766000     -1.608012000
6       -5.711619000     -2.535061000      2.680334000
6        0.692802000     -6.302524000     -1.680183000
6        0.770069000     -6.261669000      2.659668000
6       -1.455295000     -5.067432000     -0.086194000
6       -3.676681000     -3.878999000     -0.072789000
6       -5.812239000     -0.038148000     -0.088199000
6       -5.796577000     -2.571717000     -0.048919000
6        0.709587000     -6.258722000     -0.133861000
6       -0.627660000     -6.220159000      2.052207000
6       -0.694012000     -6.236852000     -2.279374000
6       -2.817362000     -4.990002000      2.064052000
6       -2.882032000     -5.022390000     -2.246812000
6       -5.095073000      1.215515000      2.054853000
6       -5.168796000      1.164599000     -2.315234000
6       -5.120025000     -1.275757000      2.078838000
6       -5.201109000     -1.331196000     -2.231609000
6       -5.020729000     -3.761530000      2.098179000
6       -5.104171000     -3.809986000     -2.190156000
6        5.109742000     -3.753013000      0.320547000
6        5.150424000     -1.266102000      0.267790000
6        5.087996000      1.219014000      0.249067000
6        5.782956000     -2.552641000     -1.831106000
6        5.837625000     -2.489472000      2.400095000
6        5.770558000     -0.031520000     -1.868408000
6        5.826335000      0.023104000      2.364979000
6        5.825004000     -2.537212000     -0.276395000
6        5.814526000     -0.005411000     -0.313260000
6        5.129445000     -3.729784000      1.872789000
6        5.039317000     -3.773218000     -2.353315000
6        5.151936000     -1.243671000      1.848712000
6        5.067272000     -1.302317000     -2.327203000
6        5.106571000      1.240843000      1.802628000
6        5.014645000      1.167196000     -2.425167000
\end{verbatim}

\subsubsection{Cross-plane (short) lonsdaleite}
\begin{verbatim}
1        1.491255000      5.361753000      2.524059000
1       -0.999518000      5.327577000      2.703161000
1       -3.141556000      4.067160000      2.860482000
1       -5.236874000      2.839405000      2.863104000
1       -6.531482000      0.690105000      3.128188000
1       -6.537509000     -1.845171000      3.217524000
1       -5.324105000     -3.985129000      3.256045000
1       -3.104898000     -5.233673000      3.324557000
1       -0.923399000     -6.465905000      3.288794000
1        1.568431000     -6.429188000      3.090547000
1        3.694432000     -5.215864000      2.775462000
1        3.666684000      4.115892000      2.474512000
1        4.907706000      1.952888000      2.541597000
1        4.913233000     -0.545334000      2.593855000
1        4.931532000     -3.036737000      2.674411000
1       -7.007383000      0.390234000     -7.704356000
1       -5.705717000      2.551713000     -7.674359000
1       -3.617303000      3.774536000     -7.907871000
1       -1.467008000      5.034612000     -8.004154000
1        1.030582000      5.061883000     -8.043281000
1        3.197897000      3.814760000     -8.114283000
1        3.230479000     -5.513201000     -7.910060000
1        1.091827000     -6.733485000     -7.983872000
1       -1.412207000     -6.770874000     -7.964138000
1       -3.590113000     -5.533113000     -7.884349000
1       -5.795835000     -4.282581000     -7.694478000
1       -7.010742000     -2.144101000     -7.658185000
1       -6.867581000      0.460667000     -3.380376000
1       -5.577858000      2.615198000     -3.350284000
1       -3.458434000      3.828953000     -3.626250000
1       -1.311780000      5.094812000     -3.759916000
1        1.167876000      5.140992000     -3.736448000
1        3.356059000      3.898897000     -3.803147000
1        3.358885000     -5.386018000     -3.549670000
1        1.243529000     -6.641220000     -3.616809000
1       -1.257723000     -6.674772000     -3.389720000
1       -3.428515000     -5.451629000     -3.338667000
1       -5.622754000     -4.175721000     -3.278799000
1       -6.856394000     -2.059543000     -3.331649000
1        1.096564000      5.089372000     -5.966395000
1       -1.375666000      5.086487000     -5.738111000
1       -3.556212000      3.800776000     -5.636480000
1       -5.684136000      2.576490000     -5.701392000
1       -6.940229000      0.421204000     -5.516998000
1       -6.933613000     -2.098816000     -5.436076000
1       -5.720331000     -4.221828000     -5.498460000
1       -3.531082000     -5.500788000     -5.552329000
1       -1.350001000     -6.756015000     -5.598145000
1        1.152491000     -6.731124000     -5.578251000
1        3.286908000     -5.464301000     -5.788752000
1        3.262049000      3.851084000     -6.034065000
1        1.255979000      5.214096000     -1.776011000
1       -1.212707000      5.161983000     -1.537676000
1       -3.357858000      3.871218000     -1.391151000
1       -5.456673000      2.664784000     -1.449850000
1       -6.759195000      0.533814000     -1.177246000
1       -6.771384000     -1.976951000     -1.121982000
1       -5.623303000     -4.123585000     -1.129726000
1       -3.356481000     -5.432819000     -1.193019000
1       -1.147788000     -6.667781000     -1.261566000
1        1.349205000     -6.568222000     -1.260125000
1        3.429793000     -5.325196000     -1.566953000
1        3.452028000      3.977241000     -1.832338000
1        4.725778000      1.855291000     -1.725297000
1        4.690914000     -0.717661000     -1.637503000
1        4.672518000     -3.162150000     -1.558042000
1        4.586254000     -3.225409000     -3.776381000
1        4.582922000     -0.757606000     -3.858047000
1        4.584888000      1.737301000     -3.942903000
1        4.535785000     -3.313326000     -5.767785000
1        4.526877000     -0.802741000     -5.816007000
1        4.522715000      1.712783000     -5.903531000
1        4.431175000      1.648616000     -8.165591000
1        4.465230000     -3.331759000     -8.031173000
1       -6.658510000      0.609619000      0.944228000
1       -5.384342000      2.756375000      0.895857000
1       -3.277093000      3.966719000      0.596787000
1       -1.106177000      5.255269000      0.439915000
1        1.375623000      5.281629000      0.451451000
1        3.556279000      4.041730000      0.395941000
1        3.559329000     -5.277101000      0.667542000
1        1.429703000     -6.542717000      0.694839000
1       -1.071197000     -6.563559000      0.944217000
1       -3.242572000     -5.305534000      1.007646000
1       -5.426940000     -4.045752000      1.077588000
1       -6.660534000     -1.923685000      0.997449000
1        4.803046000     -3.137213000      0.417095000
1        4.802316000     -0.634540000      0.328418000
1        4.806946000      1.897608000      0.277815000
1        4.441310000     -0.840866000     -8.084954000
1       -5.250447000     -3.286808000     -9.043282000
1       -3.005035000     -4.436407000     -9.135717000
1       -0.826711000     -5.665653000     -9.207199000
1        2.638360000     -4.547083000     -9.264256000
1        2.652179000     -2.121235000     -9.233307000
1        2.625802000      0.356157000     -9.289232000
1        2.568574000      2.792775000     -9.405585000
1        0.409383000      4.044999000     -9.342755000
1       -1.685390000      2.860100000     -9.182898000
1       -3.484356000      1.600901000     -9.184413000
1       -5.670142000      1.616324000     -9.149345000
1       -5.674228000     -0.908296000     -9.061686000
1       -3.524198000     -0.882217000     -9.142332000
1       -1.659038000     -2.125342000     -9.108017000
1        0.500399000     -3.354270000     -9.163661000
1        0.485417000     -0.875994000     -9.168759000
1        0.462010000      1.608080000     -9.224132000
1       -1.666848000      0.363614000     -9.117677000
1        0.994179000      4.417383000      3.927537000
1        3.158184000      3.169361000      3.872079000
1        3.212577000      0.728290000      3.885043000
1        3.229820000     -1.750606000      3.961943000
1        3.220191000     -4.179369000      4.124115000
1       -0.242157000     -5.291924000      4.415685000
1       -2.422474000     -4.071367000      4.464298000
1       -4.666791000     -2.923958000      4.503562000
1       -5.094741000     -0.542159000      4.431254000
1       -5.078974000      1.988689000      4.380587000
1       -2.898525000      1.965048000      4.236645000
1       -1.103229000      3.220061000      4.012554000
1        1.043196000      1.981093000      3.938520000
1        1.070581000     -0.504651000      4.020743000
1        1.083275000     -2.983863000      4.147469000
1       -1.082577000     -1.761602000      4.217299000
1       -2.945457000     -0.504784000      4.341316000
1       -1.086958000      0.729344000      4.083659000
6       -1.159938000      0.685310000      2.981361000
6       -1.174459000      3.199150000      2.912528000
6       -1.158426000     -1.819949000      3.116485000
6        0.996140000      1.953217000      2.834705000
6        0.970696000      4.444114000      2.827797000
6        1.015404000     -3.051122000      3.047207000
6        3.185149000     -1.796201000      2.860884000
6        3.174227000      0.705268000      2.783163000
6        3.150705000      3.193547000      2.771936000
6        3.186387000     -4.276349000      3.028458000
6       -3.273354000      1.937495000      3.199539000
6       -3.285923000     -0.569199000      3.293389000
6       -0.462551000     -0.584867000      2.500549000
6       -0.472257000      1.927834000      2.406849000
6       -0.477314000      4.429693000      2.342433000
6       -0.437185000     -3.049572000      2.600613000
6        1.709001000     -1.817587000      2.434879000
6        1.702935000      0.690951000      2.344449000
6        1.696937000      3.209067000      2.287452000
6        1.745451000     -4.285831000      2.543305000
6        3.873940000     -3.072341000      2.375527000
6        3.858586000     -0.560478000      2.278059000
6        3.854972000      1.954394000      2.224310000
6       -2.652754000      0.659723000      2.603432000
6       -2.645022000      3.151075000      2.505884000
6       -2.627003000     -1.804114000      2.689898000
6       -0.379902000     -5.510649000      3.345432000
6       -2.565286000     -4.275464000      3.390631000
6       -4.815475000      1.934457000      3.314421000
6       -4.844153000     -0.568891000      3.358480000
6       -4.766576000     -3.051396000      3.414342000
6       -1.211681000     -4.401413000      2.713536000
6       -3.394133000     -3.163720000      2.774926000
6       -5.496116000      0.680657000      2.753232000
6       -5.508619000     -1.851632000      2.827749000
6        0.965835000     -5.622978000      2.653697000
6        1.008745000     -0.548968000      2.917694000
6       -1.566613000      0.439612000     -5.442586000
6       -1.574684000      2.951968000     -5.481843000
6       -1.568004000     -2.091773000     -5.376667000
6        0.610526000      1.704285000     -5.529763000
6        0.586086000      4.196043000     -5.582432000
6        0.611250000     -3.325441000     -5.410460000
6        2.794274000     -2.054964000     -5.526213000
6        2.793218000      0.458612000     -5.589002000
6        2.769098000      2.949846000     -5.645601000
6        2.782968000     -4.548304000     -5.453193000
6       -3.682126000      1.675388000     -5.366272000
6       -3.682898000     -0.831108000     -5.306201000
6       -0.906669000     -0.856968000     -7.523793000
6       -0.915828000      1.655251000     -7.570604000
6       -0.743769000      1.711698000     -3.364844000
6       -0.917298000      4.154876000     -7.640281000
6       -0.754086000      4.213840000     -3.417185000
6       -0.875162000     -3.326711000     -7.489941000
6       -0.732825000     -3.209221000     -3.218893000
6        1.271064000     -2.099510000     -7.580226000
6        1.385252000     -1.986250000     -3.357055000
6        1.259708000      0.409206000     -7.624236000
6        1.405695000      0.500653000     -3.420553000
6        1.253032000      2.925583000     -7.706274000
6        1.418532000      3.018293000     -3.492685000
6        1.308868000     -4.567608000     -7.558220000
6        1.444513000     -4.455498000     -3.281455000
6        3.438473000     -3.352053000     -7.639267000
6        3.555785000     -3.239273000     -3.397961000
6        3.418872000     -0.839580000     -7.677722000
6        3.555360000     -0.740923000     -3.469334000
6        3.409599000      1.670143000     -7.760337000
6        3.564029000      1.766680000     -3.541662000
6       -3.103658000      0.386646000     -7.509542000
6       -2.897205000      0.445203000     -3.230789000
6       -3.091308000      2.877952000     -7.547001000
6       -2.910320000      2.940297000     -3.284278000
6       -3.067048000     -2.081218000     -7.456348000
6       -2.885046000     -1.994366000     -3.153360000
6       -0.852141000     -0.830543000     -5.952314000
6       -0.859315000      1.685829000     -6.001520000
6       -0.860231000      4.182622000     -6.085225000
6       -0.829876000     -3.308430000     -5.926361000
6        1.323591000     -2.074158000     -5.999899000
6        1.320131000      0.440133000     -6.054804000
6        1.311544000      2.956300000     -6.126656000
6        1.352211000     -4.549834000     -5.982370000
6        3.483070000     -3.327566000     -6.075722000
6        3.473304000     -0.809712000     -6.124099000
6        3.468160000      1.706669000     -6.204186000
6       -3.042100000      0.414281000     -5.930465000
6       -3.035734000      2.900362000     -5.988400000
6       -3.010299000     -2.054391000     -5.893107000
6       -1.479474000      0.435656000     -3.859970000
6       -1.647195000      0.385630000     -8.013084000
6       -1.505743000      2.962727000     -3.899416000
6       -1.660985000      2.899235000     -8.081448000
6       -1.490487000     -2.021571000     -3.803831000
6       -1.642870000     -2.124218000     -8.003979000
6        0.640185000     -0.764673000     -3.911358000
6        0.524057000     -0.856165000     -8.064640000
6        0.674805000      1.738822000     -3.952009000
6        0.510590000      1.646266000     -8.120748000
6        0.643955000      4.225681000     -4.038767000
6        0.482983000      4.133099000     -8.248403000
6        0.662657000     -3.270166000     -3.834956000
6        0.532342000     -3.359880000     -8.059780000
6        2.829438000     -2.002421000     -3.947637000
6        2.704177000     -2.103620000     -8.131982000
6        2.838406000      0.493074000     -4.017185000
6        2.686908000      0.396038000     -8.189103000
6        2.829160000      2.984181000     -4.102643000
6        2.659075000      2.879317000     -8.312351000
6        2.832736000     -4.492738000     -3.907363000
6        2.701089000     -4.586830000     -8.166419000
6       -3.606268000      1.692174000     -3.794282000
6       -3.773753000      1.630818000     -8.120359000
6       -3.618107000     -0.804276000     -3.738659000
6       -3.782022000     -0.879760000     -8.069353000
6       -0.806960000     -5.836940000     -5.319909000
6       -2.990241000     -4.582597000     -5.262045000
6       -5.231176000      1.665572000     -5.298263000
6       -5.243475000     -0.824533000     -5.263061000
6       -5.169585000     -3.313491000     -5.209774000
6       -1.650598000     -4.680986000     -7.477514000
6       -1.498942000     -4.582983000     -3.174919000
6       -3.834166000     -3.439866000     -7.417868000
6       -3.677829000     -3.354426000     -3.117007000
6       -5.943400000      0.402215000     -7.421504000
6       -5.801450000      0.473512000     -3.100262000
6       -5.951903000     -2.128654000     -7.360780000
6       -5.792919000     -2.041408000     -3.048582000
6        0.526394000     -5.906513000     -7.537377000
6        0.672410000     -5.801368000     -3.210217000
6       -1.606319000     -4.670643000     -5.897822000
6       -3.782027000     -3.415207000     -5.840903000
6       -5.894819000      0.424721000     -5.866274000
6       -5.895586000     -2.096011000     -5.800011000
6        0.565353000     -5.899798000     -5.977945000
6       -0.742431000     -5.776861000     -3.767863000
6       -0.873281000     -5.824166000     -8.118950000
6       -2.927392000     -4.541877000     -3.709712000
6       -3.056745000     -4.582904000     -8.044982000
6       -5.166708000      1.686510000     -3.753302000
6       -5.319657000      1.622879000     -8.107468000
6       -5.185476000     -0.794520000     -3.683269000
6       -5.340199000     -0.878376000     -8.012378000
6       -5.105739000     -3.278116000     -3.649920000
6       -5.256724000     -3.358178000     -7.944614000
6       -1.305070000      0.506897000     -1.136293000
6       -1.378667000      3.038634000     -1.256608000
6       -1.503715000     -2.092179000     -0.997527000
6        0.811193000      1.834011000     -1.327799000
6        0.767089000      4.313930000     -1.381836000
6        0.753550000     -3.175478000     -1.159099000
6        2.936833000     -1.918717000     -1.311462000
6        2.998362000      0.596033000     -1.387299000
6        2.961223000      3.075732000     -1.443193000
6        2.937562000     -4.410069000     -1.213792000
6       -3.460834000      1.749398000     -1.107352000
6       -3.496118000     -0.744233000     -1.021770000
6       -0.655249000      1.745371000     -1.779236000
6       -0.682204000      4.261552000     -1.875009000
6       -0.696918000     -3.160835000     -1.637370000
6        1.414639000     -1.906987000     -1.769518000
6        1.508910000      0.568088000     -1.837830000
6        1.497234000      3.075517000     -1.923067000
6        1.498935000     -4.400382000     -1.717247000
6        3.614515000     -3.191398000     -1.852133000
6        3.629506000     -0.693343000     -1.924732000
6        3.662281000      1.830561000     -2.002074000
6       -2.786071000      0.456406000     -1.658060000
6       -2.831418000      2.973129000     -1.742964000
6       -2.840331000     -1.990848000     -1.572148000
6       -0.632838000     -5.725284000     -1.000610000
6       -2.829475000     -4.490988000     -0.952870000
6       -5.023537000      1.752581000     -1.030608000
6       -5.062122000     -0.723118000     -0.962213000
6       -5.026490000     -3.227461000     -0.887351000
6       -1.445574000     -4.576366000     -1.601521000
6       -3.660221000     -3.347161000     -1.546010000
6       -5.723883000      0.515955000     -1.556425000
6       -5.731653000     -1.997492000     -1.483203000
6        0.733268000     -5.757935000     -1.661673000
6        0.838065000     -0.594415000     -1.198982000
6        0.623174000     -0.806703000     -5.484339000
6       -0.547399000      1.871389000      0.839822000
6       -0.554865000      4.373101000      0.789001000
6       -0.529537000     -3.106226000      1.046170000
6        1.620247000     -1.869585000      0.856079000
6        1.628908000      0.643571000      0.778795000
6        1.620493000      3.158092000      0.708955000
6        1.648876000     -4.344272000      0.972343000
6        3.780461000     -3.129153000      0.815626000
6        3.778382000     -0.612944000      0.726237000
6        3.781753000      1.908691000      0.669106000
6       -2.726449000      0.597960000      1.023830000
6       -2.722828000      3.087000000      0.950697000
6       -2.706445000     -1.855740000      1.137022000
6       -1.285674000      0.585474000      0.399355000
6       -1.311566000      3.109997000      0.318995000
6       -1.327198000     -1.931795000      0.475623000
6        0.878211000     -0.618716000      0.320968000
6        0.876963000      1.877286000      0.247293000
6        0.842751000      4.365442000      0.162501000
6        0.857900000     -3.147084000      0.407207000
6        3.044385000     -1.885015000      0.259515000
6        3.060977000      0.632225000      0.183527000
6        3.032414000      3.121453000      0.103117000
6        3.028112000     -4.378213000      0.327233000
6       -3.413742000      1.829141000      0.455024000
6       -3.427393000     -0.672554000      0.535708000
6       -1.300029000     -4.464820000      1.130894000
6       -3.474808000     -3.212862000      1.192180000
6       -5.586340000      0.619038000      1.201638000
6       -5.593727000     -1.905748000      1.267597000
6        0.870456000     -5.693211000      1.097224000
6       -0.556287000     -5.668793000      0.555664000
6       -2.734750000     -4.410192000      0.604053000
6       -4.964948000      1.820791000      0.512962000
6       -4.990890000     -0.665172000      0.616141000
6       -4.917199000     -3.155472000      0.678954000
6       -0.544329000     -0.644192000      0.935638000
7       -0.758414000     -0.775638000     -3.448086000
\end{verbatim}

\subsubsection{Cross-plane (long) lonsdaleite}
\begin{verbatim}
1       -7.010307000      0.390851000     -7.737513000
1       -5.760529000      2.562484000     -7.684079000
1       -3.604052000      3.798388000     -7.878451000
1       -1.424880000      5.088510000     -7.969194000
1        1.052001000      5.109213000     -7.903659000
1        3.249616000      3.850610000     -7.930288000
1        3.227948000     -5.469453000     -7.778036000
1        1.111967000     -6.723988000     -7.901998000
1       -1.383231000     -6.765886000     -7.773779000
1       -3.577900000     -5.524763000     -7.731217000
1       -5.773927000     -4.241129000     -7.677307000
1       -6.993179000     -2.125436000     -7.726242000
1       -6.860484000      0.477108000     -3.471180000
1       -5.497770000      2.593737000     -3.367663000
1       -3.509611000      3.775945000     -3.685507000
1       -1.348793000      5.116600000     -3.740674000
1        1.131672000      5.187705000     -3.675679000
1        3.356808000      3.913308000     -3.706187000
1        3.311064000     -5.366120000     -3.511670000
1        1.214496000     -6.657187000     -3.581819000
1       -1.274354000     -6.784632000     -3.403683000
1       -3.520894000     -5.514041000     -3.372447000
1       -5.763565000     -4.189245000     -3.337954000
1       -6.897505000     -2.022629000     -3.395027000
1        1.100328000      5.144087000     -5.907083000
1       -1.372356000      5.126146000     -5.731604000
1       -3.559826000      3.790921000     -5.636642000
1       -5.633699000      2.575157000     -5.713187000
1       -6.920774000      0.439164000     -5.530497000
1       -6.946589000     -2.060427000     -5.509003000
1       -5.796574000     -4.222745000     -5.553078000
1       -3.562525000     -5.533640000     -5.582617000
1       -1.331804000     -6.803633000     -5.609412000
1        1.167037000     -6.691719000     -5.542391000
1        3.254960000     -5.427458000     -5.752048000
1        3.292255000      3.886675000     -5.937037000
1        1.182369000      5.203359000     -1.696476000
1       -1.292947000      5.165328000     -1.509683000
1       -3.449505000      3.885378000     -1.450366000
1       -5.618735000      2.655452000     -1.457251000
1       -6.864647000      0.487790000     -1.261871000
1       -6.847084000     -2.020419000     -1.179606000
1       -5.646300000     -4.143185000     -1.192033000
1       -3.437273000     -5.436756000     -1.191582000
1       -1.221467000     -6.677155000     -1.217066000
1        1.275374000     -6.606825000     -1.226209000
1        3.371879000     -5.350439000     -1.510428000
1        3.380206000      3.945948000     -1.728905000
1        4.603659000      1.781546000     -1.595001000
1        4.587042000     -0.710742000     -1.574180000
1        4.596530000     -3.178995000     -1.529886000
1        4.473388000     -3.167327000     -3.740894000
1        4.490478000     -0.791774000     -3.796588000
1        4.548492000      1.736927000     -3.820292000
1        4.476486000     -3.256321000     -5.684418000
1        4.488420000     -0.811363000     -5.731905000
1        4.530098000      1.741181000     -5.802597000
1        4.482759000      1.697556000     -8.035363000
1        4.472712000     -3.315619000     -7.906003000
1       -6.766349000      0.576887000      1.111537000
1       -5.492396000      2.725879000      0.933784000
1       -3.399911000      3.963570000      0.656768000
1       -1.231338000      5.244392000      0.566596000
1        1.265895000      5.235532000      0.567840000
1        3.424719000      3.997907000      0.537141000
1        3.478300000     -5.293659000      0.752643000
1        1.383018000     -6.510970000      0.732622000
1       -1.130480000     -6.579960000      0.970304000
1       -3.332795000     -5.335155000      0.998334000
1       -5.546726000     -4.074753000      1.028500000
1       -6.790573000     -1.949513000      1.031063000
1        4.724534000     -3.134559000      0.498676000
1        4.694102000     -0.646475000      0.491384000
1        4.692014000      1.847809000      0.470792000
1       -3.716987000     -3.475719000      2.274417000
1       -2.657164000     -1.603583000      2.212763000
1       -5.431150000      0.595135000      2.266027000
1       -2.762230000      3.035975000      2.017972000
1       -0.627710000      4.286274000      1.918005000
1        1.495898000      3.101877000      1.832217000
1        3.591847000      1.857481000      1.848490000
1        3.603995000     -0.624989000      1.877708000
1        3.702340000     -3.081597000      1.932720000
1        0.705258000     -5.943900000      2.239403000
1       -1.516987000     -4.705022000      2.249162000
1       -0.421635000     -2.855308000      2.154566000
1        1.734277000     -4.027267000      2.093583000
1        1.551202000     -1.877191000      1.944183000
1        1.521766000      0.649534000      1.879953000
1       -0.618399000      1.889177000      1.921819000
1       -2.815276000      0.608803000      2.066889000
1       -0.617347000     -0.623336000      2.007628000
1       -5.566219000     -1.906984000      2.301215000
1        4.470662000     -0.823477000     -7.962893000
1        1.020595000      5.052558000    -10.170350000
1       -1.471966000      5.064142000    -10.057679000
1       -3.648555000      3.783892000     -9.989835000
1       -5.762531000      2.534389000    -10.088995000
1       -7.034265000      0.376689000    -10.102725000
1       -7.043153000     -2.141627000     -9.926147000
1       -5.790700000     -4.261235000     -9.903582000
1       -3.588190000     -5.516938000     -9.929098000
1       -1.401634000     -6.758461000     -9.967989000
1        1.121767000     -6.695394000     -9.858723000
1        3.214234000     -5.493405000    -10.046198000
1        3.185402000      3.809542000    -10.196421000
1        4.480367000     -3.333725000     -9.952687000
1        4.457142000     -0.838125000    -10.043261000
1        4.453062000      1.664497000    -10.119871000
1       -5.769645000      0.347842000    -11.331387000
1       -3.103412000      0.363031000    -11.251334000
1       -3.084521000      2.795449000    -11.339485000
1       -0.931164000      4.052499000    -11.396187000
1        1.196001000      2.876930000    -11.364771000
1        3.284188000      1.625533000    -11.437920000
1        3.296594000     -0.875092000    -11.368881000
1        3.377688000     -3.338020000    -11.326727000
1        1.405960000     -4.269775000    -11.337194000
1        0.369442000     -6.190685000    -11.351602000
1       -1.832799000     -4.920046000    -11.292501000
1       -4.021638000     -3.693728000    -11.256401000
1       -5.886457000     -2.147428000    -11.257936000
1       -2.923131000     -1.852041000    -11.284377000
1       -0.938301000     -0.869930000    -11.226015000
1       -0.952352000      1.622909000    -11.282151000
1        1.202015000      0.379100000    -11.297859000
1        1.222594000     -2.101547000    -11.249912000
1       -0.771317000     -3.056142000    -11.289439000
6       -1.502394000      0.468621000     -5.487077000
6       -1.564125000      2.991487000     -5.483724000
6       -1.749458000     -2.386905000     -5.215318000
6        0.621831000      1.762365000     -5.521337000
6        0.590641000      4.243505000     -5.540625000
6        0.560789000     -3.304326000     -5.394635000
6        2.723643000     -2.019213000     -5.481664000
6        2.786405000      0.495400000     -5.536283000
6        2.780102000      2.988455000     -5.568184000
6        2.740819000     -4.519415000     -5.411295000
6       -3.637456000      1.661948000     -5.390681000
6       -3.663023000     -0.846150000     -5.329004000
6       -0.866965000     -0.790233000     -7.556654000
6       -0.883974000      1.710753000     -7.592414000
6       -0.754916000      1.734516000     -3.384503000
6       -0.887051000      4.196347000     -7.621831000
6       -0.789723000      4.237400000     -3.394256000
6       -0.877451000     -3.310086000     -7.486202000
6       -0.812760000     -3.300913000     -3.203948000
6        1.272303000     -2.071390000     -7.569419000
6        1.221228000     -1.953392000     -3.342174000
6        1.299460000      0.461133000     -7.613201000
6        1.336300000      0.535074000     -3.418621000
6        1.294825000      2.979511000     -7.663117000
6        1.393053000      3.058408000     -3.442449000
6        1.308322000     -4.542920000     -7.527944000
6        1.376601000     -4.466463000     -3.258131000
6        3.428780000     -3.322910000     -7.566911000
6        3.447223000     -3.219785000     -3.354720000
6        3.429979000     -0.808414000     -7.609078000
6        3.460641000     -0.738890000     -3.416744000
6        3.447589000      1.712902000     -7.669620000
6        3.513502000      1.772012000     -3.455769000
6       -3.063783000      0.407358000     -7.515925000
6       -2.841710000      0.379018000     -3.266951000
6       -3.060472000      2.905619000     -7.539030000
6       -2.929415000      2.919807000     -3.315324000
6       -3.035504000     -2.073498000     -7.454422000
6       -2.971016000     -2.060536000     -3.167812000
7       -0.760792000     -0.639752000     -6.099655000
6       -0.839635000      1.748503000     -6.021827000
6       -0.849694000      4.222700000     -6.069962000
6       -0.862734000     -3.335988000     -5.922383000
6        1.266041000     -2.035278000     -5.987544000
6        1.340474000      0.499624000     -6.048201000
6        1.331554000      3.010067000     -6.087334000
6        1.323184000     -4.517604000     -5.956627000
6        3.430480000     -3.288767000     -6.012741000
6        3.441463000     -0.784561000     -6.061234000
6        3.480242000      1.740741000     -6.120838000
6       -2.988770000      0.402303000     -5.947992000
6       -3.009174000      2.911635000     -5.993979000
6       -3.030656000     -2.089950000     -5.899465000
6       -1.332623000      0.419045000     -3.928959000
6       -1.620466000      0.437877000     -8.083696000
6       -1.520133000      2.984476000     -3.906170000
6       -1.636241000      2.952909000     -8.129294000
6       -1.715800000     -2.359446000     -3.873594000
6       -1.624880000     -2.103935000     -8.038347000
6        0.490586000     -0.624581000     -3.965436000
6        0.564435000     -0.811026000     -8.101059000
6        0.665134000      1.801698000     -3.936574000
6        0.556661000      1.703987000     -8.142255000
6        0.622406000      4.268287000     -3.991237000
6        0.528816000      4.198305000     -8.219827000
6        0.576780000     -3.276924000     -3.822582000
6        0.536116000     -3.348649000     -8.079625000
6        2.705677000     -1.973538000     -3.910649000
6        2.727737000     -2.073171000     -8.117071000
6        2.788481000      0.513322000     -3.968470000
6        2.740612000      0.444881000     -8.172183000
6        2.815013000      3.012537000     -4.021379000
6        2.724679000      2.940187000     -8.244785000
6        2.766662000     -4.484353000     -3.872022000
6        2.710346000     -4.568329000     -8.128575000
6       -3.556707000      1.645846000     -3.836271000
6       -3.759229000      1.663330000     -8.076673000
6       -3.615880000     -0.836470000     -3.775580000
6       -3.750061000     -0.863659000     -8.042977000
6       -0.827666000     -5.869914000     -5.308050000
6       -3.029165000     -4.614862000     -5.280919000
6       -5.189722000      1.654336000     -5.324183000
6       -5.232749000     -0.827727000     -5.302619000
6       -5.222229000     -3.332012000     -5.254773000
6       -1.645388000     -4.684533000     -7.487879000
6       -1.569822000     -4.683990000     -3.140885000
6       -3.826083000     -3.435263000     -7.455624000
6       -3.779703000     -3.414529000     -3.112167000
6       -5.947067000      0.412450000     -7.444944000
6       -5.807267000      0.463671000     -3.145219000
6       -5.936120000     -2.109456000     -7.419654000
6       -5.838251000     -2.047422000     -3.096481000
6        0.533004000     -5.889849000     -7.494243000
6        0.617772000     -5.831133000     -3.183026000
6       -1.624227000     -4.713956000     -5.908652000
6       -3.833984000     -3.443752000     -5.877540000
6       -5.880526000      0.428037000     -5.893192000
6       -5.902280000     -2.088297000     -5.853257000
6        0.556547000     -5.877379000     -5.942276000
6       -0.794931000     -5.852525000     -3.747620000
6       -0.861469000     -5.850140000     -8.096533000
6       -3.000208000     -4.601001000     -3.710471000
6       -3.056472000     -4.608736000     -8.055894000
6       -5.121527000      1.654309000     -3.778960000
6       -5.321168000      1.645378000     -8.085739000
6       -5.189575000     -0.807219000     -3.719388000
6       -5.319007000     -0.850382000     -8.026598000
6       -5.191957000     -3.313050000     -3.681982000
6       -5.239657000     -3.342925000     -8.019984000
6       -1.452094000      0.492558000     -1.195390000
6       -1.475802000      3.034975000     -1.263120000
6       -1.471740000     -1.993856000     -1.087796000
6        0.698026000      1.801069000     -1.312182000
6        0.679519000      4.303515000     -1.319371000
6        0.680981000     -3.231047000     -1.142040000
6        2.849892000     -1.947525000     -1.284408000
6        2.859260000      0.553613000     -1.336402000
6        2.870153000      3.050243000     -1.351836000
6        2.865417000     -4.439873000     -1.165385000
6       -3.593934000      1.755490000     -1.154452000
6       -3.588209000     -0.760632000     -1.061781000
6       -0.784848000     -0.766985000     -1.629843000
6       -0.757606000      1.757831000     -1.791102000
6       -0.760239000      4.265750000     -1.847176000
6       -0.759302000     -3.228073000     -1.645888000
6        1.359573000     -1.960437000     -1.757868000
6        1.384104000      0.531816000     -1.831160000
6        1.416840000      3.065496000     -1.866661000
6        1.434888000     -4.440169000     -1.688300000
6        3.535177000     -3.209149000     -1.810899000
6        3.528702000     -0.709908000     -1.874973000
6        3.552069000      1.797487000     -1.912215000
6       -2.908036000      0.474534000     -1.694124000
6       -2.923156000      2.975601000     -1.774319000
6       -2.910964000     -1.998742000     -1.618370000
6       -0.694777000     -5.742234000     -0.959188000
6       -2.899754000     -4.506665000     -0.933397000
6       -5.156591000      1.752777000     -1.046644000
6       -5.156044000     -0.740649000     -0.999868000
6       -5.086140000     -3.236354000     -0.915112000
6       -1.516400000     -4.603337000     -1.565374000
6       -3.701476000     -3.354561000     -1.538301000
6       -5.813788000      0.498326000     -1.597685000
6       -5.803066000     -2.018771000     -1.529450000
6        0.668790000     -5.791268000     -1.632477000
6        0.653865000     -0.710762000     -1.240086000
6        0.585019000     -0.727200000     -5.513394000
6       -0.680508000      1.841512000      0.816413000
6       -0.681321000      4.328359000      0.819963000
6       -0.631951000     -3.132760000      1.102377000
6        1.502413000     -1.903289000      0.837349000
6        1.487095000      0.597579000      0.773680000
6        1.479808000      3.102184000      0.729992000
6        1.555191000     -4.353040000      1.055394000
6        3.679504000     -3.139126000      0.834607000
6        3.632482000     -0.645968000      0.777041000
6        3.629912000      1.854385000      0.749038000
6       -2.846796000      0.575032000      0.960160000
6       -2.838461000      3.055719000      0.919459000
6       -2.821256000     -1.892093000      1.156154000
6       -1.418996000      0.574712000      0.403925000
6       -1.431145000      3.088760000      0.321745000
6       -1.427397000     -1.969012000      0.516838000
6        0.760328000     -0.670466000      0.348038000
6        0.746099000      1.834870000      0.262250000
6        0.726878000      4.336323000      0.235011000
6        0.753785000     -3.189139000      0.436381000
6        2.936016000     -1.912614000      0.291909000
6        2.926095000      0.591380000      0.241137000
6        2.912941000      3.083777000      0.202005000
6        2.942749000     -4.403090000      0.393370000
6       -3.538996000      1.810362000      0.421579000
6       -3.540292000     -0.708275000      0.507583000
6       -1.399051000     -4.484913000      1.175520000
6       -3.595048000     -3.244321000      1.203782000
6       -5.668239000      0.578888000      1.191374000
6       -5.705548000     -1.936397000      1.209525000
6        0.797332000     -5.695900000      1.172182000
6       -0.622843000     -5.677876000      0.594442000
6       -2.827818000     -4.432930000      0.614906000
6       -5.084754000      1.801279000      0.506396000
6       -5.101240000     -0.694735000      0.576103000
6       -5.028954000     -3.182860000      0.645180000
6       -0.671357000     -0.667432000      0.901438000
6       -1.647634000      0.404367000     -9.649888000
6       -1.665226000      2.926212000     -9.713642000
6       -1.651910000     -2.124376000     -9.623016000
6        0.522880000      1.669907000     -9.711520000
6        0.499779000      4.167869000     -9.777418000
6        0.521012000     -3.367016000     -9.656823000
6        2.715925000     -2.093963000     -9.693521000
6        2.710130000      0.416826000     -9.749098000
6        2.689362000      2.911368000     -9.801834000
6        2.702986000     -4.587660000     -9.691316000
6       -3.779240000      1.646053000     -9.647662000
6       -3.778886000     -0.877591000     -9.610469000
6       -0.925043000     -0.855823000    -10.121717000
6       -0.935050000      1.652260000    -10.177860000
6       -0.934756000      4.138027000    -10.299196000
6       -0.900432000     -3.319560000    -10.224979000
6        1.248040000     -2.099383000    -10.145740000
6        1.237876000      0.405190000    -10.194089000
6        1.231818000      2.916032000    -10.263535000
6        1.281936000     -4.555679000    -10.279460000
6        3.418560000     -3.346891000    -10.227618000
6        3.382630000     -0.847182000    -10.271365000
6        3.378507000      1.660900000    -10.342171000
6       -3.105651000      0.382331000    -10.147765000
6       -3.102091000      2.864581000    -10.240603000
6       -3.076720000     -2.085456000    -10.217312000
6       -0.869802000     -5.848045000     -9.651564000
6       -3.065711000     -4.601857000     -9.606930000
6       -5.329690000      1.629411000     -9.646005000
6       -5.340224000     -0.866784000     -9.606329000
6       -5.260877000     -3.353287000     -9.580189000
6       -1.669017000     -4.673047000    -10.231201000
6       -3.856034000     -3.434619000    -10.198477000
6       -5.943075000      0.378265000    -10.244978000
6       -5.969536000     -2.133373000    -10.160246000
6        0.516123000     -5.898202000    -10.302111000
6        0.532353000     -0.841691000     -9.656765000
\end{verbatim}

\bibliography{Lonsdaleite,NVLons}